\documentclass[]{raa}           
\usepackage{graphicx,times}
\usepackage{natbib}
\usepackage{amssymb,amsmath}
\usepackage{xspace}
\usepackage[T1]{fontenc}
\bibpunct{(}{)}{;}{a}{}{,}

\usepackage[a4paper=true,pagebackref=true]{hyperref}
\hypersetup{pdftitle = The title of my PDF, pdfauthor = My name, pdfsubject= The subject, pdfkeywords = keyword1 keyword2 keyword3} 
\hypersetup{colorlinks = true, linkcolor = green, anchorcolor = red, citecolor = blue, filecolor = red, pagecolor = red, urlcolor = red}

\newcommand{\uvec}[1]{ \widehat{\mathbf{#1}} }

\newcommand{\parb}{ \frac{\partial\uvec{B}}{\partial\mathbf{r}} } 
\newcommand{\kms}{~km~s$^{-1}$\xspace} 
\begin{document}

\title{Magnetic Flux Ropes in the Solar Corona: Structure and Evolution toward Eruption $^*$
\footnotetext{\small $*$ Supported by the National Natural Science Foundation of China.}
}

 \volnopage{ {\bf 2020} Vol.\ {\bf 20} No. {\bf 10} id.165}
   \setcounter{page}{1}

   \author{Rui Liu\inst{1,2,3}  }

   \institute{CAS Key Laboratory of Geospace Environment, Department of Geophysics and Planetary Sciences, University of Science and Technology of China, Hefei, Anhui 230026, China;  {\it rliu@ustc.edu.cn} \\
   \and
   CAS Center for Excellence in Comparative Planetology, Hefei, Anhui 230026, China \\
   \and 
   Mengcheng National Geophysical Observatory, University of Science and Technology of China, Mengcheng, Anhui 233500, China \\	
	\vs \no
	{\small Received 2020 May 10; accepted 2020 June 28}
	}

\abstract{Magnetic flux ropes are characterized by coherently twisted magnetic field lines, which are ubiquitous in magnetized plasmas. As the core structure of various eruptive phenomena in the solar atmosphere, flux ropes hold the key to understanding the physical mechanisms of solar eruptions, which impact the heliosphere and planetary atmospheres. Strongest disturbances in the Earth's space environments are often associated with large-scale flux ropes from the Sun colliding with the Earth's magnetosphere, leading to adverse, sometimes catastrophic, space-weather effects. However, it remains elusive as to how a flux rope forms and evolves toward eruption, and how it is structured and embedded in the ambient field. The present paper addresses these important questions by reviewing current understandings of coronal flux ropes from an observer's perspective, with emphasis on their structures and nascent evolution toward solar eruptions, as achieved by combining observations of both remote sensing and in-situ detection with modeling and simulation. It highlights an initiation mechanism for coronal mass ejections (CMEs) in which plasmoids in current sheets coalesce into a `seed' flux rope whose subsequent evolution into a CME is consistent with the standard model, thereby bridging the gap between microscale and macroscale dynamics.
\keywords{magnetic fields --- magnetic reconnection --- Sun: magnetic fields --- Sun: corona --- Sun: coronal mass ejections (CMEs) --- Sun: flares --- Sun: filaments, prominences}
}

   \authorrunning{R. Liu }            
   \titlerunning{Magnetic Flux Ropes in the Solar Corona}  
   \maketitle

%
\section{Introduction}  \label{sect:intro}
Large-scale ordered magnetic fields are ubiquitous in plasmas permeating the universe \citep{Schrijver&Zwaan2000, Beck2012, Blackman2015}. Among them, helical magnetic fields have attracted great interest in diverse areas: they play important roles in fundamental physical processes such as magnetic reconnection and particle acceleration \citep[e.g.,][]{Shibata&Tanuma2001,Drake2006,Daughton2011}; they are important agents in shaping the dynamics of the solar corona \citep[e.g.,][]{Rust&Kumar1996}, of the heliosphere \citep[e.g.,][]{Burlaga1981}, and of the Earth's magnetotail \citep[e.g.,][]{Slavin2003}, in coupling the interplanetary and planetary magnetic fields \citep[e.g.,][]{Russell&Elphic1979}, and in propelling astrophysical jets with scales up to thousands of light years \citep[e.g.,][]{Marscher2008}. Additionally, according to the theory of plasma relaxation, a system with a fixed amount of magnetic helicity is destined to relax into a force-free, minimum-energy state of helical fields to the largest scale available \citep{Taylor1974,Taylor1986,Blackman2015}.

\begin{figure}[ht!]
	\centering
	\includegraphics[width=\hsize]{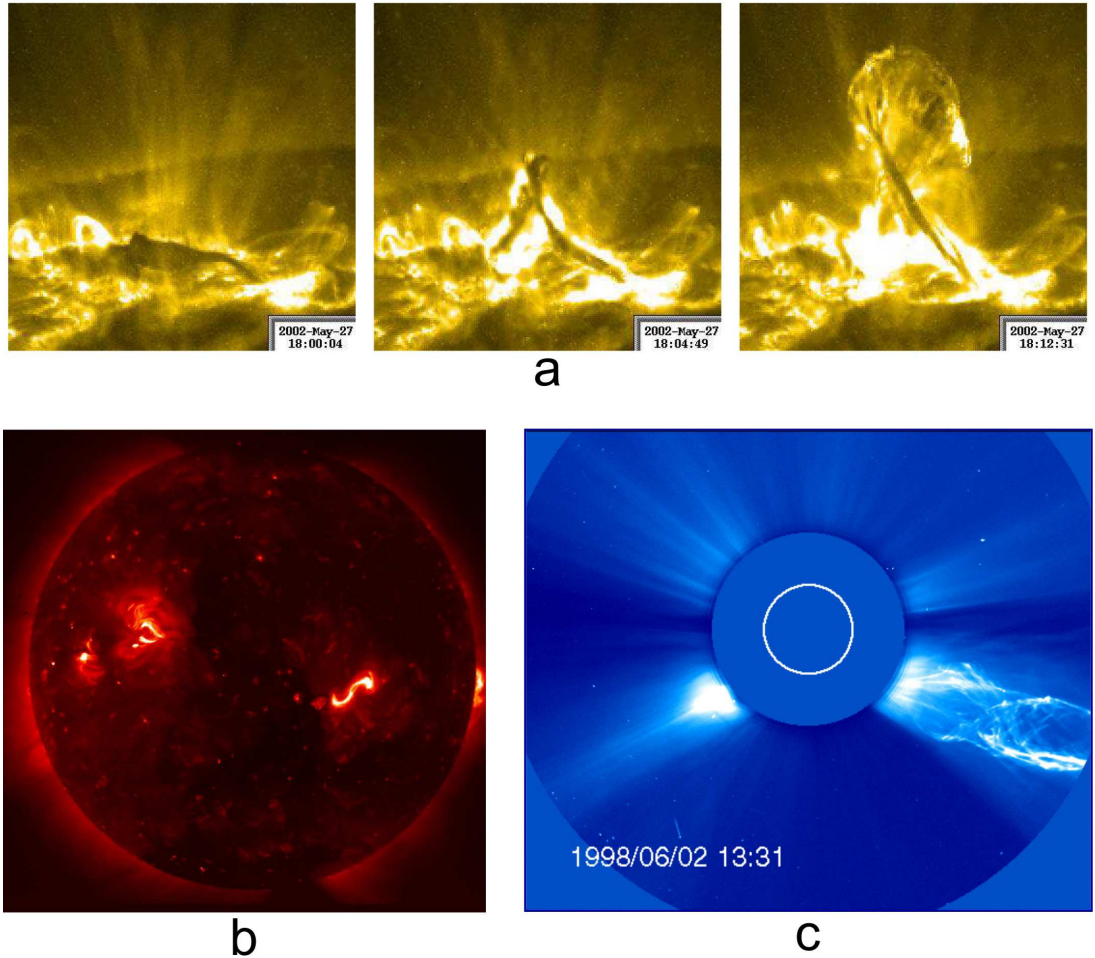} 
	\caption{Exemplary helical structures observed in the solar atmosphere. \textbf{a)} Evolution of a filament observed at 195~{\AA} by the Transition Region and Coronal Explorer (TRACE). The low-lying, dark filament first transforms into an inverted $\gamma$ and then an inverted $\delta$ shape through a rotation and writhing of the filament spine, which is a hallmark of the helical kink instability \citep{Ji2003,Torok&Kliem2005,Gilbert2007}. \textbf{b)} A full-Sun X-ray image from the X-Ray Telescope (XRT) on-board Hinode. A bright S-shaped structure is visible in the southwest quadrant. \textbf{c)} A coronal mass ejection (CME) observed by the Large Angle and Spectrometric Coronagraph (LASCO) on-board the Solar and Heliospheric Observatory (SOHO). A occulting disk obscures bright light from the photosphere and the white circle in the center indicates the size and location of the solar disk. \label{fig:helical}} 	
\end{figure}

Particularly, helical magnetic fields are observed to be systematically present in the solar atmosphere and to exhibit certain recurring patterns, e.g., the spiral shapes of sunspot fibrils \citep{Hale1927}, helical-shaped filaments \citep[e.g., Figure~\ref{fig:helical}a;][]{Rust&Kumar1994,Pevtsov2003,Gilbert2007}, sigmoidal-shaped coronal X-ray or EUV emissions \citep[e.g., Figure~\ref{fig:helical}b;][]{Rust&Kumar1996,Canfield1999,Sterling2000}, and interplanetary magnetic clouds \citep[MCs;][]{Rust1994}. Interpretations of the sense of magnetic helicity in these observed structures have revealed a hemispheric helicity rule whereby patterns of negative helicity occur predominantly in the northern solar hemisphere, and those of positive helicity in the south \citep{Pevtsov&Balasubramaniam2003,Pevtsov2014}. 

Often the term ``magnetic flux rope''  or ``flux rope'' is used to refer to a group of helical field lines collectively winding around a common axis. The proximity of the Sun makes the solar atmosphere an ideal laboratory to study the physics of flux ropes. A prodigious amount of data at multi-wavelengths, high cadence, and high resolution have been systematically collected for over the last fifty years. However, to explain the genesis of such an organized, coherent structure in the solar corona is a long-standing challenge, largely owing to the fact that we are still unable to properly measure the three-dimensional distribution of coronal magnetic fields. Further, in contrast to the coherency observed in coronal flux ropes, their footpoints `anchored' in the dense photosphere are subject to turbulent shuffling motions due to the convection and granulation whose temporal and spatial scales are much smaller than those of coronal flux ropes \citep{Stein2012}. 

The size of coronal flux ropes spans quite a few orders of magnitude: flux ropes associated with coronal mass ejections (CMEs) are comparable in size as the Sun ($10^{7-8}$ km) and can retain their coherency when propagating through the Earth and beyond \citep{Webb&Howard2012}; mini flux ropes in coronal jets may span only a few to tens of arcsecs \citep[$10^{3-4}$ km;][]{Patsourakos2008,Sterling2015}; plasma blobs of scales $10^{4-5}$ km flowing intermittently along ray-like structures in the wake of CMEs \citep{Lin2008} or above helmet streamers \citep{Sheeley2009,Rouillard2010,Rouillard2011} are believed to be small flux ropes formed and ejected through magnetic reconnection in the rays. Similarly, interplanetary flux ropes have a diverse size distribution. Magnetic clouds (MCs) typically lasts one day (or about 0.1 AU) at the Earth's orbit, as compared with much smaller flux ropes whose durations range from tens of minutes to a few hours \citep{Cartwright&Moldwin2008,ChenY2019}. 

A flux rope's magnetic twist implies that it possesses field-aligned electric currents inside the rope in the low-$\beta$ coronal environment. It has been debated whether coronal flux tubes are isolated and therefore current-neutralized \citep{Melrose1995, Parker1996, Melrose1996, Melrose2017}. In case of neutralization, the current flowing in the corona as expected from a twisted or sheared magnetic flux tube, also known as `direct current', is completely canceled by a `return current' that flows in the opposite direction around the tube, supposedly at its surface, which shields the ambient field from the direct current, therefore suppressing any current-driven instabilities. However, both observation \citep[e.g.,][]{Georgoulis2012,Cheng&Ding2016,LiuY2017} and numeric modeling \citep[e.g.,][]{Torok2014,Dalmasse2015} are against current neutralization. In case of non-neutralization, two mechanisms might be at work to produce the twisted fields: they can be twisted by photospheric and sub-photospheric flow motions \citep{Klimchuk&Sturrock1992,Torok&Kliem2003, Yan2015, Dalmasse2015}, or transported into the corona through the emergence of current-carrying flux tubes \citep{Leka1996, Longcope&Welsch2000, Fan2001, Torok2014}. With the measurements of photospheric transverse magnetic fields becoming more reliable, it has been revealed that electric currents tend to be non-neutralized in flare- and CME-producing active regions \citep{Wheatland2000, Georgoulis2012, LiuY2017, Kontogiannis2017}, especially when magnetic shear is present around polarity inversion lines (PILs). Although the controversy has not been completely settled, many studies support that current-driven instabilities and current-channel interactions are important triggering mechanisms for solar eruptions. 

In this review, we will first introduce how to identify flux ropes by quantifying magnetic connectivty and magnetic twist (\S~\ref{sec:quant}), and recapitulate key observational and modeling results relevant to flux ropes in the solar corona, particularly in solar eruptions (\S~\ref{sec:eruption}). We then focus on what we have learned about how a flux rope forms and evolves toward eruption in the corona (\S~\ref{sec:formation}), and how a flux rope is structured (\S~\ref{sec:struc}), including its twist profile (\S~\ref{subsec:struc:twist}) and boundary structure (\S~\ref{subsec:struc:boundary}), as well as more complex configurations such as `double-deckers' (\S~\ref{subsec:struc:double-decker}). We also draw the reader's attention to other reviews devoted to magnetic flux ropes in the solar atmosphere as well as in interplanetary space, including, but not limited to, \citet{Russell1990,Marubashi2000,Low2001,Demoulin2008,Linton&Moldwin2009,Filippov2015,Cheng2017,ChenJ2017,Gibson2018,Wang&Liu2019}.

\section{Quantification and Identification} \label{sec:quant}
\subsection{Magnetic Topology} \label{subsec:topology}
\begin{figure}[ht!] 
	\centering
	\includegraphics[width=\hsize]{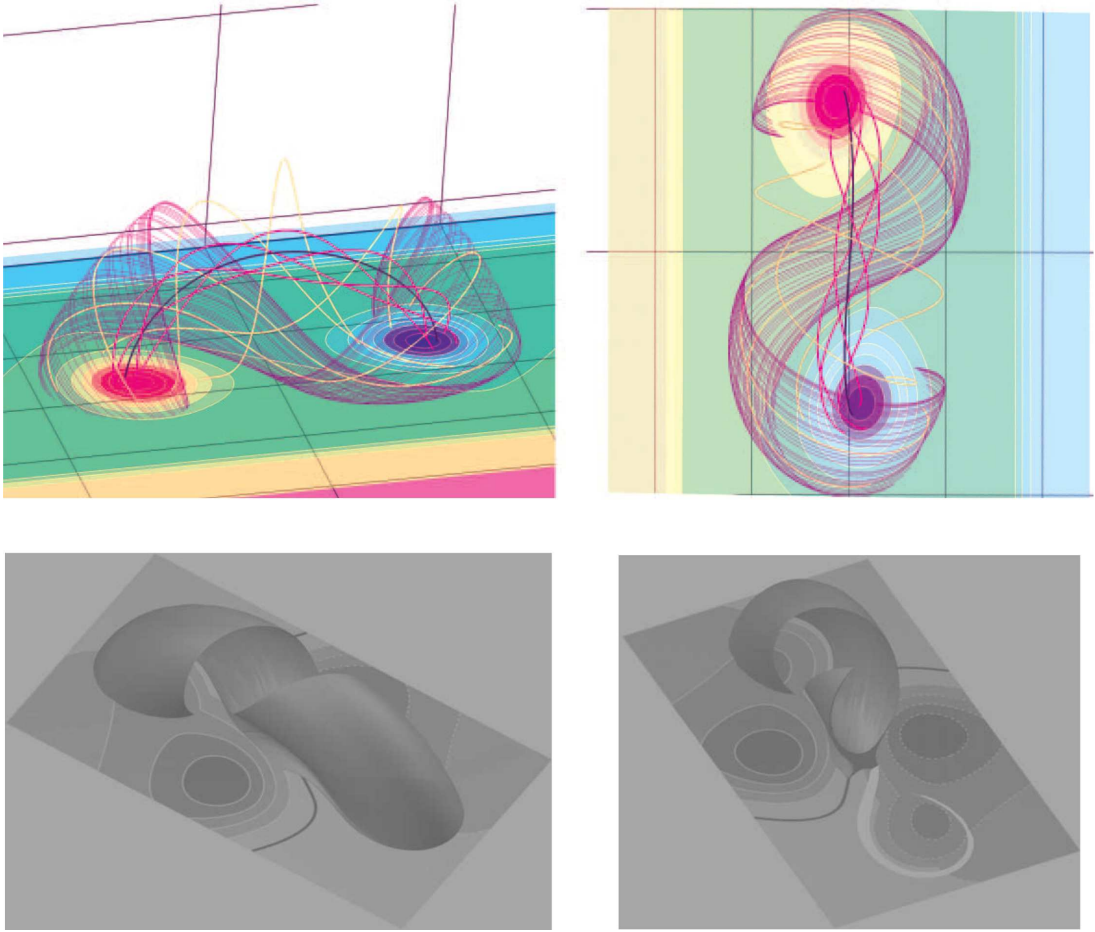} 
	\caption{Topological structures associated with a flux rope embedded in a potential field. \textbf{Top}: a bald patch separatrix surface (BPSS) in different perspectives \citep[from][]{Gibson2004}. The BPSS is made up of the field lines (magenta) tangent to the photophsere at the bald patch (BP) points, and wraps around other sample flux-rope field lines (red and yellow). \textbf{Bottom}: a hyperbolic flux tube (HFT; left) and the corresponding cross section in the center \citep[right; from][]{Titov2007}. \label{fig:topology}} 	
\end{figure}

Despite their ubiquitous presence in plasmas, flux ropes have not been quantitatively defined. The term can loosely refer to any type of helical fields in the literature. Here we adopt a qualitative definition that is generally accepted by the solar community, i.e., a group of helical field lines collectively winding around a common axis. This simple description, however, has two important implications in regard to the coherence of the structure: 1) magnetic field lines inside the flux rope share similar orientations and are anchored in similar places at the photosphere, i.e., they have similar magnetic connectivities; 2) these fields lines hence have distinct magnetic connectivities than those surrounding the rope, i.e., a magnetic boundary may be present to separate the twisted field of a flux rope from its surrounding untwisted field. 

Indeed, a flux rope whose underside is attached to the photosphere is wrapped around by a bald patch separatrix surface \citep[BPSS;][top panels in Figure~\ref{fig:topology}]{Titov&Demoulin1999,Gibson&Fan2006mfr}. Separatrix surfaces define the boundaries of topologically distinct domains, and magnetic field lines threading a BPSS are tangent to the sections of the photospheric PIL called ``bald patches'', where $(\mathbf{B}\cdot\nabla)B_z>0$ \citep{Titov1993}. On the other hand, a flux rope suspended in the corona is wrapped around by a hyperbolic flux tube (HFT; bottom panels in Figure~\ref{fig:topology}), which is composed of two intersecting quasi-separatrix layers (QSLs), thin volumes across which field lines are drastically different in terms of magnetic connectivity. The HFT displays an X-shaped cross section beneath the flux rope \citep{Titov2007,Aulanier2010}, hence is considered as the three-dimensional counterpart of the two-dimensional X-type magnetic null. When a BPSS flux rope rises in altitude, the BPSS configuration is transformed to HFT \citep[e.g.,][]{Titov2007,Aulanier2010}. Both BPSS and HFT are the preferential sites for the formation of current-sheets, which can be driven by shearing motions of magnetic-field footpoints at the photosphere \cite[e.g.,][]{Low1987,Titov2003} or induced by MHD instabilities such as the helical kink instability \citep[e.g.,][see also \S~\ref{subsec:model}]{Fan&Gibson2004,Torok2004}. 

To understand the magnetic connectivities in an active region, one typically extrapolate the photospheric field into the higher solar atmosphere, because it is still impossible to measure the full three-dimensional distribution of the magnetic field from above the photosphere to the corona. Most extrapolations invoke the force-free assumption, which neglects non-magnetic forces in the low-$\beta$ corona, and consequently the Lorentz-force must also vanish in equilibrium. Compared with potential and linear force-free field extrapolations, the nonlinear force-free field (NLFFF) model is a more realistic approach by taking the force-free parameter $\alpha=\nabla\times\mathbf{B}/\mathbf{B}$ as a function of position. To understand the evolution of an active region, one may build a series of NLFFF models \citep[e.g.,][]{Liu2016}. The force-free assumption may not be valid during the impulsive phase of the flares, when plasmas are accelerated primarily by Lorentz forces; but the flare-related changes of the coronal field can be inferred from a comparison of the NLFFF before and after the flare. For a self-consistent description of the plasma and magnetic field, however, one must relax the force-free assumption and turn to magnetohydrostatic or magnetohydrodynamic models \citep[see the review by][]{Inoue2016,Wiegelmann2017}. 

Obviously, a flux rope may exist in any but the potential model of the coronal magnetic fields, but to identify and study the rope in a quantifiable manner, one must first quantify the magnetic connectivity and the magnetic twist, which are explicated below. 

\begin{figure}[ht!] 
	\centering
	\includegraphics[width=\hsize]{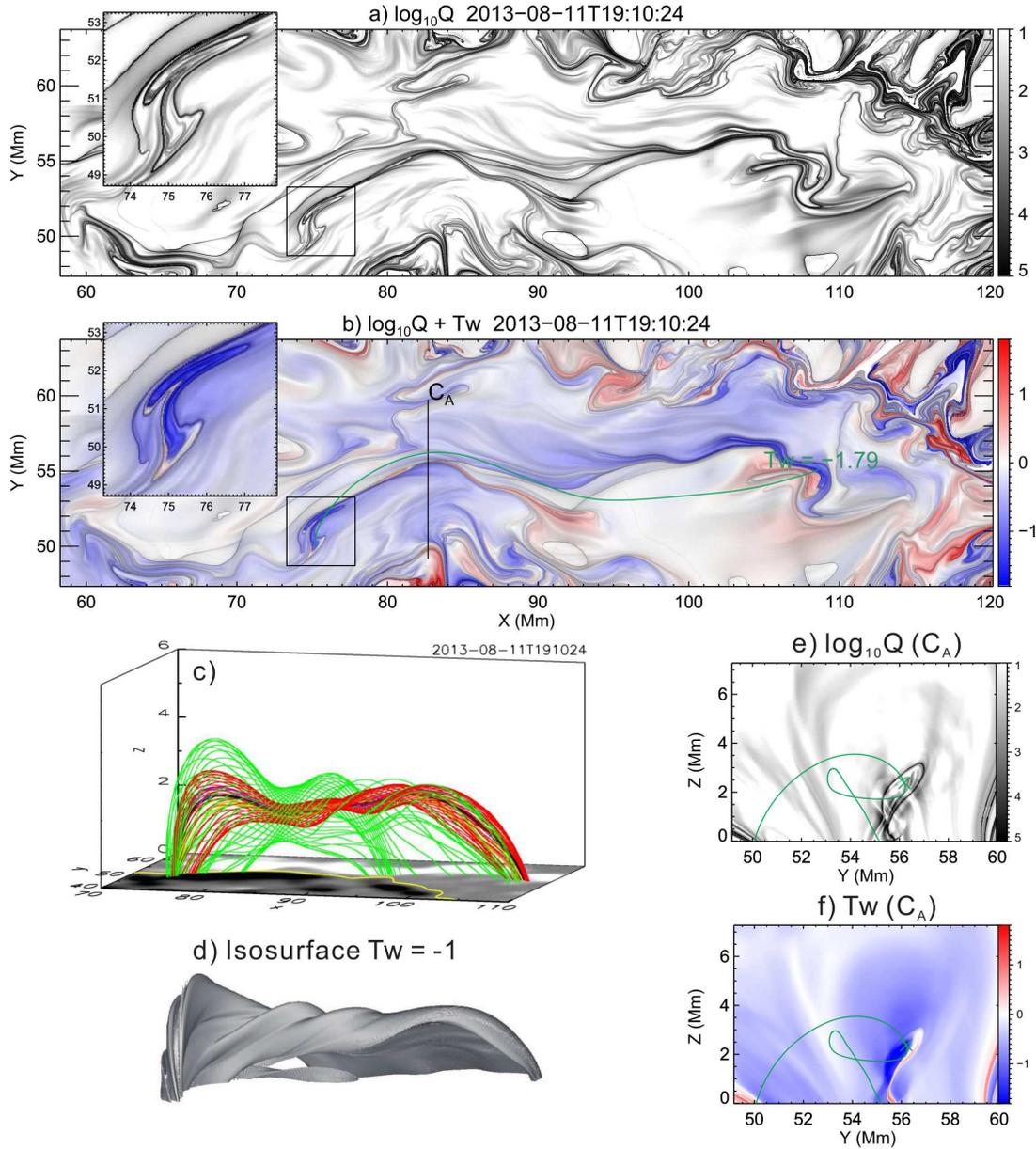} 	
	\caption{A flux rope identified in the NOAA active region 11817 before a C2.1-class flare on 2013 August 11 \citep[adapted from][]{Liu2016}. \textbf{a)} Distribution of $\log Q$ above 1 (white) and saturated at 5 (black) at the photosphere. \textbf{b)} Twist number saturated at $\pm1.8$ and blended with the same distribution of $\log Q$ as in (a). In (a) and (b) a rectangle marks the flux rope's eastern feet. This rectangular region is enlarged and redisplayed in the upper left corner. \textbf{c)} Three-dimensional perspective of the flux rope shown in field lines. Rainbow-colored field lines are in the vicinity of the axis (black). Red and green field lines are farther away from the axis. \textbf{d)} Isosurface of $T_w = -1$ viewed from the same perspective as (d). \textbf{e)} and \textbf{f)} show $\log Q$ and $T_w$ in a cutting plane C$_A$ (marked in (b)), respectively. A representative twisted field line (dark green) of $T_w=-1.79$ is projected onto both C$_A$ (e \& f) and the photosphere (b). The symbol `x' indicates where the field line threads the cutting plane. \label{fig:tw+q}}
\end{figure}
\newpage

\subsection{Quantifying Magnetic Connectivity}

Magnetic connectivities can be quantified by the squashing factor $Q$ of elemental magnetic flux tubes \citep{Demoulin1996qsl,Titov2002,Titov2007}, which is defined through the mapping between two footpoints of a field line that threads twice a plane, usually the photosphere, i.e., $\Pi_{12}:  \mathbf{r}_1(x_1,\ y_1)\mapsto \mathbf{r}_2(x_2,\ y_2)$. With the Jacobian matrix of the mapping 
\begin{equation} \label{eq:d12} 
D_{12}=\left[\frac{\partial \mathbf{r}_2}{\partial \mathbf{r}_1}\right]= 
\begin{pmatrix}
\partial x_2/\partial x_1 &\partial x_2/\partial y_1 \\
\partial y_2/\partial x_1 & \partial y_2/\partial y_1
\end{pmatrix} \equiv 
\begin{pmatrix}
a & b \\
c & d
\end{pmatrix},
\end{equation}
the squashing factor associated with the field line is given as follows 
\begin{equation} \label{eq:q}
Q\equiv \frac{a^2+b^2+c^2+d^2}{|B_{n,1}(x_1,y_1)/B_{n,2}(x_2,y_2)|},
\end{equation}
where $B_{n,1}(x_1,y_1)$ and $B_{n,2}(x_2,y_2)$ are the components normal to the plane of the footpoints, and their ratio is equivalent to the determinant of $D_{12}$. High-$Q$ structures (typically $Q\ge100$), where the field-line mapping has a steep yet finite gradient, are referred to as quasi-separatrix layers (QSLs), whereas $Q\rightarrow\infty$ at topological structures \citep{Titov2002}. It is helpful to visualize these complex three-dimensional structures by calculating $Q$ in a three-dimensional volume box. This can be done by stacking up maps of $Q$ in uniformly spaced cutting planes \citep{Liu2016}. To calculate such a $Q$-map in each cutting plane, the chain rule of the Jacobian is employed \citep{Pariat&Demoulin2012}, e.g., if the above-mentioned field line threads a cutting plane at $\mathbf{r}_c(x_c,y_c)$, then
\begin{equation} \label{eq:qcs}
D_{12}=\left[\frac{\partial {\bf r}_2}{\partial {\bf r}_1}\right] =
\left[\frac{\partial {\bf r}_2}{\partial {\bf r}_c}\right] \times \left[\frac{\partial {\bf r}_c}{\partial {\bf r}_1}\right], 
\end{equation}
where $[\partial\mathbf{r}_c/\partial\mathbf{r}_1]$ is given by its inverse,
\begin{equation}
\left[\frac{\partial {\bf r}_c}{\partial {\bf r}_1}\right] = \frac{1}{|B_{n,c}(x_c,y_c)/B_{n,1}(x_1,y_1)|} 
\begin{pmatrix}
\partial y_1/\partial y_c & -\partial x_1/\partial y_c \\
-\partial y_1/\partial x_c & \partial x_1/\partial x_c
\end{pmatrix},
\end{equation}
so that each point on the cutting plane can be assigned a $Q$ value (see \citealt{Tassev&Savcheva2017} and \citealt{Scott2017} for an alternative implementation). To eliminate spurious high-$Q$ structures introduced by field lines touching the cutting plane, i.e., $B_{n,c}(x_c,y_c)\rightarrow 0$, an optimal solution is to apply Eq.~\ref{eq:qcs} to a local plane perpendicular to the field line under question \citep{Titov2007,Liu2016}.

\subsection{Quantifying Magnetic Twist} \label{subsec:twist}
Magnetic twist number measures how many turns a magnetic field line winds about the axis. For an idealized axisymmetric flux rope, this can be given by $B_\phi/rB_z$ in radians per unit length in cylindrical coordinates $(r, \phi, z)$, with $z$ along the rope axis. In practice, this approach cannot be directly applied to quantifying the intrinsic twist or `knottedness' \citep{Moffatt1969} in flux ropes that do not possess a clearly defined axis. In general, suppose that $\mathbf{x}(s)$ is a smooth, non-self-intersecting curve parametrized by the arclength $s$, and that $\mathbf{y}$ a second such curve surrounding $\mathbf{x}$ to form a ribbon, \citet[][their Eq.~12]{Berger&Prior2006} gave the number of turns that $\mathbf{y}$ makes about $\mathbf{x}$, 
\begin{equation}
\mathcal{T}_g = \frac{1}{2\pi}\int_{\mathbf{x}(s)}\uvec{T}(s)\cdot\uvec{V}(s)\times \frac{d\uvec{V}(s)}{ds}\,ds, \label{eq:Tg}
\end{equation}
which is considered as the general definition of twist number; here $\uvec{T}(s)=d\mathbf{x}/ds$ is the unit tangent vector to the axis curve, and $\uvec{V}(s)$ is a unit vector normal to $\uvec{T}(s)$ and pointing to $\mathbf{y}$ at the point $\mathbf{y}(s)=\mathbf{x}(s)+\varepsilon\uvec{V}(s)$, so that $\mathbf{y}$ is also parameterized by $s$ along the axis curve $\mathbf{x}(s)$.

\citet[][their Eq.~16]{Berger&Prior2006} gave an alternative twist number $\mathcal{T}_w$ to approximate $\mathcal{T}_g$ in the vicinity of $\mathbf{x}$ ($\varepsilon\ll1$) in a magnetic field, 
\begin{equation}  \label{eq:Tw}
\mathcal{T}_w=\frac{\mu_0}{4\pi}\int_L\frac{\mathbf{J}\cdot\mathbf{B}}{B^2}\,dl.  
\end{equation}
In a force-free field, $\nabla\times\mathbf{B}=\alpha \mathbf{B}$, so that 
\begin{equation} \label{eq:Tw_alpha}
\mathcal{T}_w=\frac{1}{4\pi}\int_L \alpha\,dl. 
\end{equation} 
In particular, for a cylindrically symmetric flux tube of length $L_z$, it is well known that the twist number about the axis $z$ is
\begin{equation}  \label{eq:N}
\mathcal{N}(r) = \frac{1}{2\pi}\frac{L_zB_\phi(r)}{rB_z(r)}. 
\end{equation}

\citet{Liu2016} concluded that $\mathcal{T}_g$ is the generalization of $\mathcal{N}$, and $\mathcal{T}_w$ approaches $\mathcal{T}_g$ in the vicinity of the axis of a nearly cylindrically symmetric flux tube, but deviates otherwise. Below we derive succinctly the relations among these three twist numbers; readers are referred to Appendix C in \citet{Liu2016} for details. 

\subsubsection{Relations among $\mathcal{T}_g$, $\mathcal{T}_w$, and $N$} \label{subsubsec:Tg&Tw}
To clarify the relationship between $\mathcal{T}_g$ (Eq.~\ref{eq:Tg}) and $\mathcal{T}_w$ (Eq.~\ref{eq:Tw}), we need express $d\uvec{V}/ds$ in terms of physical quantities, in this case, the magnetic field. Obviously $\uvec{T}(s) = \uvec{B}(\mathbf{x}(s)) = \mathbf{B}/B$ for magnetic field lines. The distance between $\mathbf{x}$ and $\mathbf{y}$ at point $s$, $\delta\mathbf{r}(s) = \varepsilon(s)\uvec{V}(s) = \mathbf{y}(s)-\mathbf{x}(s)$, changes at a rate 
\begin{equation*}
\frac{d}{ds}\delta\mathbf{r} = \frac{d\mathbf{y}}{ds}-\frac{d\mathbf{x}}{ds}.
\end{equation*} 
Given the arclength $s^\prime$ and unit tangent vector $\uvec{T}^\prime$ at $\mathbf{y}$, we can rewrite $d\mathbf{y}/ds = d\mathbf{y}/ds^\prime \cdot ds^\prime/ds = \uvec{T}^\prime \cdot ds^\prime/ds = \uvec{B}(\mathbf{y}(s)) \cdot ds^\prime/ds$. For $\varepsilon\ll 1$, $ds^\prime \approx ds$, so that 
\begin{equation*}
\frac{d}{ds}\delta\mathbf{r} \simeq \uvec{B}(\mathbf{x}(s) + \delta\mathbf{r}(s)) -\uvec{B}(\mathbf{x}(s))\simeq \delta\mathbf{r} \cdot \frac{\partial\uvec{B}}{\partial\mathbf{r}}.
\end{equation*} 
Thus, 
\begin{equation}
\frac{d \uvec{V}}{ds} \simeq \uvec{V} \cdot \frac{\partial\uvec{B}}{\partial\mathbf{r}} - \frac{1}{\varepsilon}\frac{d\varepsilon}{ds}\uvec{V}. \label{eq:dvds}
\end{equation}
Inserting Eq.~\ref{eq:dvds} into the local density of $\mathcal{T}_g$, we have 
\begin{equation*}
\frac{d\mathcal{T}_g}{ds} = \frac{1}{2\pi} \uvec{T} \cdot \uvec{V} \times \frac{d\uvec{V}}{ds} \simeq \frac{1}{2\pi} \uvec{T}\cdot \uvec{V} \times \left(\uvec{V}\cdot\parb \right). \label{eq:dTgds}
\end{equation*}
Splitting $\partial\mathbf{B}/\partial\mathbf{r}$ into symmetric and antisymmetric parts, it can be derived that  
\begin{equation*}
\left[ \uvec{V}\cdot\parb \right]_i 
= \frac{1}{B} \left(\mathcal{S}_{ij}\widehat{V}_j + \frac{1}{2}\mu_0\epsilon_{jik}\widehat{V}_j J_k\right) - \frac{\widehat{T}_i}{B}\widehat{V}_j\frac{\partial B}{\partial x_j}.
\end{equation*}
Here $\mu_0\, \mathbf{J}=\nabla\times\mathbf{B}$ and $\mathcal{S}_{ij}\equiv [\mathbb{S}]_{ij}$ denotes the symmetric part of $\partial\mathbf{B}/\partial\mathbf{r}$. Generally, $\mathbb{S} \cdot \uvec{V} = c_1\, \uvec{T} + c_2\, \uvec{V} + c_3 \, \uvec{T} \times \uvec{V}$, where the coefficients $c_1$, $c_2$, and $c_3$ depend on both $\mathbb{S}$ and $\uvec{V}$. With some vector calculus, it turns out that only the $c_3$ term of $\mathbb{S} \cdot \uvec{V}$ and the antisymmetric part of $\partial\mathbf{B}/\partial\mathbf{r}$ remain: 
\begin{equation}
\frac{d\mathcal{T}_g}{ds} \simeq \frac{c_3}{2\pi B} + \frac{\mu_0J_\parallel}{4\pi B}, \label{eq:dTgds_approx}
\end{equation}
where all quantities are taken at the axis field line $\mathbf{x}(s)$. In contrast, $\mathcal{T}_w$ (Eq.~\ref{eq:Tw}) is evaluated at the field line of interest, $\mathbf{y}(s)$. Thus, 
\begin{equation}
\lim_{\varepsilon\to0}\mathcal{T}_w(\varepsilon) = \mathcal{T}_g - \int_{\mathbf{x}(s)} \frac{c_3}{2\pi B}\,ds, \label{eq:Tg&Tw} 
\end{equation} 
which specifies two conditions for $\mathcal{T}_w$ to reliably approximate $\mathcal{T}_g$: first, the field line must be sufficiently close to the axis such that $J_\parallel/B$ on $\mathbf{x}$ and $\mathbf{y}$ are approximately equal; second, the contribution from $\mathbb{S}$ proportional to $c_3$ can be negligible, in other words, the flux rope must possess certain degree of coherence. For example, in cylindrical symmetry, $B_r = 0$, $B_\phi=B_\phi(r)$, and $B_z=B_z(r)$, all elements of $\mathbb{S}$ vanish identically except
\begin{equation*}
S_{r\phi}\equiv\frac{1}{2}r\frac{\partial}{\partial r} \left(\frac{B_\phi}{r}\right),
\end{equation*} 
which vanishes at the axis, too, for a smooth distribution of $J_\parallel(r)$. This is the case for both a constant-$\alpha$ force-free flux rope \citep{Lundquist1950} and a uniformly twisted flux rope\citep{Gold&Hoyle1960}. Therefore the smaller the ratio of the two terms in Eq.~\ref{eq:dTgds_approx}, $2c_3/\mu_0J_\parallel$, locally the closer a flux rope approaches to cylindrical symmetry. 

Now apply Eq.~\ref{eq:Tg} directly to a cylindrical flux tube: for all $s$, $\uvec{T}=\uvec{e}_z$, $\uvec{V} = \uvec{e}_r$, and 
\[\frac{d\uvec{V}}{dz} = \frac{d\phi}{dz}\,\uvec{e}_\phi.\] 
From the field-line equation in cylindrical coordinates, $dr/B_r = rd\phi/B_\phi = dz/B_z = ds/B$, we recover the classical formula for $\mathcal{N}$ (Eq.~\ref{eq:N}), 
\begin{equation*}
\mathcal{T}_g = \frac{1}{2\pi}\int d\phi = \frac{1}{2\pi}\int \frac{B_\phi(r)}{rB_z(r)}dz = \mathcal{N}(r).
\end{equation*}

\subsubsection{Application of $\mathcal{T}_w$}
$\mathcal{T}_g$ is pertinent to strict stability analyses, but it depends on the precise determination of the axis, which is both non-trivial and demanding for numerical magnetic fields. On the other hand, $\mathcal{T}_w$ can be computed straightforwardly for any field lines. Approaching with caution, one can combine a map of $\mathcal{T}_w$ and the corresponding map of squashing factor $Q$ to conveniently identify and characterize flux ropes. 

$\mathcal{T}_w$ is also useful in locating a flux rope's axis, a necessary requirement for computing $\mathcal{T}_g$. One can see that from Eq.~\ref{eq:Tw} the radial profile of $J_\parallel/B$, or $\alpha(r)$ in force-free fields, determines where $\mathcal{T}_w$ peaks in the cross section of flux ropes. For a flux rope with some degree of cylindrical symmetry, $\mathcal{T}_w$ reaches a local extremum at the axis, unless $J_\parallel/B$ is uniform around the axis. For example, $\mathcal{T}_w$ matches $\mathcal{N}$ at the axis in either a constant-$\alpha$ force-free flux rope \citep{Lundquist1950} or a uniformly twisted flux rope \citep{Gold&Hoyle1960}; but away from the axis, $\mathcal{T}_w$ overestimates (underestimates) $\mathcal{N}$ in the former (latter) case \citep[Appendix C in][]{Liu2016}. This is further checked against an approximately force-free Titov-D\'emoulin flux-rope equilibrium \citep{Titov&Demoulin1999}, using two different toroidal current density $J_\mathrm{t}(r)$, one roughly uniform, the other strongly peaked at $r=0$. $\mathcal{T}_w$ and $\mathcal{N}$ are found to agree to within 5\% at the axis. $\mathcal{T}_w$ reaches the maximum (minimum) at the axis with the peaked (uniform) $J_\mathrm{t}(r)$ \citep{Liu2016}. 

\citet{Liu2016} ran a tomography scan of a flux rope identified in the NLFFF by computing $\mathcal{T}_w$ maps in vertical cutting planes throughout the rope and tracing in each map a field line from the peak-$|\mathcal{T}_w|$ point. These field lines coincide within the limits of numerical accuracy over nearly the whole rope axis. This is further confirmed by cutting the rope perpendicularly at where it runs horizontally (e.g., at the apex point). The in-plane field vectors display a rotational pattern centered at the identified axis point, and that current density is enhanced normal to the cutting plane; both features are consistent with the existence of a flux rope \citep[][their Figure 4]{Liu2016}. Outlined by high-$Q$ lines, the flux rope displays a rather compact and vertically elongated cross section (Figure~\ref{fig:tw+q}(e \& f)). Tracing field lines from points following this shape in the cutting plane (red lines in Figure~\ref{fig:tw+q}c) or plotting the isosurface of $|\mathcal{T}_w|=1$ (Figure~\ref{fig:tw+q}d) demonstrates the three dimensional configuration of this flux rope.  

To summarize, one can define a coherent flux rope as a three-dimensional volume of enhanced $|\mathcal{T}_w|$ as enclosed by QSLs or BPSS. In the cross section of a flux rope with approximate cylindrical symmetry, the axis is located at the local extremum of the $|\mathcal{T}_w|$ map, unless $J_\parallel(r)/B(r)$ is uniformly distributed. This criteria has helped identify flux ropes of various configurations in various extrapolations  \citep[e.g.,][]{WangH2015,WangH2017,Yang2016,LiuL2017,Zhu2017,Awasthi2018,Su2018} or MHD models \citep[e.g.,][]{Guo2017,Jiang2018} of coronal magnetic fields. 

\subsubsection{Field Line Helicity}
An alternative quantity to characterize flux ropes is the field line helicity, which is given by an integral along a magnetic field line of length $L$ 
\begin{equation}
\mathcal{A}(L) = \int_L \frac{\mathbf{A}\cdot\mathbf{B}}{B}\,dl
\end{equation}
where $l$ is the field-line arclength and $\mathbf{B}=\nabla\times\mathbf{A}$. However, the vector potential $\mathbf{A}$ is gauge-dependent and nonlocal, so is $\mathcal{A}$ \citep{Yeates+Hornig2016}. Alternatively, $\mathcal{A}$ can be calculated as the limit in the infinitesimal tubular volume $D_\epsilon$ around the magnetic field line, which possesses magnetic flux $\Phi_\epsilon$ and an infinitesimal radius $\epsilon$ \citep{Berger1988},
\begin{equation}
\mathcal{A}(L)=\lim_{\varepsilon\to0}\frac{1}{\Phi_\epsilon}\int_{D_\epsilon}\mathbf{A}\cdot\mathbf{B}\,dV.
\end{equation}
Integrating over all field lines $L\supset D$ gives the total helicity $H = \int_D \mathbf{A}\cdot\mathbf{B}\,dV$ in the volume $D$. Thus, the field line helicity effectively describes how $H$ is distributed within the coronal volume, and flux ropes can be identified as concentrations of high field-line helicity in the corona \citep{Yeates+Hornig2016,Lowder+Yeates2017}. Naturally, $\mathcal{A}$ is correlated with $\mathcal{T}_w$ \citep{Yeates+Hornig2016}. This is because the helicity within the infinitesimal tubular volume $D_\epsilon$ can be written as $H_{\epsilon} = \Phi_\epsilon\mathcal{A} \simeq \Phi_\epsilon^2 \mathcal{T}_w$, if we neglect the contribution from the writhe assuming that the flux tube is not highly kinked, and if we consider only the self-helicity assuming that the field line is isolated. The same argument applies to calculating the helicity of a flux rope by its twist \citep{Guo2013}.  With these assumptions, one has 
\begin{equation}
\mathcal{A}\simeq \mathcal{T}_w\Phi_\epsilon.
\end{equation}
However, since $\mathcal{A}$ is nonlocal, it remains an open question whether $\mathcal{A}$ can precisely quantify flux ropes, which often have a definite boundary.

\section{Magnetic Flux Ropes in solar eruptions} \label{sec:eruption}
Solar flares, filament/prominence eruptions, and coronal mass ejections (CMEs) in the solar atmosphere are the most spectacular phenomena in the solar system. Colloquially, when these events occur together, as they frequently do, we refer to them as solar storms. A typical storm releases more than $10^{32}$ ergs of energy, as it ejects up to $10^{16}$ g of plasma into interplanetary space with speeds often exceeding 1000 km~s$^{-1}$, heats local coronal plasmas to temperatures in excess of 10 MK, and accelerates particles up to GeV energies. 

Magnetic field plays a dominant role in solar storms, because in the solar atmosphere where most of the disturbances take place, the typical plasma $\beta$ is less than 0.1. It has been a consensus that solar eruptive phenomena draw energy from highly stressed magnetic fields in the corona \citep{Forbes2000}. The magnetic field $\mathbf{B}$ can be always decomposed into a current-free, potential component $\mathbf{B}_p$ and a current-carrying, non-potential component $\mathbf{B}_c$, so that the magnetic energy $E_m$ in a volume $V$ can be written as \citep{Sakurai1981}
\begin{equation}
E_m=\int_V \frac{B^2}{8\pi}\,dV=\frac{1}{8\pi}\int_VB_p^2\,dV+\frac{1}{2c}\int_V \mathbf{A}_c\cdot\mathbf{J}\,dV,
\end{equation}
where $\mathbf{B}_c=\nabla\times\mathbf{A}_c$ and $\mathbf{J}=\frac{c}{4\pi}\nabla\times\mathbf{B}_c$, because $\nabla\times\mathbf{B}_p=0$. In the solar atmosphere, the first term is the energy of the potential field produced by sub-surface currents, which is inaccessible to the coronal plasma. The free energy powering solar eruptions can only be contained in the second term carrying electric currents above the surface. Indeed, the gradual buildup of free energy over days or even weeks prior to eruptions in active regions is typically manifested as the development of strong-field, strong-gradient, highly-sheared polarity-inversion lines \citep[PILs;][]{Toriumi&Wang2019}. Obviously the field around such a PIL carries significant electrical currents because a current-free field is perpendicular to the PIL. It is debatable whether such electric currents represent the presence of a flux rope before eruption (see \S~\ref{subsec:formation:theory}). 

\subsection{Observation} \label{subsec:observation}
From observational perspective, flux ropes are clearly present ``after'' CMEs, as evidenced, in particular, by in-situ detected magnetic clouds at 1~AU \citep[][see also Figure~\ref{fig:boundary}b]{Burlaga1981}, which possess a stronger, smoothly rotating magnetic field and a lower ion temperature than the ambient solar wind. Considerable efforts have been invested into developing flux rope models to characterize magnetic clouds, based on in-situ measurements of magnetic field and plasma parameters along the single-point traversing path made by spacecrafts through magnetic clouds. The approach varies from parametric fitting with different flux-rope solutions \citep[e.g.,][]{Burlaga1988,WangY2015}, including the famous linear force-free Lundquist solution \citep{Lundquist1950} with twist increasing from the axis to the boundary and the nonlinear force-free Gold-Hoyle solution with uniform twist \citep{Gold&Hoyle1960}, to the Grad-Shafranov reconstruction based on the magnetohydrostatic theory \citep{Hu&Sonnerup2002,Hu2017rev}. The geometry of models ranges from symmetric cylinder \citep[e.g.,][]{Burlaga1988}, asymmetric cylinder \citep[e.g.,][]{Mulligan&Russell2001} to torus \citep[e.g.,][]{Marubashi&Lepping2007}. \citet{Cane&Richardson2003} found that 100\% of the interplanetary counterparts of CMEs (ICMEs) detected during solar minimum were magnetic clouds, but the fraction reduces to $<20$\% during solar maximum. Some authors argue that all ICMEs contain a flux rope \citep[e.g.,][]{Gopalswamy2013,Xie2013,Hu2014}. \citet{Awasthi2018}, on the other hand, found that a complex ICME originates from a system of multiple flux ropes braiding about each other. 

Near the Sun, about 1/3 of CMEs exhibit a three-part structure of a bright loop front ahead of an emission-depleted cavity embedding a bright core, which is often attributed to prominence material \citep{Illing&Hundhausen1986}. Recent observations, however, demonstrate that the CME core can also arise from the eruption of a flux rope void of prominence material \citep{Howard2017,Veronig2018,Gou2019,Song2019}. The three-part structure can often be traced back to a coronal streamer with a teardrop-shaped cavity underneath. It was recognized early that the cavity rather than the prominence core drove the CME \citep{Hundhausen1987}. Generally, concave-upward or circular or helical features that appear before or during CMEs are believed to be consistent with the flux-rope geometry, and such CMEs are referred to as ``flux rope CMEs'' \citep{Dere1999,Krall2007,Vourlidas2013}. Taking into account loop-CMEs that exhibit a bright loop font followed by emission depletion and jet-CMEs that contain helical structures, \citet{Vourlidas2013} estimated the occurrence rate of flux-rope CMEs to be 41\%, which is close to the occurrence rate (35\%) of magnetic clouds among ICMEs \citep{Chi2016}. In particular, a streamer may gradually swell into a slow CME with a three-part structure, leaving the streamer significantly depleted in its wake. Such ``streamer blowout'' CMEs exhibit the flux rope morphology at a much higher rate (61\%) than regular CMEs \citep{Vourlidas&Webb2018}.

Below we focus on filaments and sigmoids, which are of most frequently observed CME progenitors and of most trusted flux-rope indicators on the Sun. Many observational features of filaments and sigmoids can be naturally explained by a flux-rope model. A caveat to keep in mind is that most of these features, if not all, can also be accommodated by a sheared magnetic arcade consisting of weakly twisted field lines, which wind less than a full turn about a central axis. 

\subsubsection{Filaments}  \label{subsubsec:filament}
\begin{figure}[ht!]
	\centering
	\includegraphics[width=0.8\hsize]{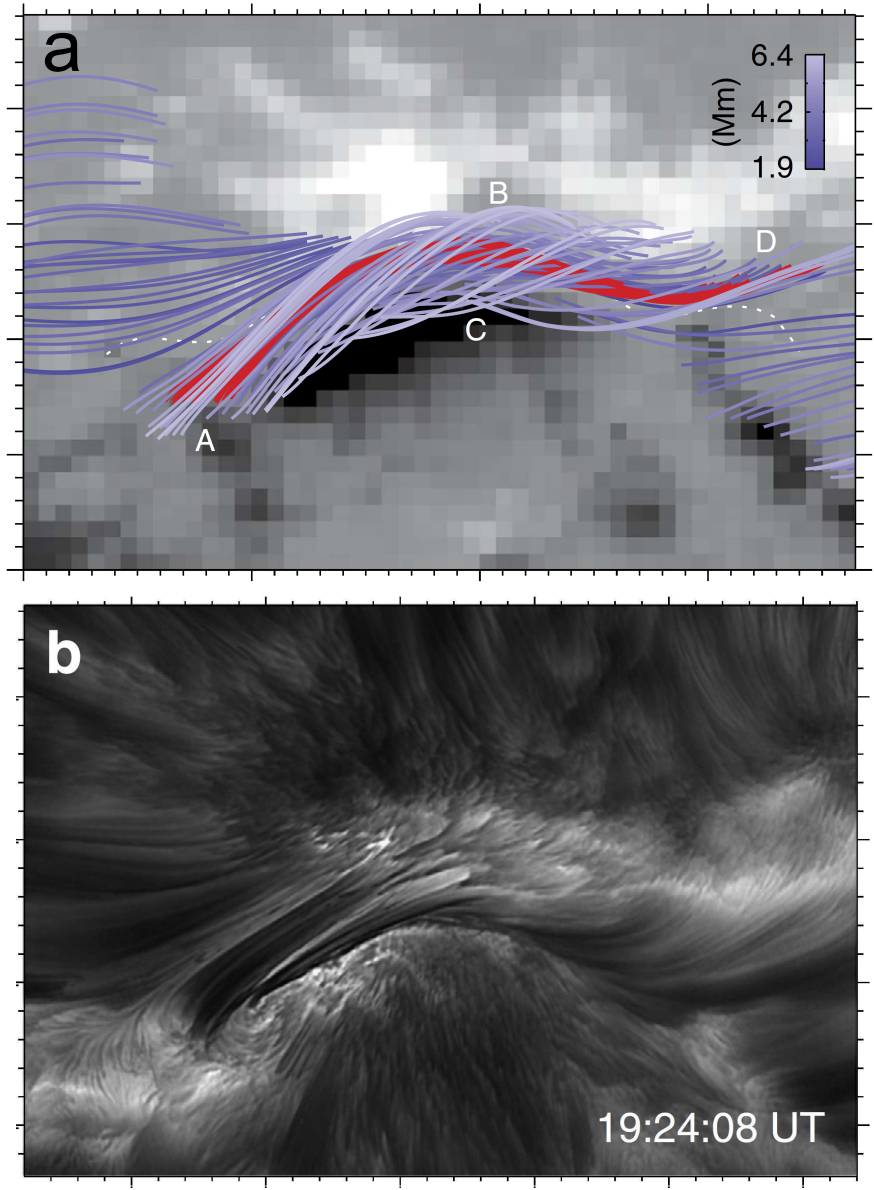}
	\caption{Extrapolated flux rope in relation to an active-region filament \citep[from][]{WangH2015}. \textbf{a)} Twisted field lines (red) surrounded by less twisted field lines (purple) from a NLFFF extrapolation model of the NOAA active region 11817 before a C2.1-class flare on 2013 August 11. The field lines are projected upon the photospheric $B_z$ distribution. \textbf{b)} The filament located within the same active region is observed with a spatial resolution as high as 60~km at the $H\alpha$ line center by the 1.6-m Goode Solar Telescope. \label{fig:filament}}	
\end{figure}

\begin{figure}[ht!]
	\centering
	\includegraphics[width=0.8\hsize]{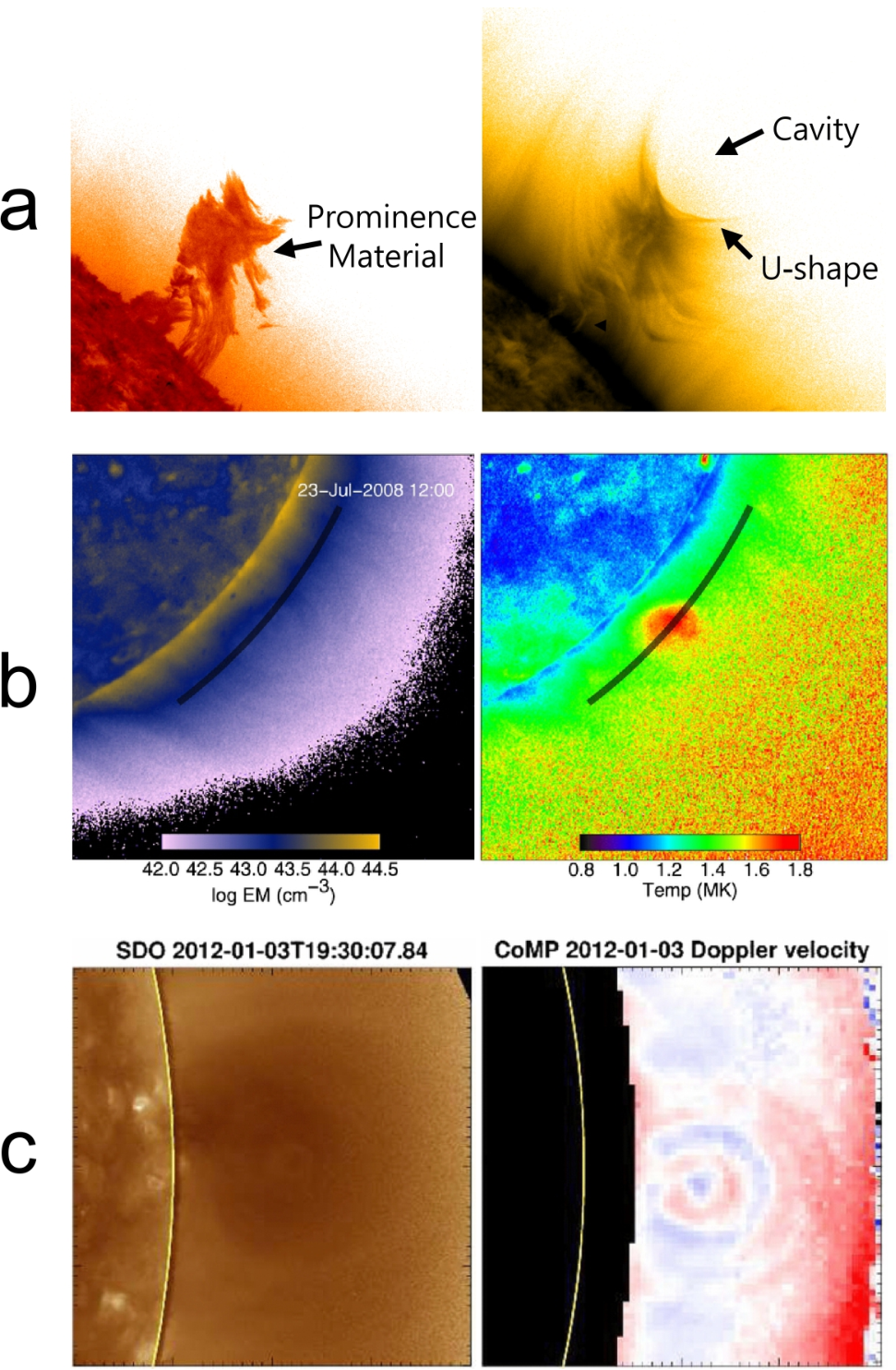}
	\caption{Structures of coronal cavities. \textbf{a)}: A cavity observed by SDO/AIA at 304~{\AA} (left) and at 171~{\AA} (right; negative image) on 2010 June 13 \citep[from][]{Regnier2011}. \textbf{b)}: A cavity observed by Hinode/XRT on 2008 July 23 \citep[from][]{Reeves2012}. The maps of emission measure (left) and temperature (right) are derived using the XRT Thin-Be/Ti-poly filter pair. \textbf{c)}: A coronal cavity observed by SDO/AIA at 193~{\AA} on 2012 January 3 (left) and the corresponding Doppler velocity pattern obtained by the Coronal Multi-Channel Polarimeter (CoMP) \citep[right; from][]{Bak-Steslicka2013}. \label{fig:cavity}}	
\end{figure}

Solar filaments are composed of dense (10$^{11-12}$ cm$^{-3}$) and cold plasma (10$^4$~K) suspended in the tenuous (10$^{8-9}$ cm$^{-3}$) and hot (10$^6$~K) corona, hence appear dark in chromospheric lines such as H$\alpha$ against the solar disk, but in emission above the limb, hence termed prominences \citep[see comprensive reviews by][]{Tandberg-Hanssen1995,Martin1998,Mackay2010,Labrosse2010,Parenti2014,Gibson2018}. Filaments are located in filament channels, where the chromospheric fibrils in H$\alpha$ are aligned with the PIL \citep{Gaizauskas1998}. These fibrils are thought to give the direction of the magnetic field in the chromosphere. Similarly, filament threads are most likely aligned with magnetic field \citep[][see also Figure~\ref{fig:filament}]{Lin2005}. In EUV, a dark corridor termed ``EUV filament channel'' is well extended in width beyond the H$\alpha$ filament. This is explained by Lyman continuum absorption of EUV radiation ($\lambda<912$~{\AA}) and ``volume blocking'', an additional reduction in EUV intensity because the cool plasma occupying the corridor does not emit any EUV radiation \citep{Anzer+Heinzel2005}. Above the limb, the dense filament material is seen at the bottom of a cavity, which is density-depleted and overarched by coronal loops (Figure~\ref{fig:cavity}). These observations imply that the filament traces only a portion of a much larger, tunnel-like structure that extends from the photosphere throughout into the low corona.

Three distinct magnetic configurations have been proposed for filaments, namely the empirical wires \citep{Martin&Echols1994, Lin2008}, the sheared magnetic arcade \citep{Kippenhahn&Schluter1957}, and the twisted flux rope \citep{Kuperus&Raadu1974}. The empirical wire model assumes that a filament is composed of field-aligned fine threads. It differs from the other two in the absence of magnetic dips. Present either at the top of a sheared arcade or the bottom of a flux rope, magnetic dips are essential in supporting dense filament material against gravity, but become less indispensable when filament material is highly dynamic \citep{Karpen2006}. The flux-rope model is appealing in that its helical windings provide for filament material both the support against gravity and the thermal insulation from the hot corona. Besides, it demonstrates structural and morphological similarities with coronal cavities \citep[\S3.3 in][see also Figure~\ref{fig:cavity}]{Gibson2018} as well as consistency with many active-region filaments \citep[e.g.,][see also Figure~\ref{fig:filament}]{Dudik2008,Canou&Amari2010,Sasso2014,Liu2014,Liu2016}. Further, it explains the inverse-polarity configuration observed often in quiescent filaments \citep{Leroy1984, Bommier&Leroy1998}, i.e., the magnetic field traversing the filament is directed from negative to positive polarity. The sheared arcade model generally implies a normal-polarity configuration, which is more often found in active-region filaments than quiescent filaments. In reality, magnetic configurations of filaments can be complicated. For example, \citet{Guo2010rope} found that a flux rope and a sheared arcade match two sections of a filament separately. A mixture of normal- and inverse-polarity dips is found in numerical experiments \citep{Aulanier2002}. To explain a `double-decker' filament that was resolved stereoscopically and later erupted partially, \citet{Liu2012} proposed two possible configurations, either a double flux rope or a single flux rope atop a sheared arcade. See \S\ref{subsec:struc:double-decker} for more details. 

The pattern of filament chirality provides an additional modeling constraint. By definition, a filament is \emph{dextral} (\emph{sinistral}) if its axial magnetic field points right (left) when a hypothetical observer is standing at the positive-polarity side of the PIL. It is believed that a dextral (sinistral) filament has right-bearing (left-bearing) barbs, a bundle of filament threads extruding out of the filament spine in a way similar to right- or left-bearing exit ramps off a highway. The majority of filaments in the northern (southern) hemisphere indeed have right-bearing (left-bearing) barbs and are overarched by left-skewed (right-skewed) coronal arcades, corresponding to the dominantly negative (positive) helicity in the same hemisphere \citep{Martin1998, Pevtsov2003, Yeates2007}. However, it is argued that the correspondence between the filament chirality and the bearing sense of barbs works only for filaments supported by flux ropes and the correspondence is reversed for sheared arcades, if the sheared arcade possesses the same sign of helicity as the flux rope \citep{Guo2010rope,Chen2014}. Alternatively, \citet{Chen2014} proposed that a filament is dextral (sinistral) if during the eruption the conjugate sites of plasma draining are right-skewed (left-skewed) with respect to the PIL. Employing this new criteria, it is found that the hemispheric rule of filament chirality is significantly strengthened \citep{Ouyang2017,ZhouZ2020}.

In equilibrium, dense filament plasmas may only trace a portion of magnetic field lines, but when disturbed, they flow dominantly along field lines in a low-$\beta$ plasma environment, thereby providing clues on the magnetic configuration of the filament \citep[e.g.,][]{Su&vanBallegooijen2013,Awasthi2019,Awasthi&Liu2019} or how it interacts with the surrounding field \citep[e.g.,][]{Liu2018}. It is also possible the observed flows represent motions of the magnetic structure itself instead of being along stationary field lines \citep{Williams2009,Okamoto2016}. It becomes even more elusive to determine the nature of mass motions in so-called tornado filaments \citep{Li2012,WangW2017apj}. Counterstreaming flows along the filament spine \citep{Schmiede1991,Zirker1998,LinY2003,WangH2018} seem to be in favor of the sheared arcade model \citep[e.g.,][]{Luna2012apjl,Alexander2013,ZhouY2020}; but within the cavity, swirling motions in plane of sky projection \citep{Wang&Stenborg2010, Li2012, Panesar2013, WangW2017apj} and line-of-sight flows that forming a bullseye pattern \citep[][see also Figure~\ref{fig:cavity}c]{Bak-Steslicka2016} are reminiscent of the nested toroidal flux surfaces of a flux rope's cross section. Writhing deformations (e.g., Figure~\ref{fig:helical}a; see also \S~\ref{subsubsec:KI}) as well as unwinding motions \citep[e.g.,][]{Yan2014,Xue2016} observed in erupting filaments also indicate the presence of magnetic twist.

Large-amplitude oscillations in filaments are also employed to probe the filament magnetic field. Often activated by shock waves impacting filaments side-on, transverse oscillations perpendicular to the filament spine are modeled by a damped harmonic oscillator with magnetic tension serving as the restoring force \citep{Hyder1966,ZhouY2018}. Longitudinal oscillations along the spine, on the other hand, are often activated by a subflare \citep[e.g.,][]{Jing2003,Jing2006} or a jet \citep[e.g.,][]{Luna2014,Awasthi2019} at one end of the filament, or, occasionally by a shock wave propagating along the filament spine \citep[e.g.,][]{Shen2014}. Various restoring forces have been considered since the discovery of this phenomenon \citep{Jing2003}, e.g., the magnetic pressure gradient \citep{Vrvsnak2007}, the gas pressure gradient \citep{Jing2003,Vrvsnak2007}, and the projected gravity in a magnetic dip \citep{Jing2003,ZhangQM2012,Luna+Karpen2012}. The first two forces have implications that are seldom observed, either predicting motions perpendicular to the local magnetic field \citep[however, see][]{ZhangQM2017} or requiring a temperature difference of several million Kelvins \citep{Vrvsnak2007}. The simple pendulum model, however, appears self-consistent and can provide magnetic parameters such as the curvature of the field-line dip and the minimum field strength \citep{Luna+Karpen2012,Luna2014,ZhouY2017,ZhouY2018}. \citet{Awasthi2019} investigated mass motions driven by a surge initiated at one end of an active-region filament, and found that the filament material predominately exhibits rotation about the spine, which is evidenced by antisymmetric Doppler shifts about the spine, and longitudinal oscillations along the spine, featuring a dynamic barb extending away from the H$\alpha$ spine until the transversal edge of the EUV filament channel. Combined together, the composite motions are consistent with a double-decker structure comprising a flux rope atop a sheared-arcade system (see also \S~\ref{subsec:struc:double-decker}).

At the base of quiescent prominences, dome-like structures termed ``bubbles'' appear dark in H$\alpha$ and Ca II but bright in EUV \citep{Berger2010}. Sometimes they are observed to emerge from underneath and expand into quiescent prominences. The arc-shaped boundary of bubbles is an active location for the formation of rising 'plumes' \citep{Berger2010, Berger2011,Berger2017}, which are suggested to be a probable source of mass supply into the prominence, as part of a ``magneto-thermal convection'' cycle to compensate the downward draining of prominence material \citep{Berger2011}. It is generally agreed that the bubble is filled with low-density plasma, which complicates the further plasma diagnostics, since most of the light we see may be coming from the foreground and background coronal plasma shining through the bubble. \citet{Gunar2014} argued that the apparent brightening is due to the prominence-corona-transition-region outside the bubble, which explains the presence of cool prominence material in the lines of sight intersecting the bubble. It is believed that the origin of prominence bubbles is associated with emergence of magnetic flux from below. Particularly, it is argued that bubbles could be formed due to perturbations in the prominence field by emerging parasitic bipoles from below \citep{Dudik2012,Gunar2014}. The absence of dips in the arcade field lines of the bipoles explains why a bubble is devoid of filament material \citep{Dudik2012}. Using He I D3 spectropolarimetric observations \citet{Levens2016} estimated that the magnetic field strength is higher inside the bubble than outside in the prominence. Thus, the rise of bubbles may be driven by the relatively large magnetic pressure rather than hot plasma inside the bubble. Naturally a magnetic separatrix or quasi-separatrix layer would form to separate the bubble from the surrounding prominence field, and the generation of plumes could be caused by reconnection at the separatrix \citep{Dudik2012,Gunar2014,Shen2015}, while the dynamic behaviors of plumes are consistent with the magnetic Rayleigh-Taylor and Kelvin-Helmholtz instabilities \citep{Ryutova2010,Berger2017}. The generation of large mushroom-headed plumes, a signature of the Kelvin-Helmholtz instability progressing into the non-linear explosive stage \citep{Ryutova2010}, seems conducive to the bubble formation and expansion \citep{Awasthi&Liu2019}. Besides plumes, plasma inside bubbles can be dynamic. \citet{Awasthi&Liu2019} observed a prominence bubble with a disparate morphology in the H$\alpha$ line-center compared to line-wings. Combining Doppler maps with flow maps in the plane of sky reveals a complex yet organized flow pattern inside the bubble, which is interpreted to be outlining a flux rope undergoing kink instability (see also \S~\ref{subsubsec:KI}). Likely related to rising plumes, \citet{Zhang2014} found that prior to eruption a prominence is perturbed multiple times by underlying chromospheric fibrils that rise upward and merge into the prominence, whose subsequent eruption can be interpreted by a `flux feeding' mechanism (see \S~\ref{subsec:struc:double-decker}).

\subsubsection{Sigmoids} \label{subsubsec:sigmoid}
\begin{figure}[ht!]
	\centering
	\includegraphics[width=\hsize]{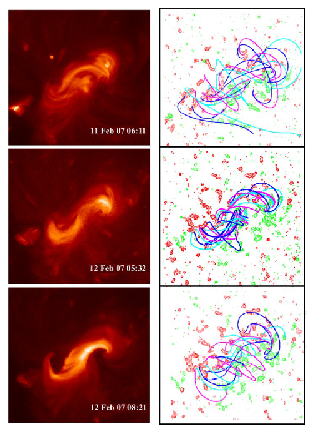}
	\caption{Evolution of a sigmoid \citep[from][]{Savcheva&vanBallegooijen2009}. \textbf{Left:} SXR images obtained by the X-Ray Telescope (XRT) on-board Hinode at different times, displaying a  transformation from a double-J (top) to S (middle) configuration. When the sigmoid erupts (bottom), coronal dimmings can be seen inside the hooked parts of the S shape. \textbf{Right:}: selected field lines from NLFFF modeling with a flux-rope insertion method; in the background, red and green indicate positive and negative polarity, respectively. \label{fig:sigmoid}}	
\end{figure}

Coronal sigmoids are S-shaped structures emitting in soft X-rays (SXR) and extreme ultraviolet (EUV) \citep[Figure~\ref{fig:sigmoid};][]{Sakurai1992,Canfield1999,Moore2001}. Typically the central part of a sigmoid is approximately aligned with the photospheric PIL, indicating the concentration of magnetic stresses and electric currents, hence they are described either by a flux rope \citep{Rust&Kumar1996,Titov&Demoulin1999, McKenzie&Canfield2008,Savcheva&vanBallegooijen2009} or by a highly sheared magnetic arcade \citep{Moore2001}. Sigmoids are first discovered by the Soft X-Ray Telescope on-board Yohkoh, and soon recognized as an important CME progenitor \citep{Canfield1999}. In the wake of eruption, the sigmoid evolves into a post-flare cusp or arcade, known as the sigmoid-to-arcade evolution \citep{Sterling2000}. The majority of sigmoids are found to be forward (reverse) S-shaped in the southern (northern) hemisphere, following the hemispheric helicity rule, i.e., positive (negative) helicity is preferred in the southern (northern) hemisphere \citep{Pevtsov2014}.  

Some sigmoids can be visible for days, but become bright only shortly prior to or during the early impulsive phase of flares. In a minority of SXR sigmoids, a continuous S shape is observed to appear through a transition from a double J shape prior to the onset of eruption (Figure~\ref{fig:sigmoid}). This transition is consistent with flux rope formation from a sheared arcade through reconnection \citep{Green&Kliem2009,Green2011,Green&Kliem2014}. Flux-rope models, however, suggest that the sigmoidal emission does not trace out the axis of the erupting flux rope, but highlight the formation of a sigmoidal current sheet at its underside \citep[][see also \S~\ref{subsec:topology}]{Titov&Demoulin1999,Fan&Gibson2003, Kliem2004, Savcheva&vanBallegooijen2009}. 

More recently, the complexity in the structure and evolution of coronal sigmoids are further revealed by high-resolution SXR images obtained with the X-Ray Telescope
on-board Hinode and high-resolution, high-cadence EUV images obtained with the the Atmospheric Imaging Assembly (AIA) onboard the Solar Dynamic Observatory (SDO). AIA is equipped with EUV filters sensitive to hot plasmas, such as the 94~{\AA} (\ion{Fe}{18}, $\log T = 6.85$) and 131~{\AA} (\ion{Fe}{21}, $\log T = 7.05$) passbands. \citet{Tripathi2009sigmoid} found that a double J-shaped hot arcs ($T>2$ MK) coexists with an S-shaped cold loop ($T\approx1-1.3$ MK) bridging the gap of the double J structure in a decaying active region. This can be explained by the cooling of plasma carried by flux shells during earlier reconnection. \citet{Liu2010tc} observed that an EUV sigmoid transforms continuously from a pair of opposing J-shaped loops first to a continuous S-shaped loop and then to a semi-circular eruptive structure (Figure~\ref{fig:cartoon}(c1--c3)). The transformation in the latter stage, which is likely related to the rise of a faint, nearly linear feature in SXRs \citep{Moore2001,McKenzie&Canfield2008,Green2011} and is also recognized as the appearance of a ``hot channel'' in EUV \citep{Zhang2012}, indicates the formation of a highly coherent but unstable structure, most likely a flux rope. The hot channel may appear different from different viewing angles, e.g., a blob if viewed along the axis \citep{Cheng2011blob,Nindos2015}. Hundreds of hot channels have been detected by SDO/AIA in recent years \citep{Zhang2015,Nindos2015}, and detailed investigations into a few cases show that the hot channel forms either prior to the eruption onset, during a flare precursor or a slow-rise phase leading up to eruption \citep{Liu2010tc, Zhang2012, Cheng2013driver, Cheng2013successive, Joshi2017}, or during confined flares preceding the eruptive flare \citep{Patsourakos2013}.

\subsubsection{Filament-Sigmoid Relationship} \label{subsubsec:relation}
It is difficult to understand the relationship between filaments and sigmoids that are aligned along the same PIL. To be consistent with the hemispheric helicity rule, it is suggested that a sigmoid is represented by sigmoidal field lines threading the current sheet formed at the BPSS or HFT underneath an upwardly kinked flux rope, whose sense of S shape is opposite to that of the sigmoid \citep{Titov&Demoulin1999,Fan&Gibson2003, Kliem2004,  Savcheva&vanBallegooijen2009}. This scenario may find support in observations of the apex rotation during filament eruptions. Associated with a forward (reverse) S-shaped sigmoid, the filament apex rotates clockwise (counterclockwise) if viewed from above, suggesting that the filament is embedded in a right-handed (left-handed) flux rope \citep{Green2007,ZhouZ2020}. However, it is unclear whether the flux rope already takes on an S shape opposite to the sigmoid before the eruption, because the cold filament and the hot sigmoid almost always display similar S shapes \citep[e.g.,][]{Pevtsov2002,Cheng2014double,ZhouZ2017,ZhouZ2020}. The filament can indeed reverse the sense of its S shape through apex rotations during the eruption \citep{Rust&Labonte2005, Romano2005,Green2007, ZhouZ2020}. This may indicate that the sign of magnetic writhe is maintained when the originally low-lying flux rope rises to high altitudes during the eruption \citep{Torok2010}.  

However, this does not explain why the filament often survives the eruption, showing no significant changes beneath the post-flare arcade, which replaces the sigmoid during the eruption. In some cases one can clearly see the undisturbed filament underneath the sigmoid and between the two flare ribbons during the whole process of the eruption \citep{Pevtsov2002,LiuC2007,Cheng2014double}, which excludes the possibility that a filament may reform immediately after the eruption. One possibility is that magnetic reconnection occurs within the flux rope but above the embedded filament, which leads to a CME without a corresponding filament eruption \citep{Gilbert2000,Gibson&Fan2006partial,Liu2007}.  

It is also possible that the low-lying filament and the high-lying sigmoid is only part of a much more complex structure, e.g., a double-decker (see \S~\ref{subsec:struc:double-decker}). \citet{Regnier2002} and \citet{Regnier&Amari2004} employed NLFFF extrapolations to investigate the magnetic configuration of an S-shaped filament (height 30 Mm) and a sigmoid (height 45 Mm) aligned along the same PIL in the NOAA active region 8151. They found both structures have the similar S shape and can be described by a twisted flux tube with a twist number short of unity, but the electric current density is positive in the right-handed filament and negative in the left-handed sigmoid. In addition, a right-handed flux rope with a twist number slightly exceeding unity is present at a higher altitude (60 Mm). Thus, the structure is essentially consistent with one of the two double-decker configurations, comprising a flux rope atop a sheared arcade \citep[see \S~\ref{subsec:struc:double-decker};][]{Liu2012}. \citet{Cheng2014double} invoked the other double-decker configuration, i.e., a double flux rope, to explain the sigmoid eruption on 2012 July 12: a low-lying flux rope associated the sigmoid and a co-spatial filament forms in two groups of sheared arcades a half day before the eruption; a high-lying flux rope in the form of an S-shaped hot channel appears 2 hrs before the eruption. Although only the low-lying rope is identified in the NLFFF model, the hot channel is verified to be a flux rope when it is detected in situ as a magnetic cloud three days later. 

In partial eruptions, the filament-sigmoid association becomes more complicated. That a sigmoid splits during eruption or a sigmoid immediately reforms after eruption is often considered to be the signature of a BPSS flux rope whose lower part of the rope is tied to the dense photosphere \citep{Gibson&Fan2006partial,Green2011}. For a sigmoid forming in the underside of an HFT flux rope, it is expected that the whole flux rope is ejected and that the sigmoid ceases to exist after the eruption \citep{Aulanier2010,Fan2010,Fan2012}. When a filament is involved, however, multiple factors may affect whether a full, partial, or failed filament eruption will occur; these include how much the magnetic dips are filled with filament material and where these material is located relative to where the rope breaks into two due to internal magnetic reconnections \citep{Gibson&Fan2006partial,Gilbert2000,Gilbert2007}. The brightening within a filament body and its subsequent splitting during eruptions is perhaps the most convincing observational signature of internal reconnections \citep{Liu2008,Tripathi2009partial,Cheng2018}, which may take place at a central vertical current sheet that forms as the rope writhes and expands upward, specifically, between the BPSS field lines and the ambient field lines, including those in the flux rope and the surrounding magnetic arcades \citep{Gibson&Fan2006mfr}. \citet{Liu2008} reported that a discontinuous sigmoid becomes a continuous S lying above the filament at the eruption onset, yet both the filament and sigmoid still bifurcate during the eruption, suggesting that reconnections take place both above and within the filament. Alternatively, internal reconnections may naturally occur at the HFT within a double-decker configuration (see \S~\ref{subsec:struc:double-decker}) or between braided flux ropes \citep[e.g.,][]{Awasthi2018}. 

\subsection{Modeling} \label{subsec:model}
The eruptions in the solar atmosphere exhibit distinctly diverse patterns across a vast range of spatio-temporal scales, from CMEs in the shape of stellar-sized bubbles, to localized flares within active regions that harbor sunspots, to collimated jets down to the resolution limit of modern telescopes. Although the complexity and diversity of eruptive phenomena makes it impossible to build a `universal' model that is capable of explaining all observational aspects in all events, the standard or CSHKP flare model \citep{Carmichael1964, Sturrock1966, Hirayama1974, Kopp&Pneuman1976} is successful in explaining the major characteristics of two-ribbon eruptive flares, which provides a solid basis to understand the flare-CME coupling \citep[see reviews by][]{Forbes2000,Priest&Forbes2002,Forbes2006}: destabilized by an unspecified trigger, a flux rope starts to rise and stretch the overlying magnetic field, also termed ``strapping field'', which serves to constrain the rope; as a result, a vertical current sheet develops in the wake of the rope, where successive magnetic reconnections add layers of plasma and magnetic flux to the rope and simultaneously produce the growing flare loop system whose footpoints correspond to the separating flare ribbons in the lower atmosphere. By converting magnetic fluxes of the strapping field to those of the flux rope, the role of flare reconnection is therefore twofold: it reduces the downward-pointing but increases the upward-pointing Lorentz forces exerted upon the flux rope, making it rise and expand faster, which in turn enhances the plasma inflow toward the current sheet and therefore the reconnection rate. This positive feedback creates a snowballing CME propagating into the outer corona. 

More recently, aided by nonlinear force-free field or MHD modeling of the coronal magnetic field, it has been demonstrated that H$\alpha$ and UV flare ribbons often coincide with the footprints of QSLs \citep[e.g.,][]{Demoulin1997,Demoulin2006, Liu2014, Liu2016SR, Liu2018, Su2018,Jiang2018,Janvier2014,Janvier2016}. In particular, the footprints of QSLs wrapping around the flux rope correspond to a pair of J-shaped ribbons of high electric current densities, with their hooks surrounding the rope legs \citep{Janvier2014,Janvier2016,WangW2017}. In a data-driven MHD simulation, \citet{Jiang2018} tracked  the footpoints of field lines newly reconnected at the HFT below the flux rope, where the QSL intersects itself. They found that the location of these footpoints not only match the observed flare ribbons as well as the boundary of the rope's feet at far ends of the ribbons, but their evolution also emulates the temporal separation of the flare ribbons. Motivated by these observational and modeling results, it has been proposed that the two-dimensional standard model can be extended to three dimensions to address the shape, location, and dynamics of flares with a double J-shaped ribbons \citep{Aulanier2013, Aulanier2012, Janvier2013, Janvier2015}.  

The mainstream models of solar eruptions are in the `storage and release' category, i.e. the free magnetic energy is quasi-statically accumulated in the corona on time-scales of days to weeks, and then rapidly released during the eruption on time-scales of minutes to hours \citep[see the reviews by][]{Chen2011,Forbes2006,Forbes2010}. The key parameters and detailed processes that govern the evolution leading up to an eruption are not fully understood, nor does the pre-eruption magnetic configuration. But in all storage-and-release models, the core erupting structure is a flux rope, no matter it is embedded in the pre-eruption configuration or formed by magnetic reconnection on the fly. In the former scenario, including the standard flare model \citep{Kopp&Pneuman1976}, the pre-existent flux rope may originally emerge from below the photosphere into the corona \citep[e.g.,][]{Fan2001,Roussev2012,Torok2014}, or form in the low corona by slow magnetic reconnection in a sheared magnetic arcade \citep{vanBallegooijen&Martens1989}, which is driven by the gradual evolution of the magnetic field in the photospheric boundary \citep[e.g.,][]{Amari2014}. In the latter scenario, the initial state typically contains sheared arcades and a new flux rope forms via magnetic reconnection during the course of the eruption \citep{Antiochos1999,Moore2001,Karpen2012}. 

The models of solar eruptions can also be categorized as either ideal or resistive according to the initiation and driving mechanism. In resistive models, magnetic reconnection is responsible for the onset and growth of the eruptive structure in time, as well as for the formation of the flux rope during the eruption \citep[e.g.,][]{Antiochos1999,Moore2001}. In ideal models, the coronal magnetic field reaches a critical point where a loss of equilibrium \citep[also known as a `catastrophe'; e.g.,][]{Lin&Forbes2000} or a loss of stability \citep[e.g.,][]{Torok&Kliem2005,Kliem&Torok2006} occurs, leading to the eruption. Both catastrophe and instability can lead to the formation of a vertical current sheet underneath the flux rope, as in the standard flare model, but magnetic reconnection at the current sheet is only invoked as a byproduct in these ideal models. The two most frequently cited ideal MHD instabilities are the torus and the kink instability, both are driven by electric currents. It is argued that the catastrophe and the torus instability are equivalent descriptions for the eruption onset condition of a flux rope \citep{Demoulin&Aulanier2010,Kliem2014instability}. Below we first focus on the torus instability (\S~\ref{subsubsec:TI}) and the kink instability (\S~\ref{subsubsec:KI}), both involving a single flux rope, and then touch on instabilities involving interacting flux ropes or current systems (\S~\ref{subsubsec:interact}).

\subsubsection{Torus Instability} \label{subsubsec:TI}
The torus instability arises in a competition between the upward `hoop' force, which is the Lorentz force between a toroidal current and its own poloidal field, and the downward `strapping' force, which is the Lorentz force between the same toroidal current and an external potential field perpendicular to the axis of the torus \citep{Bateman1978,Chen1989,Kliem&Torok2006}. The torus instability, also termed lateral kink instability, sets in against expansion if the external field decreases fast enough in the direction of the major axis of the torus. The rate at which the field decreases with height is quantified by the decay index $n=-d\ln B/d\ln h$, where $h$ is the height above the photosphere. The threshold value of the instability $n_\mathrm{crit}$ is derived to be 1.5 for a toroidal current channel \citep{Kliem&Torok2006}, while for a straight current channel, $n_\mathrm{crit}\gtrsim1$ \citep{Demoulin&Aulanier2010}.

In comparison with the prototypical magnetic configuration of a CME progenitor, i.e., a flux rope is embedded in a sheared external field, which becomes less sheared at higher altitudes, approaching a potential field, one can see that both the toroidal component of the external field and the poloidal current of the flux rope are missing in the idealized, analytical models. Hence, it is not surprising that in numerical studies \citep[e.g.,][]{Fan&Gibson2007, Kliem2013, Zuccarello2015}, $n_\mathrm{crit}$ is found to vary in a relatively wide range [1.4--2.0]. Additionally, the analysis in \citet{Kliem&Torok2006} applies to a slender flux rope. When it comes to a flux rope of finite size, it is unclear whether the decay index should be computed at the rope apex or axis \citep{Zuccarello2016}. In calculating the decay index, it is difficult to decouple the external magnetic field from the field induced by the electric currents inside the flux rope. The usual practice is to use the potential field extrapolated from the vertical component of the photospheric magnetic field to approximate the strapping field. This is because potential fields are expected to be perpendicular to PILs along which filament channels that host flux ropes are formed. But it is unclear as to how good the approximation is, especially when PILs are curved, and what role is played by the sheared, nonpotential component of the external field. Also, one may need take into account nonradial expansion that is frequently observed in filament eruptions \citep{Liu2009}; one solution is to compute the decay index along the eruption direction \citep{Duan2019}.  

In a laboratory experiment, \citet{Myers2015} recognized four separated regimes in the parameter space spanned by the torus instability parameter, decay index $n$, and the kink instability parameter, safety factor $q$, which is approximately the inverse of twist number (see \S\ref{subsec:twist}). Besides the expected `stable', `eruptive', and `failed kink' states, they found a new `failed torus' regime, in which a torus-unstable but kink-stable flux rope fails to erupt, due to the Lorentz force between the rope's poloidal current and the external toroidal field. When similar scatter plots are made for solar eruptions, it is unclear whether a failed torus regime actually exists \citep{Jing2018,ZhouZ2019,Duan2019}. \citet{ZhouZ2019} found that significant rotational motion, which may be caused by the helical kink instability, tends to be associated with failed filament eruptions that are normally judged to be torus unstable. However, the rotation driven by the helical kink instability cannot be easily distinguished from that caused by a Lorentz force resulting from the misalignment between the flux rope's toroidal current and the external toroidal field \citep{Isenberg&Forbes2007,Kliem2012}. Thus, it remains obscure as to what parameters can differentiate failed from successful CMEs.

Despite the simplifications and obscurities, how the magnetic field overlying an eruptive structure decays with height is indeed found to play an important role in regulating eruptive behaviors \citep[e.g.,][]{Torok&Kliem2005, LiuY2008, Guo2010confined, Cheng2011torus,Zuccarello2014, WangD2017, Baumgartner2018, Amari2018}. Two types of $n(h)$ profiles emerge in observation \citep{Guo2010confined,Cheng2011torus,WangD2017}: 1) $n$ increases monotonically as the height increases; and 2) the $n(h)$ profile is saddle-like, exhibiting a local minimum $n_b$ at a height higher than the critical height $h_\mathrm{crit}$ corresponding to $n_\mathrm{crit}=1.5$, which is approximately half of the distance between the centroids of opposite polarities in active regions \citep{WangD2017}. The saddle-like $n(h)$ profile is found exclusively in active regions of multipolar field configuration, despite that the majority cases of monotonously growing $n(h)$ also originate from multipolar field. Supposedly the saddle-like profile provides a potential to confine an eruptive structure if the local minimum $n_b$ at the bottom of the saddle is significantly below $n_\mathrm{crit}$. Indeed \citet{WangD2017} found that $h_\mathrm{crit}$ is significantly higher for confined flares than for eruptive ones, and that $n_b$ is significantly smaller in confined flares than that in eruptive ones. In a data-driven MHD simulation, \citet{Guo2019simulation} computed the decay index along the eruption path of a flux rope, and found that the rope starts to rise rapidly at the same height as the decay index crosses the canonical critical value of 1.5. Similarly, recent studies on the height-time profiles of eruptive filaments \citep{Vasantharaju2019, Zou2019,Myshyakov&Tsvetkov2020,Cheng2020} and of hot channels \citep{Cheng2020} generally concluded that the decay index at the height where the slow rise transitions to fast rise is close to the threshold of the torus instability.

The torus instability can be triggered or suppressed by magnetic reconnection that modifies the overlying field. This effect is often invoked to explain sympathetic eruptions, a sequence of eruptions that occur at different places within a relatively short time interval. The distanced regions can be connected by magnetic reconnection of large-scale magnetic field, as suggested by observational investigations \citep{LiuC2009, Zuccarello2009,Schrijver&Title2011,Schrijver2013, Jiang2011, Titov2012, Shen2012, Yang2012, Joshi2016, WangR2016, WangD2018,Zou2019twostep} and corroborated by numerical simulations \citep{Ding2006,Torok2011,Lynch+Edmondson2013}. Unlike flare reconnection in active regions, \citet{WangD2018} found that magnetic reconnection of the large-scale field in the quiet-Sun corona is subtly manifested through serpentine flare ribbons extending along chromospheric network, coronal dimmings, apparently growing hot loops, and contracting cold loops. The reconnection continually strengthens the strapping field of one filament that undergoes a failed eruption, but weakens the strapping field of the other that later erupts successfully. 

\subsubsection{Kink Instability} \label{subsubsec:KI}
In a cylindrical flux rope of radius $a$ and length $L$, the safety factor $q(r) = rB_z(r) / LB_\theta(r)$, which is related to the twist angle through $\Phi=2\pi / q$, is key to the flux-rope stability. The external kink instability occurs when $q(a)<1$, which exceeds the Kruskal-Shafranov limit of one field-line turn about the flux-rope axis \citep{Kruskal&Tuck1958,Shafranov1958}. The kink is associated with the $m = 1$ mode in the Fourier decomposition of linear perturbations in terms of $\exp(im\theta + ikz)$. This mode helically displaces the central axis as well as the surroundings, hence is also termed the helical kink instability. Taking into account the stabilizing effect of line-tying --- both footpoints of any coronal flux tubes anchor firmly in the dense photosphere --- \citet{Hood&Priest1981} showed analytically that the critical value of twist angles for a force-free, uniform-twist equilibrium of infinite radial extent is $\Phi=2.49\pi$. The actual critical twist value turns out to be rather sensitive to equilibrium details \citep[e.g.,][]{Einaudi&vanHoven1983, Baty&Heyvaerts1996, Baty2001, Torok&Kliem2003}, including the embedding of the flux rope in the external field, the radial twist profile, the radius of the rope, e.g., the critical twist number is larger for a thinner flux rope \citep[see also][]{WangY2016}, as well as the weight of prominence material at the bottom of the flux rope \citep{Fan2018}. 

The external kink instability has been proposed as a trigger for magnetic reconnection responsible for coronal heating \citep{Browning2008,Hood2009} or flare heating \citep{Hood&Priest1979,Pariat2009,Srivastava2010}. The invoke of this instability in solar eruptions is mainly motivated by the dramatic development of helical eruptive structures \citep[e.g., ][]{Ji2003, Romano2003, Rust&Labonte2005, Williams2005, Alexander2006, Liu2007,Patsourakos2008, Liu&Alexander2009, Cho2009, Karlicky&Kliem2010, Kumar2012, Yang2012, Kumar&Cho2014}, typically when a filament rises and rotates into an inverted $\gamma$ or $\delta$ shape in projection \citep[Figure~\ref{fig:helical}a;][]{Gilbert2007}. These events exhibit not only a winding of the filament threads about the axis \citep[e.g.,][see also Figure~\ref{fig:filament}b]{WangH2015}, arguing for the existence of considerable twist, but also an overall helical shape, indicating a writhed axis. This combination therefore strongly indicates the helical kink instability of a flux rope, whereby magnetic twist (winding of magnetic field lines around the rope axis) is abruptly converted to magnetic writhe (winding of the axis itself). Such a conversion reduces the bending of the field lines as well as the magnetic energy of the flux rope, resulting in a rotation of the rope apex \citep{Kliem2012}. 

Meanwhile, a vertical current sheet is formed underneath as the rising rope stretches the overlying field, as predicted by the standard flare model, and a helical current sheet wrapping around and passing over the rope is formed through the helical displacement \citep{Torok2004}. These current sheets have two important observational consequences. First, field lines that thread either of the current sheets are sigmoidal in projection \citep{Titov&Demoulin1999,Fan&Gibson2003,Kliem2004,Gibson2004}, whose orientation agrees with the chirality of sigmoids \citep{Rust&Kumar1996,Pevtsov1997,Green2007,ZhouZ2020}. A coronal sigmoid may be produced because the plasma located in/near the current sheets is heated by current dissipation or magnetic reconnection. Second, when a flux rope is kinked into the inverted $\gamma$ or $\delta$ shape, its two legs are forced to interact with each other, producing a hard X-ray or microwave source at the crossing point of the inverted $\gamma$ or $\delta$ in addition to the footpoint sources \citep{Alexander2006,Liu&Alexander2009,Karlicky&Kliem2010,Kliem2010}. This can be explained by magnetic reconnection at the current sheets between the two approaching legs. The so-called leg--leg reconnection may break up the flux rope, with the upper part of the original rope evolving into a CME \citep{Cho2009,Kliem2010}. \citet{Kliem2010} found in MHD simulations that sections of the helical current sheet are squeezed into a temporary double current sheet between the two approaching flux-rope legs, thereby facilitating fast reconnection and the formation of moving plasmoids through subsequent island coalescence. These plasmoids are believed to propagate along the helical current sheet to, and merge at, the top of the flux rope, where they are observed as a compact microwave source rising rapidly with the erupting rope \citep{Karlicky&Kliem2010,Kliem2010}. 

However, it is debatable whether the helical kink instability plays a significant role in solar eruptions. A doubt on the sufficiency of magnetic twist in active regions \citep{Leamon2003} was apparently settled by examining localized active-region flux ropes \citep{Leka2005}, but there are more issues for consideration, in addition to the inaccuracy of magnetic twist derived from NLFFF extrapolations or imaging observations. First of all, the helical kink instability may quickly saturate \citep[e.g.,][]{Torok&Kliem2005}, therefore it is often associated with failed eruptions rather than successful ones leading up to CMEs. Second, eruptive structures with a clear writhing feature are relatively rare, which raises a question as to how often the helical kink instability triggers eruptions. Third, helical patterns are often present only during eruptions \citep[e.g.,][]{Vrsnak1991, Vrsnak1993, Romano2003, Gary&Moore2004, Srivastava2010, Kumar2012}, which makes it difficult to determine whether the twist is accumulated prior to the eruption or built up in the course of the eruption. Magnetic reconnection in the vertical current sheet beneath the flux rope indeed contributes a significant amount of magnetic flux to the CME \citep{Lin2004,Qiu2007}. The observed kink might be a byproduct of the eruption. Finally, the shear component of the ambient field can cause writhing motions of a flux rope in a similar manner as the kink mode \citep{Isenberg&Forbes2007}. This effect is difficult to be excluded unless the writhing is extremely strong with an apex rotation significantly larger than 90 degrees \citep{Kliem2012}. Alternatively, a rotation could also result from a relaxation of magnetic writhe, whose direction is opposite to that driven by the kink instability \citep[e.g.,][their Figure 4]{Alexander2006}, or from a relaxation of magnetic twist, probably due to magnetic reconnections between the flux rope and the ambient field \citep{Vourlidas2011,Xue2016,Li2016}. 

The internal kink instability, on the other hand, is associated with the existence of singular radial positions at $q(r_s) = 1$ for $0 < r_s < a$ \citep[\S~9.4 in][]{Goedbloed2019}. In other words, if magnetic twist is large enough within the flux rope, the core would become kink unstable, with perturbations confined inside the flux rope boundary. \citet{WangW2017} found that a flux rope forms a highly twisted core with a less twisted envelope during the eruption; such a twist profile may be favorable for the internal kink instability (see also \S~\ref{subsec:struc:twist}). The internal mode possesses a smaller growth rate and tends to be more energetically benign than its external counterpart, and hence is proposed for the quasi-steady heating of coronal loops \citep{Galsgaard&Nordlund1997,Haynes&Arber2007}. \citet{Awasthi&Liu2019} found that mass motions inside a prominence bubble exhibit a counter-clockwise rotation with blue-shifted material flowing upward and red-shifted material flowing downward, which could be envisaged as counter-streaming mass motions in a helically distorted field resulting from the internal kink mode $m=2$. Since the bubble roughly maintains its shape and shows no obvious sign of heating, the internal kink is preferred over the external kink. \citet{Mei2018} performed three-dimensional isothermal magnetohydrodynamic (MHD) simulations in a finite plasma-$\beta$ environment, and found that both external and internal instability compete to drive a complex evolution of a flux rope through magnetic reconnection within and around the rope. \citet{Guo2012} argued that the internal kink instability could drive internal reconnection, which may result in hard X-ray emission at the flux-rope footpoints \citep[see also][]{Liu&Alexander2009}. 

\subsubsection{Instabilities of interacting flux ropes} \label{subsubsec:interact}
Here we briefly introduce ideal MHD instabilities related to interacting flux ropes or current systems. For a more comprehensive review from modeling perspectives, readers are referred to \citet{Keppens2019}.   

Two current-carrying flux ropes that are juxtaposed would attract or repel each other depending on whether the two currents run parallel or antiparallel to each other. Like-directed current channels are related to the coalescence instability \citep{Finn&Kaw1977}, while opposing-directed current channels to the tilt instability \citep{Finn1981}. Exploiting the energy principle, \citet{Richard1990} confirmed that the tilt instability operates on the ideal MHD timescale, making it relevant in the solar context. Contrary to the torus setup, it does not require toroidal curvature of the flux ropes. Embedded in a confining external magnetic field, two flux ropes would not move directly away from each other, as usually expected, but undergo a combined rotation and separation on Alfv\'{e}nic timescales \citep{Richard1990,Keppens2014}. \citet{Keppens2014} demonstrated an interplay between the kink and tilt instability in full three dimensions: a combination of helical and tilt deformations makes the two flux ropes swirl around, and separate from, each other. Thus, the tilt-kink evolution may provide a novel route to initiate CMEs, especially for the active regions where the opposite signs of helicity coexist \citep[e.g.,][]{Regnier&Amari2004,Su2018,Awasthi2019} or are injected sequentially from below \citep[e.g.,][]{Liu2010arcade,Chandra2010,Vemareddy&Demoulin2017}.

The development of tilt and coalescence instability may trigger magnetic reconnection between flux ropes. Depending on the angle between both rope axes and on whether they carry the like or opposite signs of magnetic helicity, the two ropes may bounce, merge, slingshot, or tunnel \citep{Linton2001}. Except tunneling, evidence for these interactions is often found in observations of solar filaments. For example, filaments of the same chirality may merge at their endpoints, but those of opposite chirality do not join \citep{Schmieder2004}. The two ropes in a double-decker configuration (\S~\ref{subsec:struc:double-decker}) may coalesce into an unstable structure before eruption \citep{Zhu2015,Tian2018} or merge into a CME after successive eruptions \citep{Dhakal2018}. In observations indicating the slingshot reconnection, two adjacent filaments typically approach each other, merge at their middle sections, and then separate again, in a way similar to the classic X-type reconnection \citep{Kumar2010, Chandra2011, Torok2011,Jiang2013}. CMEs are sometimes observed to interact with each other in the corona as well as in interplanetary space \citep[see][for reviews]{Lugaz2017,ShenF2017}. \citet{ShenC2012} found two CMEs interact in such a way that the total kinetic energy increases by about 6.6\%, supposedly at the expense of magnetic energy and/or thermal energy. The interaction is hence dubbed the ``super-elastic collision''. Further analysis reveals a spectrum of collisional behaviors ranging from being perfectly inelastic to being super elastic as far as the change of total kinetic energy before and after the ``collision'' is concerned \citep{ShenF2017,Mishra2017}. But this is also where the ``collision'' analogy stops, because no interaction under scrutiny results in the separation of two CMEs. After the interaction, CMEs are found to be either coalesced into one coherent flux rope \citep[e.g.,][]{Kilpua2019cme} or in the process of coalescence \citep[e.g.,][]{Feng2019,Zhao2019} in interplanetary space, with a boundary layer formed between two flux ropes \citep[e.g.,][see also \S~\ref{subsec:struc:boundary}]{Feng2019}.

\section{Formation} \label{sec:formation}

\subsection{Theoretical Debate} \label{subsec:formation:theory}
The nature of the magnetic configuration prior to solar eruptions has been under intense debate. Relevant to the debate are two prominent classes of flare/CME models that have been developed over the years. In the first, including the standard model of solar flares, a flux rope is present prior to the eruption \citep{Kopp&Pneuman1976, Forbes&Priest1995, Lin&Forbes2000, Titov&Demoulin1999, Aulanier2012}. In the second, the initial state typically contains a sheared magnetic arcade and a new flux rope forms via magnetic reconnection during the course of the eruption \citep{Antiochos1999,Moore2001,Lynch2004,Karpen2012,Dahlin2019}. On the other hand, sheared magnetic arcades can evolve continually toward the flux-rope configuration, driven by ubiquitous turbulent flows and flux cancellation in the photosphere \citep{vanBallegooijen&Martens1989,Amari2003b,Aulanier2010,Amari2014}. Thus, it seems to be a reasonable assumption that the longer a pre-eruption structure evolves, the more likely a coherent flux rope or at least its `seed' is present in the structure. One may envisage a spectrum of pre-eruptive configurations, with a pure magnetic sheared arcade or a pure flux rope at two extremes of the spectrum, but a `hybrid' state in the middle. Indeed, with pre-eruption photospheric field measurements as the boundary condition, coronal magnetic field models, including force-free, magnetohydrostatics, and magnetohydrodynamic models, frequently generate a coherent flux rope \citep[e.g., see the review by][]{Inoue2016,Guo2017}; yet realistic photospheric boundary conditions have not been adopted in MHD simulations that rely solely on sheared magnetic arcades. 

It remains open as well how a flux rope can be formed in the corona prior to an eruption. It may be formed in the convection zone, but forced by magnetic buoyancy to emerge through the solar surface into the corona \citep{Rust&Kumar1994,Low1996}; or it can be formed in the low corona by magnetic reconnection in a sheared magnetic arcade, which is also referred to as the arcade-to-rope topology transformation \citep{vanBallegooijen&Martens1989}. 

The role of flux emergence, however, may be limited in the flux-rope formation, because it is impossible for large-scale flux ropes between active regions and in the quiet Sun to be formed by emergence. Moreover, although magnetic twist can help to suppress the fragmentation of an emerging flux tube and to enable the buoyancy instability, still a coherent flux rope cannot rise bodily into the corona at ease: the original rope axis stops essentially at photospheric layers, due to the heavy plasma trapped at the bottom concave portions of the helical field lines \citep[see the review by][]{Cheung&Isobe2014}. The upper portions of the helical field lines that expand into the corona are twisted up, as torsional \citep{Fan2009,Sturrock2015} or shear Alfv\'{e}n waves \citep{Manchester2004} transport twist from the rope's interior portion toward its expanded coronal portion, which drive photospheric rotation of the polarities and shearing/converging motions along the PIL, respectively.  

The emerged arcade may keep the arcade topology under continuous shearing of the magnetic field above the photosphere, until a loss of equilibrium occurs days later \citep{vanBallegooijen&Mackay2007}. But more often, as demonstrated by various flux-emergence simulations \citep[e.g.,][]{Manchester2004, Magara2006, Archontis&Torok2008, Archontis&Hood2010, Leake2013, MacTaggart&Haynes2014}, the sheared magnetic field lines gradually develop a J shape, a current sheet gradually builds up above the PIL, and a post-emergence flux rope is formed by magnetic reconnection at the current sheet in a way closely related to the `tether-cutting' reconnection \citep{Moore2001}, or the `flux-cancellation' reconnection \citep{vanBallegooijen&Martens1989}; both mechanisms involve reconnections between converging opposite polarities in a sheared arcade, but the latter takes place so low in the solar atmosphere that the shorter reconnected loops are small enough to be pulled under the photosphere by magnetic tension force, while the longer reconnected loops may form a flux rope with the BPSS topology (\S~\ref{subsec:topology}). Such gradual arcade shearing or gradual arcade-to-rope transformation \citep[e.g.,][]{Amari2003a, Amari2003b} can be driven by the dispersal and diffusion of photospheric flux concentrations, and by flows shearing along and converging toward the PIL. The newly formed rope is not associated with the axis of the sub-photospheric flux tube any more; it can find a stable equilibrium or erupt readily, depending on the relative strength and orientation between the emerging and preexisting field \citep[e.g.,][]{Archontis&Torok2008, Archontis&Hood2012,Leake2014}.

\subsection{Observational Exploration} \label{subsec:formation:observation}
In observation, our ability to pinpoint when and how a flux rope forms in the corona is severely hampered by two inherent difficulties: 1) the direct measurement of the coronal magnetic field has not been made a routine practice, because coronal polarization signals are weak and complicated by not only the 180$^\circ$ ambiguity but also the 90$^\circ$ (Van Vleck) ambiguity and the line-of-sight integration in the optically thin corona \citep[e.g.,][]{Rachmeler2013}; and 2) morphologically speaking, the interpretation of any coronal structure is subject to the projection effect and again the line-of-sight confusion. 

In spite of the above difficulties, efforts have been made to interpret the time sequences of magnetograms as a three-dimensional representation of the emerging subsurface magnetic fields, which remains controversial because of the complexity of physics involved in the rising of flux tubes in a strongly stratified layer \citep{Cheung&Isobe2014}. In particular, two features associated with flux emergence are interpreted as a flux rope rising coherently from below the photosphere to above optical depth unity, i.e., 1) the widening and subsequent narrowing of the filament channel, also known as the `sliding-door' effect, and 2) the apparent rotation of the transverse field with respect to the PIL during the passage of the twisted tube \citep{Okamoto2008, Okamoto2009, Lites2010}. Although \citet{MacTaggart&Hood2010} were able to reproduce both effects in numerical simulations, \citet{VargasDominguez2012} pointed out two additional observable effects in simulations, which have not yet been verified in observations, i.e., the increase of unsigned flux at either side of the PIL and the shear flows driven by the emergence of flux ropes. 

Efforts have also been devoted to seeking signatures of the flux-rope formation in the pre-eruption evolution of flare/CME source regions. A key feature is the appearance of a continuous S-shaped loop in a sigmoidal active region from tens of minutes to hours before the eruption (see also \S~\ref{subsubsec:sigmoid}), in which a double-J-to-S transformation is interpreted by the conversion of the sheared-arcade field into a flux rope through flux-cancellation reconnection \citep{vanBallegooijen&Martens1989} or tether-cutting reconnection \citep{Moore2001}. Originally the overall S shape of the sigmoid comprises two opposite bundles of J-shaped loops, whose straight sections run anti-parallel to each other in the middle of the S, on opposite sides of the PIL. As the continuous S loop appears, its middle section crosses the PIL inversely, i.e., from the side of negative polarity to that of positive polarity, while its two `elbows' cross the PIL regularly \citep[e.g.,][]{Green&Kliem2009,Green2011}. The inverse PIL crossing is co-located with where canceling flux patches converge \citep[e.g.,][]{Green2011}. The continuous sigmoid may remain stable for several hours before the eruption; but the structure that erupts is often not the entire sigmoid, indicating that the flux rope either partially erupts or undergoes a further transition shortly before the CME \citep{Green&Kliem2014}. For example, \citet{McKenzie&Canfield2008} reported that before any soft X-ray flaring begins, a diffuse linear structure almost as long as the sigmoid lifts off from the middle of the sigmoid and shows slight clockwise rotation. Similarly, \citet{Green2011} observed a hot linear feature rises as part of the eruption and suggested that this feature likely traces out the field lines close to the axis of the flux rope. 

Before suddenly rising fast, coronal cavities as well as their entrained prominences typically rises slowly for hours at a speed of the order 1~\kms \citep{McCauley2015,Gibson2015}. Meanwhile they undergo subtle morphological changes, which might be associated with the transition from the BPSS to HFT topology (\S~\ref{subsec:topology}): the cavity often narrows in the bottom part and becomes increasingly more like an inverse teardrop. Teardrop-shaped cavities more likely erupt than elliptical or semicircular ones \citep{Gibson2006,Forland2013}. This `necking' process is sometimes associated with the appearance of a U-shaped `horn' in EUV, extending from the top of a prominence into the cavity above \citep[e.g., Figure~\ref{fig:cavity}a;][]{Berger2012,Schmit&Gibson2013}. The horn is thought to outline a flux rope above the prominence \citep{Berger2012,Fan2012}, while the prominence threads can be supported by numerous dips of tangled magnetic fields within a large-scale current sheet standing vertically above the PIL but underneath the rope. \citep{vanBallegooijen&Cranmer2010,Berger2012}.

In the era of SDO, the transition from the S loop to the eruptive structure is observed in greater detail. \citet{Liu2010tc} found that under persistent converging flows toward the PIL a continuous S-shaped loop with temperatures about 6 MK appears to form by connecting two opposite bundles of J-shaped cold loops. The S loop remains in quasi-equilibrium for about 50 min with its central dipped portion rising slowly at about 10 km~s$^{-1}$. About 10 minutes before the flare onset, the rising speed increases to about 60 km~s$^{-1}$, and the S loop quickly transforms into a semi-circular shape that eventually erupts as a CME. Similarly, \citet{Zhang2012} reported that a writhed sigmoidal structure as hot as 10 MK transforms toward a semi-circular shape. 

These so-called hot channels were found in a significant fraction (up to 50\%) of eruptive events \citep{Nindos2015,Zhang2015}. They are presumed to be in the form of a coherent flux rope but exhibit a variety of morphology. When viewed edge-on, a hot channel may appear to be a `plasmoid' \citep[e.g.,][]{Cheng2011blob,Patsourakos2013,Song2014,Xue2017,Gou2019}; when viewed face-on it may display tangled threads of emission that appear to wind around an axis \citep[e.g.,][]{Cheng2014track}. \citet{Song2014} reported that a plasmoid with an X shape underneath forms mysteriously from an expanding arcade during eruption and discussed the possibility that the flux rope may form on the fly. Typically the initial slow rise of a hot channel is followed by a fast acceleration phase, during which the rising channel compresses the surrounding medium into a relatively cold, leading front \citep[e.g.,][]{Cheng2013driver,Cheng2014track}, while the channels are further heated up \citep{Cheng2012}. Occasionally a hot channel is observed to entrain cold and dense plasma of an eruptive prominence \citet[e.g.,][]{Cheng2014prominence,Cheng2014track}, which might evolve into a typical three-part CME in white light. 

Generally, it is agreed that a coherent flux rope does not appear ``out of thin air'', but most likely builds up on a `seed', e.g., plasmoids or blobs, which are frequently observed in the atmosphere. Below we elaborate on the connection between the small-scale plasmoids and the large-scale CMEs and on the buildup process of CME flux ropes.
  
\subsubsection{Plasmoid \& Seed CME} \label{subsubsec:seed}
Plasmoids, mini flux ropes naturally born through tearing and coalescence instabilities in current sheets of large aspect ratios \citep{Daughton2011,Barta2011}, are believed to be a key leading to the fast reconnection in solar flares. They are continuously formed and ejected in a hierarchical, fractal-like fashion in current sheets, known as the plasmoid instability \citep{Shibata&Tanuma2001,Loureiro&Udensky2016}. Such behaviors not only influences the reconnection rate but also enhances the particle acceleration efficiency in a Fermi-like process \citep{Drake2006,Drake2013,Oka2010,Nishizuka&Shibata2013}. \citet{Uzdensky2010} and \citet{Loureiro2012} showed that due to plasmoid coalescence the distribution of both plasmoid fluxes and half-widths follow an inverse-square law in the large-Lundquist-number, plasmoid-dominated regime. They concluded that large disruptive events, i.e., ejections of ``monster'' plasmoids, are an inevitable feature of large-Lundquist-number reconnection. \citet{Nishizuka&Shibata2013} argued that the the power-law distribution of plamoid sizes via the fractal reconnection process can naturally explain the power-law spectrum in nonthermal emissions.

As an extension of the standard flare model, it is suggested that the formation and ejection of plasmoids play an essential role in flares by inducing a strong inflow into reconnection region, under high-Lundquist-number solar conditions \citep{Shibata1995,Shibata&Tanuma2001}. As it forms initially, a plasmoid staying in the current sheet reduces the reconnection rate by inhibiting inflows towards the reconnection region. Only when the plasmoid ejects out from the current sheet, can a substantial amount of magnetic flux enter the reconnection region. The reconnection outflow facilitates the ejection of the plasmoid, which in turn enhances the inflow of new magnetic flux, with faster ejection being translated to faster reconnection inflows. Through this positive feedback, the plasmoid ejection and acceleration are closely coupled with the reconnection process. Meanwhile, plasmoids break up into smaller ones and simultaneously collide with each other to make bigger ones. When these plasmoids are ejected out of the current sheet, fast reconnection occurs at various different scales in a highly time-dependent, intermittent manner. The ejection of the largest plasmoid is associated with the greatest energy release, probably corresponding to the impulsive phase of flares. It has been verified by high-resolution three-dimensional numerical simulations that small-scale plasmoids in distorted shapes are formed inside current sheets and their ejections increase the reconnection rate locally and intermittently \citep{Nishida2013,Mei2017}. A similar mechanism involving a mini flux rope is suggested for micro- and nano- flares as well as jets occurring at various altitudes on different scales  \citep{Shibata&Tanuma2001,Shibata2007,Pariat2009,Sterling2015,Wyper2017}. 

Also termed plasma blobs in the context of solar observations, plasmoids seem ubiquitous in various eruptive phenomena ranging from CMEs, flares, jets, down to small-scale reconnection events. They have been discovered to eject upward above flare loops in soft X-rays \citep[e.g.,][]{Shibata1995, Tsuneta1997, Ohyama&Shibata1997, Ohyama&Shibata1998, Shimizu2008}, in hard X-rays or microwaves \citep[e.g.,][]{Hudson2001,Kundu2001,Sui&Holman2003,Karlicky&Kliem2010}, and in EUV \citep[e.g.,][]{Reeves&Golub2011, Cheng2011blob, LiuW2013, Kumar&Cho2013, Gou2019}, which often lead up to CMEs, and also found to propagate along bright white-light rays trailing CMEs \citep[e.g.,][]{Lin2005,Chae2017} or above helmet streamers \citep[e.g.,][]{Sheeley2009}. Plasmoids are also frequently found in coronal jets \citep{Zhang&Ji2014,Ni2017}, and are revealed by subarcsecond imaging spectroscopy in UV bursts in the low atmosphere, with a size scale smaller than $0''.2$ and a time scale of seconds \citep{Rouppe2017}. 

Particularly in flares, a plasmoid is typically formed and heated up to multi-MK temperatures before the impulsive phase \citep{Ohyama&Shibata1997,Patsourakos2013,Gou2019}; it rises slowly until experiencing a strong acceleration, which is coupled to the enhanced reconnection inflow, particle acceleration, and plasma heating during the impulsive phase of the flare \citep{Ohyama&Shibata1997,LiuW2013,Gou2017}. A similar relation is found between the CME acceleration and the energy release \citep{Zhang2001,Qiu2004,Temmer2010}. Associated with the interaction or coalescence of multiple plasmoids in the current sheet, magnetic reconnection proceeds in a patchy, turbulent, and fractal fashion \citep{Shibata&Tanuma2001,Aschwanden2002,Linton&Longcope2006,Lazarian2012}. As a result, electric field in the reconnection region varies rapidly, which modulates the acceleration of electrons and ions, therefore producing bursty HXR or microwave lightcurves \citep{McAteer2007,Nishizuka2009,Nishizuka2010} as well as drifting pulsating structures (DPSs) in dynamic radio spectra \citep{Kliem2000, Karlicky&Barta2007, Karlicky&Barta2011,Nishizuka2015,Takasao2016}. Bi-directional plasmoid ejections are linked to the simultaneous detection of both negative and positive DPSs \citep{Kumar&Cho2013} and double coronal X-ray sources with their centroid separation decreasing with energy \citep{LiuW2013}. \citet{Milligan2010} studied the merging of a downward-propagating plasmoid with a looptop source in the 9--18 keV energy range. The merge may provide additional particle acceleration, resulting in enhanced nonthermal coronal emission in radio.

When the flare loop system is viewed face-on, showing cusp-shaped loops, an erupting plasmoid is found to be connected to the flare loops by a hot linear feature of temperatures about 10 MK, which is identified as the vertical current sheet in the standard flare model \citep{Reeves&Golub2011,Hannah&Kontar2013,Zhu2016,Gou2019}. \citet{LiuW2013} reported V-shaped EUV emission on the trailing edge of an erupting plasmoid, mirroring the underlying inverted V-shaped flare loops; both V are associated with distinct X-ray sources, reminiscent of two opposing Y-type null points in the standard flare model. Plasmoids are also observed to embed in, and move both upward and downward along, the linear feature \citep[e.g.,][]{Takasao2012,Liu2013,LiuW2013,Zhu2015,Kumar&Cho2013,Cheng2018cs,Gou2019}. \citet{Cheng2018cs} showed that the intensity variation along the current sheet has a power-law distribution in the spatial frequency domain, and that the intensity and velocity of the sunward outflows along the current sheet also display power-law distributions in the temporal frequency domain, which are attributed to the ongoing fragmented and turbulent magnetic reconnections. When the flare loop system is viewed side-on, showing an arcade structure, tadpole-like dark voids are observed to flow through a diffuse fan-shaped `haze' toward the flare arcade, known as surpa-arcade downflows \citep[e.g.,][]{McKenzie&Hudson1999,McKenzie&Savage2009,Chen2017}, which are interpreted to be plasmoids in the downward reconnection outflow \citep{Asai2004,LiuW2013,Liu2013}.  

A few exemplary events, e.g., the C4.9-class flare on 2010 November 3 \citep{Reeves&Golub2011,Cheng2011blob,Hannah&Kontar2013}, the X2.8-class flare on 2013 May 13 \citep{Gou2017,Gou2019}, and the X8.2-class flare on 2017 September 10 \citep{Veronig2018,Cheng2018cs,Yan2018}, are observed from such a fortuitous viewing angle that the eruptive structure emulates the standard-flare morphology with a rising flux rope connecting to the flare loops by a vertical current sheet, which casts a great light on how the flux rope forms before, and builds up during, the eruption. In the 2013-May-13 event, \citet{Gou2019} observed how plasmoids that are barely resolved merge and evolve continuously into a CME bubble within half an hour. The eruptive structure appears as a hollow ellipsoid whose bottom is connected to the top of flare loops through an extended linear feature of width $\sim\,$2$''$. The ellipsoid is visible in EUV passbands sensitive to hot plasma of 3--10 MK; it has a hot outer shell, but slightly cooler than the linear feature that is exclusively visible in 131~{\AA} (Fe XXI and XXIII; $\gtrsim\,$10~MK). Both the morphological and thermodynamic features fully agree with the standard model: hot plasmas are expected to emit not only at the vertical current sheet and the flare loops, but also in the outermost layer of the flux rope, where the magnetic field is newly reconnected \citep{Lin&Forbes2000,Lin2004}, while the inner layers of the rope have already cooled. Thus, the ellipsoid is identified as a flux rope and the linear feature as the vertical current sheet. \citet{Gou2019} showed that the current sheet is not only present during the impulsive phase of the eruption, but evolves continuously from a shorter one in the wake of a preceding confined flare, located beneath a magnetically sheared arcade. As it extends upward slowly at $\sim\,$10~\kms, the current sheet is fragmented into multiple plasmoids of widths $\sim\,$2$''$. About ten minutes before the flare onset, a leading plasmoid appears at the upper tip of the current sheet. Underneath, a chain of plasmoids move along the current sheet and merge with the leading plasmoid, which evolves into an ellipsoid of width $\sim\,$4$''$ as hot as 14--19~MK immediately before the eruption. This leading plasmoid is termed a `seed' flux rope, because it keeps a coherent shape --- a hollow ellipsoid in AIA 131~{\AA} with an aspect ratio of about 1.5 -- -as it balloons into the CME bubble; it also exhibits two legs connecting to the surface, revealing its three-dimensional nature. 

It has also been noticed that a seed flux rope may arise in the form of a hot plasmoid during one or a few confined flares at the same section of PIL, where an eruption takes place later \citep[e.g.,][]{Patsourakos2013}. The formation and initial slow rise of the rope may set up a topology (e.g., HFT) that favors coronal magnetic reconnection or lift the rope toward the critical height at which the torus instability or mechanical loss of equilibrium becomes relevant. \citet{Patsourakos2013} identified a seed flux rope during a confined flare 7 hrs before its ejection on 2012 July 19. A few observational features argue in favor of the flux-rope interpretation, i.e., the structure has a high degree of coherency, displaying a plasmoid core threaded by ``half-loop'' legs; the plasmoid core is as hot as 10 MK, consisting of intertwined threads; it is formed by a continuous addition of new outer layers; the bottom part of the core together with the underlying cusp-shaped flare loops resembles an ``X'', probably corresponding to the HFT beneath the flux rope.

\subsubsection{Buildup} \label{subsubsec:buildup}
In the standard two-dimensional or 2.5-dimensional flare model, magnetic reconnection takes place at a vertical current sheet below an erupting flux rope, and the amount of flux closed down into post-flare loops is identical to that closing up into the flux rope. How the flaring reconnection proceed in the corona is mapped by bright flare ribbons in the chromosphere, which respond instantly to the energy transported downward along field lines from the reconnection site, because the timescales of magnetic reconnection, energy transport, and heating of the lower atmosphere (a fraction of a second to a few seconds) are much shorter than the cooling time of flare ribbons (several minutes). Due to flux conservation, a connection between the coronal field undergoing reconnection and the lower-atmosphere field at the energy deposit site is established as follows \citep{Forbes&Priest1984,Forbes&Lin2000},
\begin{equation} \label{eq:rec_rate}
\frac{\partial\mathit{\Phi}_\mathrm{r}}{\partial t}=\frac{\partial}{\partial t}\int B_c\,dS_c = \frac{\partial}{\partial t}\int B_n\,dS_n,
\end{equation} 
where $\mathit{\Phi}_\mathrm{r}$ is the reconnection flux and $\partial\mathit{\Phi}_\mathrm{r}/\partial t$ gives the magnetic reconnection rate. $\mathit{\Phi}_\mathrm{r}$ is given by integrating the inflow field $B_c$ at the reconnection site over the reconnection area $S_c$ in the corona. But since the measurement of coronal field is unavailable, one can obtain $\mathit{\Phi}_\mathrm{r}$ by measuring the magnetic flux swept by flare ribbons as they separate in the lower atmosphere, i.e., by integrating the field normal component $B_n$ over the ribbon-swept area $S_n$ \citep[e.g.,][]{Qiu2004}. It is noteworthy that Eq.~\ref{eq:rec_rate} has been generalized to three dimensions \citep{Forbes&Lin2000}, although it is originally derived for two-dimensional configurations with a translational symmetry \citep{Forbes&Priest1984}. The basic assumption underlying Eq.~\ref{eq:rec_rate} is that the ribbon motion observed in the
lower atmosphere depicts the motion of a separatrix that separates two topologically distinct domains, i.e., a three-dimensional generalization of the two-dimensional concept of magnetic reconnection. If there is a translational symmetry along the ribbon, the equation is reduced to \citep{Forbes&Priest1984} 
\begin{equation}
E_c=V_\parallel\,B_n,
\end{equation}
which gives a uniform electric field $E_c$ along the vertical current sheet through the horizontal velocity $V_\parallel$ of the ribbon expansion.

Assuming that a flux rope does not change its magnetic morphology during its transit in interplanetary space, one can learn a lot about the flux rope by comparing magnetic flux reconnected in the low-corona and the flux inside the associated magnetic cloud (MC). If the flux rope contains a significant amount of pre-existent flux, one expects to find a poor correlation between the toroidal MC flux at 1~AU and the low-corona reconnection flux, at the same time a much greater poloidal MC flux than the reconnection flux. On the other hand, if the flux rope is mainly formed on the fly, one expects that the reconnection flux can largely account for the MC flux. Despite large uncertainties in estimating the MC flux, it was found that the poloidal MC flux at 1~AU is comparable to and scaled with the total reconnection flux, and that the toroidal MC flux is only a fraction of the reconnection flux \citep{Qiu2007,Hu2014}. This result has a few important implications: 1) magnetic clouds are highly twisted (see \S~\ref{subsec:struc:twist}), since the average twist can be estimated by the ratio of poloidal over toroidal flux \citep[see also][]{WangY2016}; 2) even if a flux rope is preexistent, it does not possess a significant amount of magnetic flux before eruption; 3) magnetic reconnection during flares contribute a significant amount of magnetic flux to CME flux ropes. 

It is possible to directly observe the buildup process of an erupting flux rope. When the viewing angle is favorable, limb events may display the evolution of the rope's cross section, similar to coronal cavities; disk-center events, on the other hand, display the evolution of the rope's feet. In two limb events on 2013 March 15 and 2017 September 10, the erupting flux rope is found to have a hot rim ($T\simeq10-15$~MK) enclosing a dark cavity, when both its height and cross-section area increase exponentially \citep{Gou2019,Veronig2018,Yan2018,Cheng2018cs}. This absence of plasma emission inside the flux rope is expected for a twofold reason: first, the cavity plasma is brought into the rope by earlier reconnections at the current sheet, and have since cooled down via conduction, radiation, and expansion \citep{Lin2004}; second, plasma pressure is depressed so that the total pressure, which is dominated by magnetic pressure, can be balanced with the surroundings. The hot rim, on the other hand, is produced by the most recent reconnections in the current sheet beneath, which feed poloidal flux and frozen-in plasma to the flux rope. The strong acceleration of the flux rope is associated with a rapid increase in  X-ray fluxes, as well as a sudden increase in inflow speeds exceeding 100~\kms \citep{Gou2017,Yan2018,Cheng2018cs}, suggesting an enhanced reconnection rate induced by the plasmoid ejection. The bursty HXR emission in the nonthermal range ($\ge25$ keV) suggests that the reconnection-related electric field rapidly varies with time and/or space, therefore modulating the particle acceleration \citep{Gou2017,Gou2019}. In both events \citep{Gou2019,Veronig2018}, the ellipsoid evolves into the core of the white-light CME, which challenges the traditional interpretation of the CME three-part morphology.  

The buildup of a flux rope as manifested by the evolution of its feet can be mapped by post-eruptive coronal dimmings, due primarily to the ejection of emitting plasma and the expansion of the eruptive structure \citep{Webb2000,Attrill2006,Qiu2007,Cheng&Qiu2016,Dissauer2018}, and occasionally by hard X-ray \citep{Liu&Alexander2009,Guo2012} or microwave emission \citep{ChenB2020}. Plasma diagnostics shows that dimmings are associated with outflowing material, and that the mass loss in dimming regions accounts for a significant fraction of the CME mass \citep[e.g.,][]{Harra&Sterling2001,Tian2012,Veronig2019}. Hence, post-eruptive dimmings, particularly the core dimmings that consist of a conjugated pair associated with opposite polarities of the photospheric field, are intimately related with the CME expansion and the flare reconnection. The pre-eruptive dimmings, on the other hand, are mostly associated with the slow rise and gradual expansion of the eruptive structure \citep{Qiu&Cheng2017,WangW2019}. The dimming regions may expand, shrink, or drift during the eruption \citep[e.g.,][]{Miklenic2011,WangW2017,Dissauer2018}, reflecting the interaction between the erupting flux rope and the surrounding fields \citep{Aulanier&Dudik2019}. The magnetic flux in dimming regions provides an estimate of the toroidal flux in the erupting flux rope \citep{Webb2000,Qiu2007}. \citet{Temmer2017} reported a post-flare increase in the dimming flux by more than 25\%, suggesting a continual flux contribution to the flux rope after the eruption. In a different case, \citet[][their Supplementary Figure 4]{WangW2017} found that the dimming flux slowly diminishes during the gradual phase, which is likely due to interactions between the the flux rope and the ambient field. This is further evidenced by in-situ observations \citep[][their Figure 4 and Supplementary Note]{WangW2017}: the suprathermal electron beams are dominated by an unidirectional flow rather than counterstreaming flows that are often associated with ICMEs, and the ICME's interval as determined by plasma composition extends several hours before and after the MC boundary. \citet{WangW2019} combined the pre- and post-eruptive dimmings to measure the magnetic flux and electric current through the feet of a flux rope, and found that its magnetic twist increases from $1.0\pm0.5$ to $2.0\pm0.5$ turns during a five-hour pre-eruptive dimming period, and further to about 4.0 turns~AU$^{-1}$ when it arrives at Earth. This kind of analysis, however, is subject to large uncertainties because dimming regions are often diffuse and poorly defined in EUV images. 

\begin{figure}[htp!] 
	\centering
	\includegraphics[width=\hsize]{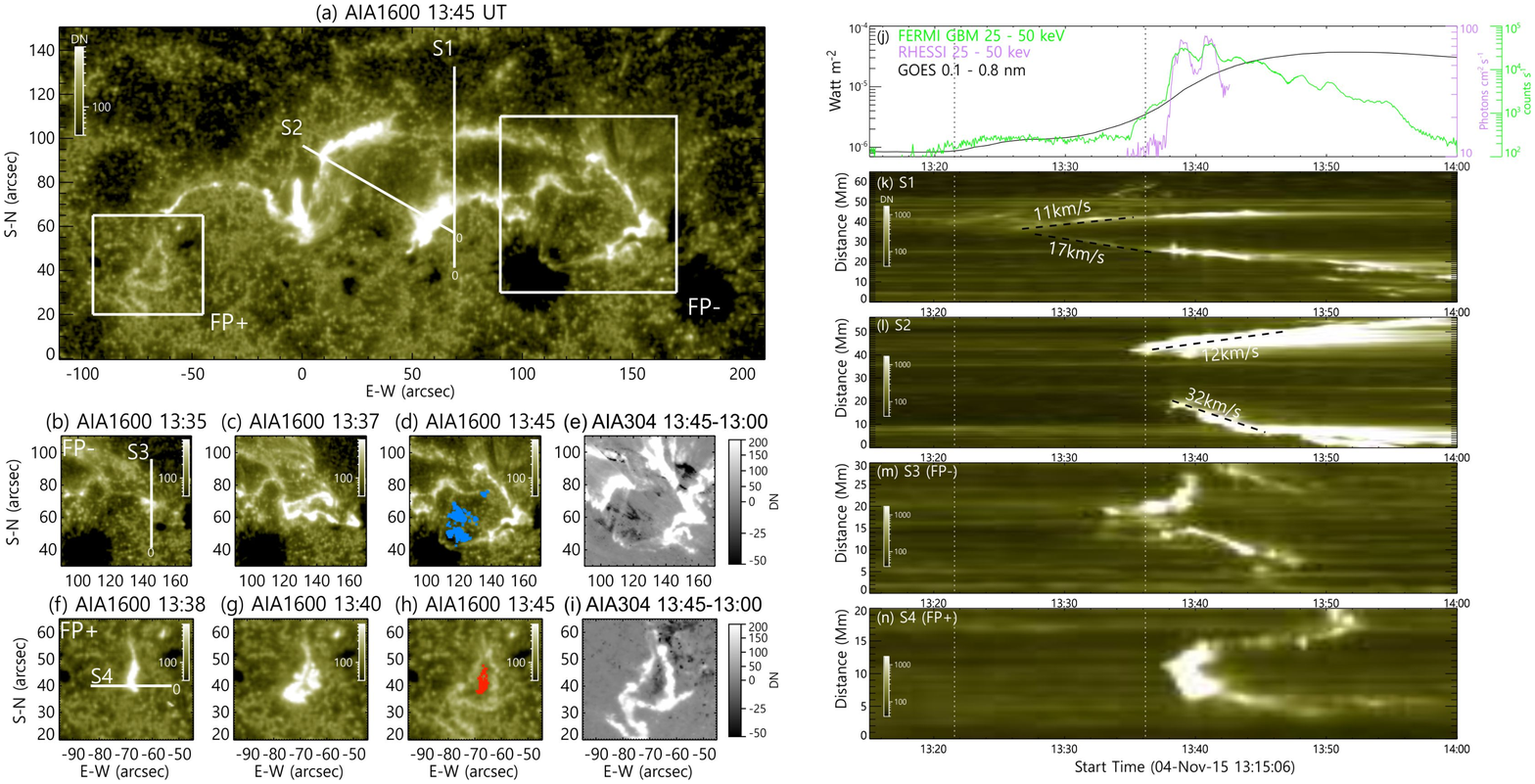}
	\caption{Formation and evolution of a flux rope's feet \citep[adapted from][]{WangW2017}. Panel \textbf{(a)} shows the flare morphology with two irregular bright rings (marked by rectangles) attached to the far ends of flare ribbons, observed by AIA 1600~{\AA}. The evolution of the feet associated with negative polarity (FP-) is shown in \textbf{(b--e)} and that associated with positive polarity (FP+) in \textbf{(f--i)}. Panels \textbf{(e and i)} show coronal dimmings in 304~{\AA} base-difference images. The dimmed pixels within FP- and FP+ are replotted in (d) and (h) in blue and red colors, respectively. Four virtual slits (S1--S4) are indicated in Panels (a), (b), and (f), with their starting points marked by `0'. The evolution of the flare ribbons as seen through S1 and S2 and the evolution of the bright rings as seen through slits S3 (for FP-) and S4 (for FP+) are shown in Panels \textbf{(k--n)}, respectively. Panel \textbf{(j)} shows 0.1--0.8 nm flux obtained by the Geostationary Operational Environmental Satellite (GOES), 25--50 keV count rate by the Gamma-ray Burst Monitor (GBM) onboard the Fermi Gamma-ray Space Telescope, and 25--50 keV photon flux observed by the Reuven Ramaty High-Energy Solar Spectroscopic Imager (RHESSI). The 1st vertical dotted line marks the flare onset with the initial appearance of flare ribbons; the 2nd line marks the beginning of the flare main phase characterized by the rapid ribbon separation and nonthermal hard X-ray production.} 
	\label{fig:fp}
\end{figure}

It has also been proposed that the footprints of the QSLs wrapping around the flux rope correspond to a pair of J-shaped ribbons of high electric current densities, and that the flux rope is anchored within the hooked parts of the ribbons \citep{Aulanier2012, Aulanier2013, Janvier2013}. However, the hooks are usually `open' and hence poorly demarcate the rope's feet. Such an `open' double-J morphology is taken as an indication that field lines of the flux rope are twisted by no more than one turn \citep{Janvier2014}. \citet{WangW2017} recognized a new flare-ribbon morphology --- double J-shaped ribbons with closed hooks in an M3.7-class flare on 2015 November 4. The eruption leads to a flux rope which passes through near-Earth spacecrafts as a typical magnetic cloud 3 days later. At first look, the flare is a classic two-ribbon one, but when the two ribbons extend to their full length along the PIL in a zipping-like fashion, co-temporal with a gradual increase in soft X-rays, an irregular bright ring takes form and expands outward from the far end of each ribbon, which is associated with coronal dimming developing inside each ring, rapid ribbon separation, and impulsive HXR bursts (Figure~\ref{fig:fp}). The conjugated pair of dimmings, which map the rope's feet, are clearly visible and fully encompassed by the bright rings in all of AIA's EUV passbands, indicating mass depletion along the rope legs. Each ring originates and expands outward from a point-like brightening, which strongly suggests that the bulk of the rope is formed on the fly. Counting magnetic flux through the feet as enclosed by the bright rings and that through the ribbon-swept area, \citet{WangW2017} reveals that the rope's core is highly twisted (up to $\sim\,$10 turns), much more than its average of $\sim\,$4 turns. We will further look into the twist profile of flux ropes in \S~\ref{subsec:struc:twist}.

\section{Structure} \label{sec:struc}

\subsection{Twist Profile} \label{subsec:struc:twist}
So far magnetic twist, like other characteristics of flux ropes on the Sun, can only be inferred indirectly from observation and modeling. Occasionally hard X-ray and microwave emission is observed at the projected crossing point of the kinked, inverted $\gamma$ or $\delta$-shaped filaments \citep{Alexander2006, Liu&Alexander2009, Cho2009, Karlicky&Kliem2010}, which suggests that the their legs approach each other and interact near the crossing point. This is possible when the flux rope is highly twisted \citep{Kliem2010,Hassanin&Kliem2016}. \citet{Vrsnak1991} measured the pitch angles of helical-shaped threads in H$\alpha$ prominences and found that the twist angles in eruptive prominences are generally larger (up to $\sim15\pi$) than stable ones. Using the same method, \citet{Romano2003} derived that the twist angle of one helical thread in a prominence is about 10$\pi$ and relaxes to about 2$\pi$ during the eruption. \citet{Srivastava2010} estimated that a coronal loop observed in the EUV 171~{\AA} passband possesses a twist angle of about 12$\pi$ according to the bright-dark alternating streaks along its long axis. \citet{Yan2014} inferred that the twist angle of a filament is at least 5$\pi$ based on the observed unwinding motions. \citet{Gary&Moore2004} found an erupting four-turn helical structure observed in the UV 1600~{\AA} passband. Since the estimated twist far exceeds the critical number of $2.5\pi$ or 1.25 turns found for a line-tying force-free flux rope with uniform twist \citep[][see also  \S~\ref{subsubsec:KI}]{Hood&Priest1981}, the helical kink instability is suggested to be involved in these eruptions. However, it is difficult to assess to what extent one can trust such estimates of twist, because they suffer inevitably from either the projection effects or the line-of-sight confusion, in addition to complex interactions between magnetic field and plasma.

Alternatively, the magnetic field in the upper atmosphere can be constructed by NLFFF extrapolation techniques and magnetohydrostatic or magnetohydrodynamic models \citep[see the review by][]{Inoue2016,Guo2017,Wiegelmann2017}, based on the pre-eruption photospheric magnetic field. Flux ropes reconstructed in strong field regions tend to be moderately twisted ($\le 2$ turns) and low lying \citep[e.g.,][]{Regnier2002, Guo2010rope, Jing2010, Inoue2011,  Liu2014, Liu2016, Chintzoglou2015,James2018}, while those in weak-field regions tend to be highly twisted and high lying \citep[e.g.,][]{Jiang2019, Guo2019reconstruction, Su&vanBallegooijen2012, Su&vanBallegooijen2013, Su2015}. In addition, the magnetic reconnection taking place at the vertical current sheet beneath an erupting flux rope will add a considerable amount of magnetic fluxes into the rope by converting overlying field to its envelope \citep{Lin2004,Qiu2007}. The post-eruption structure of the flux rope can be directly measured as a magnetic cloud by in-situ instruments on-board spacecrafts.

Revealed by various fitting and reconstruction techniques, some magnetic clouds are found to possess a large twist density, e.g., 8 turns per AU found in a magnetic cloud by \citet{Farrugia1999} and 2.4 turns per AU by \citet{Dasso2006}. \citet{WangY2016} applied a velocity-modified Gold-Hoyle model \citep{WangY2015} to 126 magnetic clouds, and found the distribution of twist density has a median value of about 5 turns per AU. Assuming that the cloud axial length ranges between 2--$\pi$ AU, \citet{WangY2016} concluded that most of magnetic clouds have a twist angle significantly larger than the theoretical kink-instability threshold of $2.5\pi$ radians, but well bounded by 2 times a cylindrical rope's aspect ratio. The total twist is subject to large uncertainties, as twist density may not be uniform along the MC field lines, whose lengths are also unknown. One way to infer the field-line length is to employ the velocity dispersion profile of in-situ energetic particles. For example, \citet{Larson1997} inferred that in a magnetic cloud the field-line length varies from about 3 AU near the edge to about 1.2 AU near the center. Expanding the study to more cases, \citet{Kahler2011JGR} found that the field-line lengths inside magnetic clouds range between 1.3 and 3.7 AU.  

Besides the local twist density, the Grad-Shafranov technique is capable of reconstructing the twist profile inside magnetic clouds \citep{Hu&Sonnerup2002,Hu2017rev}, i.e., the distribution of twist density in a cloud's cross section. Assuming a translation symmetry along the axis, this method solves the Grad-Shafranov equation in the plane perpendicular to the flux-rope axis. \citet{Hu2014} studied twist profiles inside 18 magnetic clouds and found that about half of the cases that are associated with erupting filaments have a nearly uniform and relatively low twist, while the other half exhibit high twist ($\ge5$ turns per AU) near the axis but low twist toward the edge. Further, \citet{Hu2015} estimated the field-line lengths based on Grad-Shafranov fitting results and found a good correlation with those estimated from the energetic electrons by \citet{Kahler2011JGR}. The measured field-line lengths are more consistent with \citet{Gold&Hoyle1960} than \citet{Lundquist1950} flux-rope models. 

Additional insight into the twist profile inside magnetic clouds can be obtained when a magnetic cloud reconnects with the ambient interplanetary magnetic field, by which magnetic fluxes are peeled off from the cloud \citep{McComas1994,Wei2003JGR,Dasso2006,Ruffenach2015}. \citet{WangY2018} studied a magnetic cloud observed sequentially by four spacecrafts near Mercury, Venus, Earth, and Mars, respectively, and found that the axial flux and helicity of the cloud decrease but its twist increases with increasing heliocentric distance. The imbalance in the azimuthal flux of the cloud at far distances implies that it has been eroded significantly. The erosion together with the increase in twist suggests that the cloud has a highly twisted core enveloped by a less twisted envelope.

\begin{figure}[htp!]
	\centering
	\includegraphics[width=\hsize]{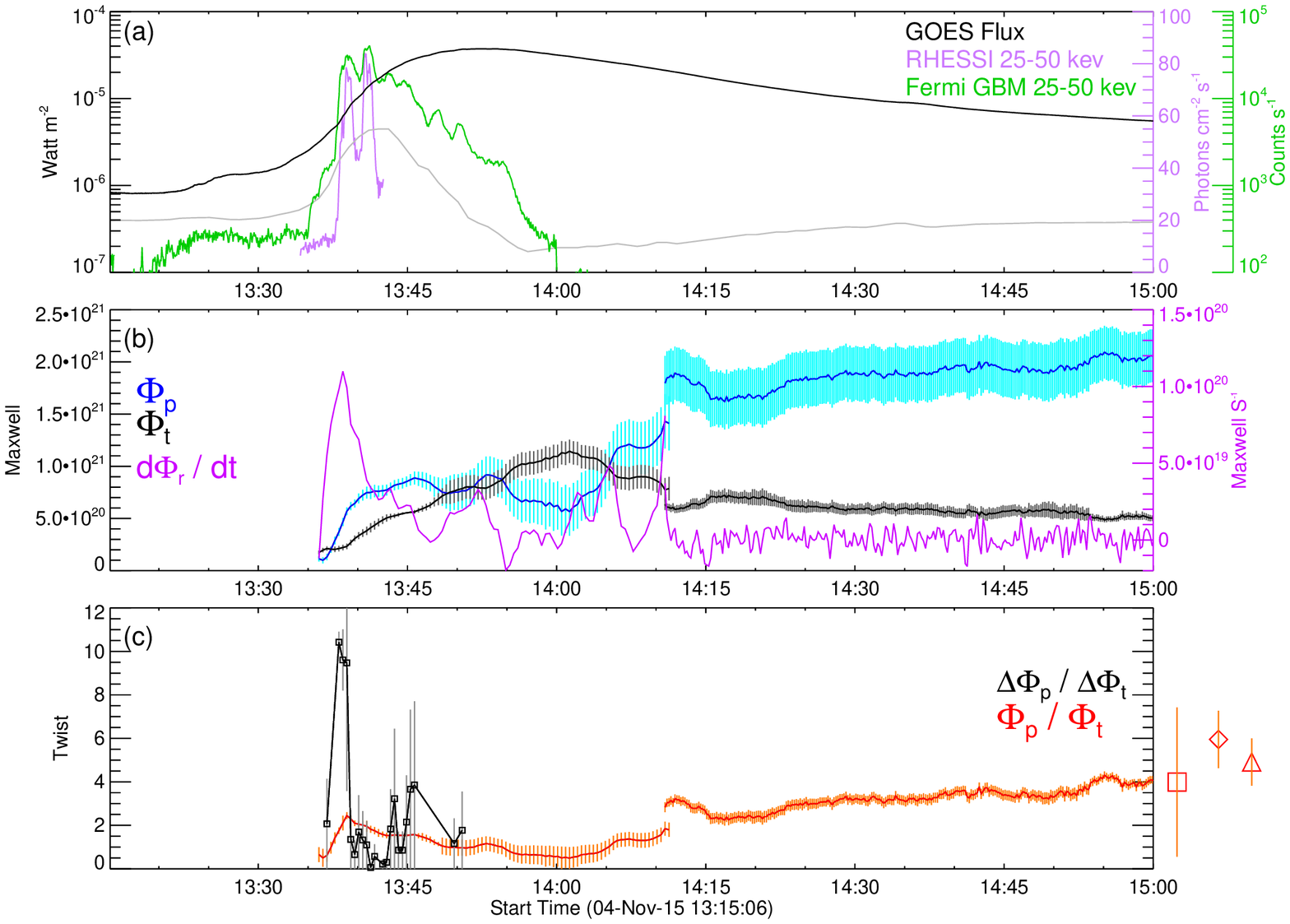}
	\caption{Temporal evolution of a flux rope in relation to the flare on 2015 November 4 \citep[from][]{WangW2017}. \textbf{a)} 0.1--0.8 nm SXR flux observed by the Geostationary Operational Environmental Satellite (GOES; black), its time derivative in an arbitrary unit (gray), and 25--50 keV HXR count rate observed by the Gamma-ray Burst Monitor (GBM) on-board the Fermi Gamma-ray Space Telescope (green) and by the Reuven Ramaty High-Energy Solar Spectroscopic Imager (RHESSI; magenta). \textbf{b)} Temporal evolution of poloidal flux $\mathit{\Phi}_\mathrm{p}$ and toroidal flux $\mathit{\Phi}_\mathrm{t}$ in the flux rope. Also shown is the time derivative of reconnection flux $\mathit{\Phi}_\mathrm{r}=\mathit{\Phi}_\mathrm{p} + \mathit{\Phi}_\mathrm{t}$. \textbf{c)} Temporal evolution of twist number in the flux rope as gauged by $\mathit{\Phi}_\mathrm{p}/\mathit{\Phi}_\mathrm{t}$ and $\Delta\mathit{\Phi}_\mathrm{p}/\Delta\mathit{\Phi}_\mathrm{t}$. Marked on the right are $\mathit{\Phi}_\mathrm{p}/\mathit{\Phi}_\mathrm{t}$ given by the Gold-Hoyle (square) and Lundquist (diamond) fittings and the Grad-Shafranov reconstruction (triangle) of the interplanetary magnetic cloud. \label{fig:ltc}. } 
\end{figure}

Efforts have been made to compare the flux content of magnetic clouds with that of coronal flux ropes estimated from dimming signatures of eruptions \citep{Webb2000,Mandrini2005,Attrill2006,Jian2006,Qiu2007}. Post-eruption dimmings are caused by mass depletion and often appear in a pair near the flaring PIL, thus mapping the feet of the eruptive flux rope. These studies indicate a dominance of the poloidal flux over the axial flux, typically by a factor of 3 \citep[e.g.,][]{Mandrini2005, Attrill2006, Qiu2007}. However, dimming regions are often diffuse, lacking in a definite boundary. \citet{WangW2017} made a first comparative study on the twist profile between a coronal flux rope and its interplanetary counterpart, taking advantage of a rare observation in which the rope's feet are clearly identified and their formation starting from two brightening points during the eruption are closely monitored. \citet{WangW2017} obtained the rope's toroidal (axial) flux $\mathit{\Phi}_\mathrm{t}$ as the magnetic flux through the footpoint regions, which are identified as the conjugated coronal dimmings completely enclosed by irregular bright rings at the far ends of flare ribbons (Figure~\ref{fig:fp}). The temporal variation of $\mathit{\Phi}_\mathrm{t}$ indicates the growth of the flux rope with time, but more importantly gives a glimpse of different `shells' that build up sequentially within the rope (Figure~\ref{fig:ltc}). The rope's poloidal flux $\mathit{\Phi}_\mathrm{p}$ can be derived from the magnetic flux swept by flare ribbons, i.e., the reconnection flux $\mathit{\Phi}_\mathrm{r}$ \citep{Forbes&Priest1984,Qiu2007}. Since the flux rope in this particular case is mainly formed during the course of the eruption, $\mathit{\Phi}_\mathrm{r}$ may account for most of the rope's flux, i.e., $\mathit{\Phi}_\mathrm{r}\approx\mathit{\Phi}_\mathrm{p}+\mathit{\Phi}_\mathrm{t}$. However, counting brightened flare area in the chromosphere gives the flux swept by both the flare ribbons and rings, denoted by $\mathit{\Phi}_\mathrm{R}$. It is important to keep in mind that the ribbons and the rings correspond to footpoints of topologically distinct magnetic structures: the rings highlight footpoints of the longer reconnected field lines that are newly assimilated by the burgeoning rope, therefore marking the boundary of the rope's feet, while the ribbons represent footpoints of the shorter reconnected field lines that pile up upon the growing flare loop system. But the rings are only slightly dimmer than, and therefore cannot be distinguished from, the ribbons; as a result, $\mathit{\Phi}_\mathrm{t}$ is counted twice in $\mathit{\Phi}_\mathrm{R}$, which leads to $\mathit{\Phi}_\mathrm{p}\approx\mathit{\Phi}_\mathrm{R}-2\mathit{\Phi}_\mathrm{t}$. 

Using $\Delta\mathit{\Phi}_\mathrm{p}(t)/\Delta\mathit{\Phi}_\mathrm{t}(t)$ to estimate the twist number at a certain shell and $\mathit{\Phi}_\mathrm{p}/\mathit{\Phi}_\mathrm{t}$ to measure the average twist number across the rope, \citet{WangW2017} inferred that the spatial distribution of magnetic twist within the flux rope is characterized by a highly twisted core (up to about 10 turns) and less twisted outer shells (down to 1--2 turns; Figure~\ref{fig:ltc}c). This is corroborated by the Grad-Shafranov reconstruction of the corresponding magnetic cloud, which exhibits high twist in the center ($\sim\,$2.7 turns per AU, or, 5.4--8.5 turns given the cloud axial length ranging between 2--$\pi$ AU) and lower twist towards the boundary. Further, \citet{WangW2017} showed similar temporal variations  between $\Delta\mathit{\Phi}_\mathrm{p}(t)/\Delta\mathit{\Phi}_\mathrm{t}(t)$ (Figure~\ref{fig:ltc}c), the nonthermal HXR emission \citep[Figure~\ref{fig:ltc}a, a proxy of both the particle and CME acceleration;][]{Temmer2010}, and the time derivative of the reconnection flux $\mathit{\Phi}_\mathrm{r}$ (Figure~\ref{fig:ltc}b, a proxy of the reconnection rate), which suggests that the instantaneous twist number may reflect the frequency of reconnections between sheared field lines, with each reconnection adding roughly one turn into the twisted field line in formation.

From theoretical perspectives, the flare ribbon morphology contains clues to the twist profile of flux ropes. Analytical models demonstrate that the photospheric footprints of the QSLs wrapping around the flux rope display two J-shaped ribbons, while the rope is anchored within the hooked parts \citep[e.g.,][]{Demoulin1996rope,Titov&Demoulin1999,Pariat&Demoulin2012}. A pair of closed hooks are associated with a flux rope of high twist numbers \citep[$\ge2$ turns;][]{Demoulin1996rope,Pariat&Demoulin2012}, in agreement with \citet{WangW2017}, while a pair of open hooks are associated with a flux rope of moderate twist numbers \citep[$\le 1$ turn;][]{Janvier2014}. Twisted field lines are believed to be produced by magnetic reconnection between sheared field lines, which converts mutual to self helicity  \citep{vanBallegooijen&Martens1989, Longcope&Beveridge2007}. 

\citet{Priest&Longcope2017} proposed that the magnetic twist builds up in the same pace as flare reconnection, which often proceeds in two phases: first the 3D ``zipper reconnection'' of sheared flux, which is associated with the extension of flare ribbons along the PIL, and then the quasi-2D ``main phase reconnection'' of unsheared flux around the flux rope, which is associated with the separation of flare ribbons away from the PIL. The zipper reconnection in a sheared arcade creates a flux rope of roughly one turn; but if starting with a preexisting flux rope under the arcade, the zipper reconnection can add substantial extra twist to the rope. Either way, the subsequent main-phase reconnection adds a layer of uniform twist of only a few turns. In this model, most of twist in the flux rope is created by zipper reconnection prior to the eruption \citep{Threlfall2018}. This is at odds with the observation made by \citet[][Figure~\ref{fig:ltc}]{WangW2017}, in which the highly twisted core is produced at around the HXR peak during the main phase rather than the zipper phase. On the other hand, motivated by \citet{Priest&Longcope2017}, \citet{Xing2020} argued that the magnetic shear angle is reflected in the geometric shear of flare ribbons, assuming a uniform distribution of magnetic flux in the area swept by flare ribbons. Applying this idea to four cases of two-ribbon flares with a preexisting flux rope but no obvious zipper ribbon motion, \citet{Xing2020} found that the preexisting rope may possess a significant amount of toroidal flux compared with that contributed by the quasi-2D reconnection during the main phase. A similar geometric argument was made by \citet{Green2011} to estimate how much flux is contributed by flux cancellation to a flux rope forming in a sigmoidal active region.

The above studies suggest that a CME flux rope builds up like an onion with nested layers of magnetic flux added sequentially as the eruption progresses, by which a non-uniform twist profile is often resulted \citep{Hu2014,WangW2017}. \citet{Awasthi2018} brought up an alternative scenario, in which multiple flux ropes or flux bundles braid about each other and signatures of internal reconnection including nonthermal electrons, flaring plasma, and bidirectional outflowing blobs are identified. In general, a force-free flux rope embedded in potential field is expected to have a non-uniform radial twist profile to match the field at the rope surface \citep[e.g.,][their Figure 2]{Torok2004}. A quasi-separatrix layer must exist between the layers of strongly different twist. As electric currents are prone to accumulate at QSLs (\S~\ref{subsec:topology}), heating and dynamic motions are expected inside flux ropes of nonuniform twist profiles.  MHD simulations of the prominence-cavity system \citep{Xia2014,Fan2018,Fan&Liu2019} also reveal that there exist different types of twisted field lines threading the cavity. These field lines also possess different thermodynamic properties, therefore giving rise to the substructures of different appearances in EUV images.  This may help understand disk-like or ring-like substructures observed in coronal flux ropes, even in a stable state, e.g., bright U-shaped ``horns'' that extend nonradially from the top of prominences \citep[Figure~\ref{fig:cavity}a;][]{Regnier2011,Schmit&Gibson2013}, hot central cores of cavities \citep[also termed ``chewy nougats'', Figure~\ref{fig:cavity}b;][]{Hudson1999,Reeves2012}, and nested ring-like Doppler patterns within cavities \citep[Figure~\ref{fig:cavity}c;][]{Bak-Steslicka2016}.

\subsection{Boundary} \label{subsec:struc:boundary}
\begin{figure}[htp!] 
	\centering
	\includegraphics[width=0.7\textwidth]{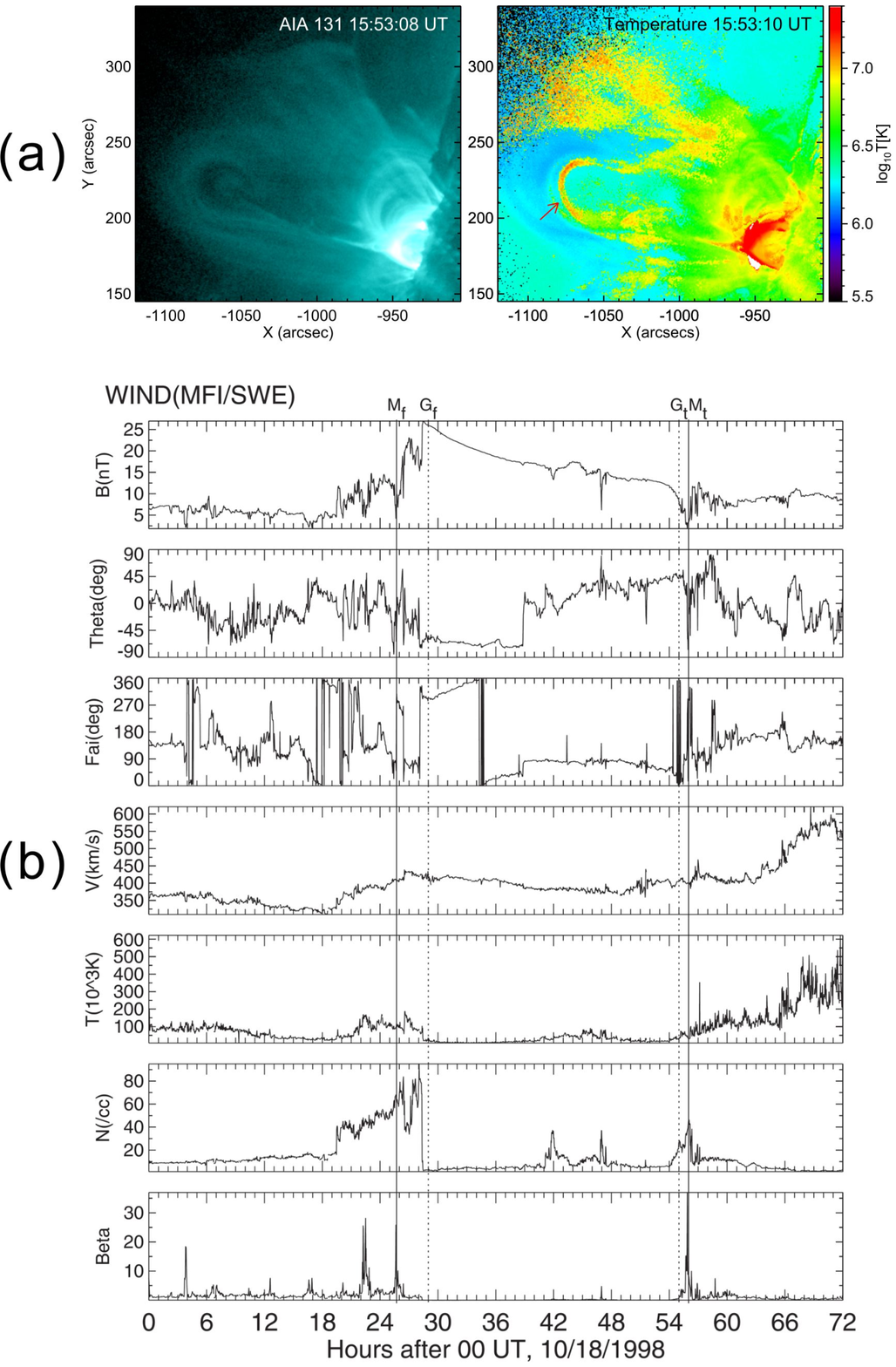} 
	\caption{Boundary of erupting flux ropes observed near the Sun and in interplanetary space. \textbf{a}: a flux rope observed by SDO/AIA at 131~{\AA} (left), which is enclosed by a hot shell of temperatures over 10 MK as shown in the temperature map \citep[right; adapted from][]{Gou2019}. \textbf{b}: an interplanetary flux rope, also known as magnetic cloud, which is enclosed by a front and a tailing boundary layer as marked by vertical lines \citep[adapted from][]{Wei2003JGR}. The boundary layers are characterized by enhanced proton temperature, proton density, and plasma $\beta$, a sudden drop in field magnitudes, and abrupt changes in field directions. \label{fig:boundary} }
\end{figure}

Both analytical models \citep[e.g.,][]{Demoulin1996rope, Titov&Demoulin1999, Pariat&Demoulin2012} and numerical experiments based on NLFFF extrapolations \citep[e.g.,][]{Guo2013,Liu2016} or MHD simulations \citep[e.g.,][]{Gibson&Fan2006mfr,Aulanier2012,Jiang2018} have demonstrated that a coherent flux rope is wrapped around by a thin volume of strong magnetic field distortion---separatrix surfaces or QSLs that separate the twisted from untwisted fields (\S~\ref{subsec:topology}). Below we elaborate on observational signatures for the boundary of flux ropes. 

For quiescent flux-rope proxies such as coronal cavities, this boundary is strongly contrasted between the dark cavity and the surrounding bright streamer (Figure~\ref{fig:cavity}). It is interesting that cavities appear to be more sharply defined prior to eruption \citep{Gibson2015}. The U-shaped horn enclosing the bottom part of the cavity (Figure~\ref{fig:cavity}a) is thought to be associated with the HFT topology \citep{Fan2012}. 

For erupting flux ropes observed above the limb, naturally the expanding and rising flux rope would stretch and compress the overlying cold loops ($T\simeq1-2$~MK), which pile up and subsequently become part of the erupting structure. \citet{Veronig2018} found that the outer front of these expanding and piled-up loops observed in EUV seamlessly matches with the CME front in white light. The stretched overlying loops often exhibit an $\Omega$ shape because the rising and expansion of the flux rope causes the overlying loops to curve in toward the current sheet beneath the erupting flux rope \citep{Cheng2011blob,Gou2017,Gou2019}. \citet{Gou2019} found that when the rope's rising speed peaks at over 500~\kms, a similar $\Omega$-shaped, thin layer as hot as 14--19~MK appears to separate the overlying loops from the flux rope (Figure~\ref{fig:boundary}a). Detailed DEM analysis shows that the layer is significantly hotter and denser than both the overlying loops and the flux rope, probably lighten up by the plasma compression as well as the current steepening and dissipation at the magnetic boundary that separates the twisted from untwisted field, which exists not only prior to but also during eruptions \citep[e.g.,][]{Jiang2008}. In the 2017 September 10 event, the erupting cavity of an inverted teardrop shape is enveloped by a hot ($T\sim13$~MK) and dense ($\mathrm{EM}\sim10^{27.5}$~cm$^{-5}$) layer during the early phase of the eruption, while the cavity itself has a lower emission measure ($\mathrm{EM}\sim10^{26}$~cm$^{-5}$) and a slightly cooler temperature \citep[$T\sim10$~MK;][]{Yan2018,Cheng2018cs}. It is interesting that the hot envelope is most prominent at the bottom part, exhibiting a U shape similar to the horn structure in quiescent cavities (Figure~\ref{fig:cavity}a). Underneath the cavity, the current sheet initially has the similar temperature as the cavity's hot envelope, but is further heated to as high as 20~MK, as the flux rope propagate into the outer corona \citep{Cheng2018cs}. 

Regarding flux ropes erupting from the disk, the footpoint boundary of the flux rope is recognized as open or closed hooks of double J-shaped flare ribbons \citep[Figure~\ref{fig:fp};][]{Janvier2014,WangW2017}. The brightening at the boundary of an erupting rope's footpoints or cross section may result either from magnetic reconnections between field lines overlying the rope, taking place at the vertical current sheet underneath the rope \citep{Lin2004}, or from reconnections between the rope and the ambient field \citep{Shiota2010,Hassanin&Kliem2016,Aulanier&Dudik2019}, taking place at the current sheets wrapping around the rope \citep[][see also \S~\ref{subsec:topology}]{Torok2004, Gibson&Fan2006mfr, Aulanier2010, Fan2012}.  

Regarding flux ropes propagating in interplanetary space, \citet{Wei2003JGR} identified a boundary layer exhibiting signatures of magnetic reconnection at both the front and tail of magnetic clouds (Figure~\ref{fig:boundary}b). The boundary layer is characterized by a local increase in proton temperature, proton density, and plasma $\beta$, a sudden drop in field magnitudes, and abrupt changes in field directions suggesting a field reversal \citep{Wei2003GRL}. \citet{Tian2010} confirmed that a significant fraction of small-scale interplanetary flux ropes show boundaries of similar plasma and magnetic-field properties as those of large-scale magnetic clouds. In addition, they identified a plasma jet of $\sim\,$30~\kms at the boundary, suggesting the presence of a Petschek-type reconnection exhaust \citep{Gosling2005}. These reconnection signatures could be relevant to the formation of small flux ropes in the solar wind, or, due to interactions between the flux rope and the interplanetary magnetic fields.

\subsection{Double-Decker Structure} \label{subsec:struc:double-decker}
\begin{figure}[htp!] 
	\centering
	\includegraphics[width=0.8\textwidth]{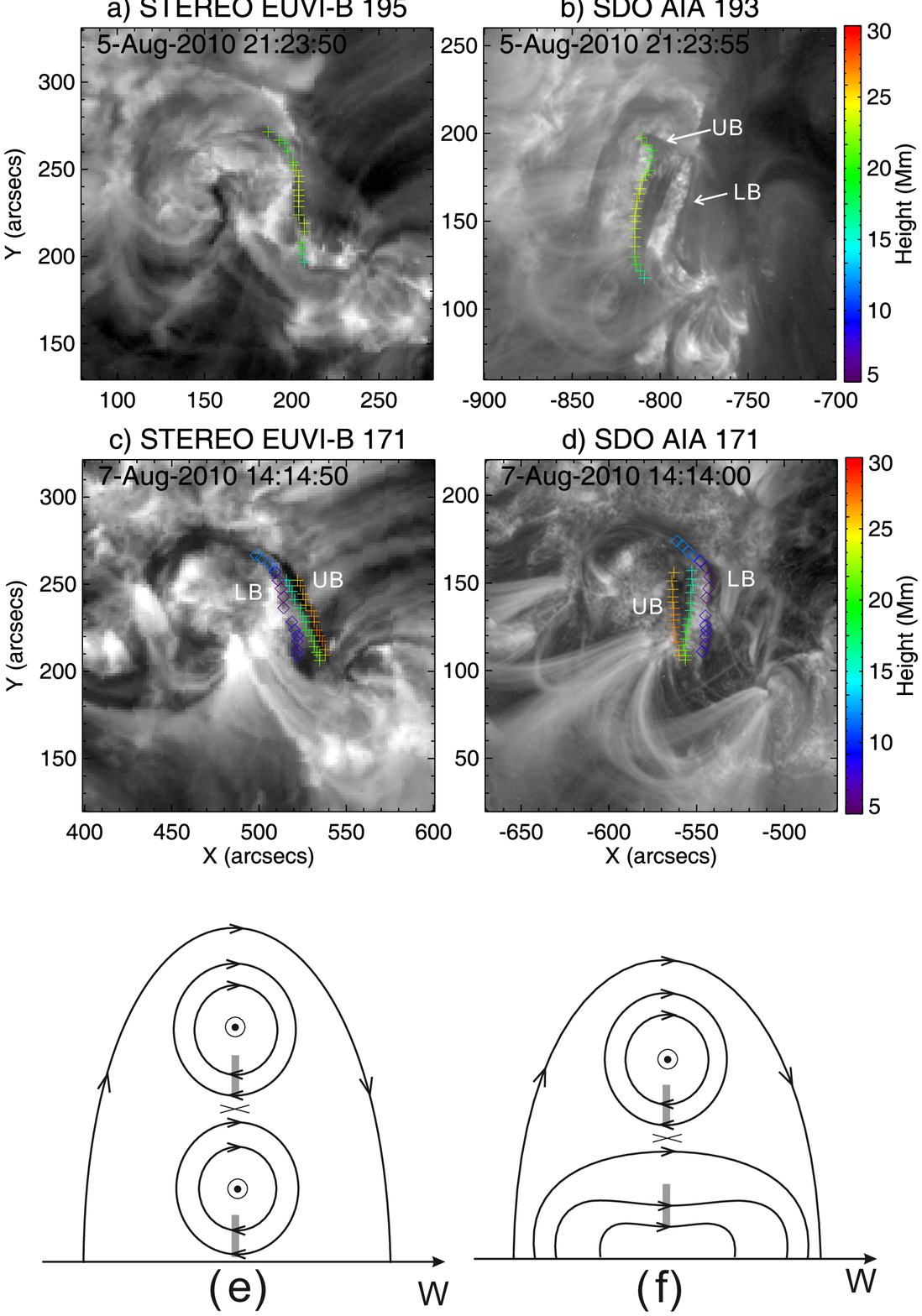} 
	\caption{Double-decker filament and possible magnetic configurations \citep[from][]{Liu2012}. \textbf{a--d)} A double-decker filament observed by both STEREO-B and SDO at different dates. Selected points on the upper branch (UB) are indicated by crosses and those on the lower branch (LB) by diamonds. The heights of these points above the solar surface in units of Mm are obtained through  stereoscopic triangulation and indicated by the color bar on the right. \textbf{e \& f)} Schematic diagrams of two possible magnetic configurations for the double-decker filament in \textbf{a--d}. The cross sections are viewed from the south, with photospheric fields of positive (negative) polarity on the east (west) of the filament. The axial field of both filament branches points out of the plane. Slabs in gray colors indicate the filament mass. The symbol `x'  between the two branches marks where current sheets might develop. \label{fig:doubledeck} }
\end{figure}

As verified stereoscopically  by \citet{Liu2012}, a `double-decker' filament consists of two vertically separated branches that are aligned along the same PIL; it can be stable for hours to days prior to eruption (Figure~\ref{fig:doubledeck}(a--d)). The associated magnetic structure has two possible configurations, a double flux rope or a single flux rope atop a sheared arcade; the two branches, possessing the same sign of helicity, are separated by an HFT \citep[][Figure~\ref{fig:doubledeck}(e \& f)]{Liu2012}. The double-decker configuration may account for a long-observed puzzling phenomenon, namely, sigmoid eruptions that are survived by a stable filament \citep[e.g.,][see also \S~\ref{subsubsec:relation}]{Pevtsov2002, LiuC2007, Cheng2014double}. This can be explained by a double decker whose lower branch embeds a filament but upper branch is avoid of filament material \citep{Cheng2014double}. As demonstrated by the NLFFF modeling of coronal magnetic field, double or even multiple flux ropes can be stacked above the same PIL \cite[e.g.,][]{Liu2016, LiuL2017, Hou2018, Awasthi2018}. The other configuration, a flux rope atop a sheared arcade, is also found in NLFFF modeling \citep{Regnier&Amari2004} and inferred from dynamic motions i.e., rotations about the spine and longitudinal oscillations along the spine in a filament disturbed by a flare surge \citep{Awasthi2019}. In these two cases, however, the two branches possess opposite rather than same signs of helicity, which makes the tilt instability relevant (see \S~\ref{subsubsec:interact}); but without the HFT in between \citep[see also][]{Jelinek2020}, the configuration is more stable in this aspect than that originally envisaged in \citet{Liu2012}.

Double-decker filaments display a wide range of eruptive behavior. Often the upper branch is ejected as a CME while the lower branch remains confined \citep[e.g.,][]{Liu2012, Zhu&Alexander2014,Cheng2014double}. Additionally the two branches may erupt successively but merge into a single CME \citep[e.g.,][]{Dhakal2018}, or merge first into as a coalesced structure before eruption \citep[e.g.,][]{Zhu2015,Tian2018}. Sometimes during the slow rise phase of the upper branch, filament threads within the lower branch intermittently brighten up, lift upward, and merge into the upper branch \citep{Liu2012,Zhu&Alexander2014}. Since filament field is dominantly horizontal \citep{Leroy1989}, these discrete episodes of mass transfer implies a flux transfer from the lower to the upper branch, also termed `flux feeding', which may destabilize the upper branch by producing a flux imbalance in the upper branch relative to the strapping field \citep[e.g.,][]{Su2011,Kliem2013}. 

Along this thought, \citet{Kliem2014} modeled the double-decker configuration with two concentric, toroidal flux ropes, and by reducing (increasing) the flux and current in the lower (upper) rope, they were able to reproduce the ejection of the upper rope with the lower rope being stable. Alternatively, \citet{Kliem2014} identified a double flux rope in MHD simulations of a filament eruption, in which the highly sheared core field is gradually energized through flux cancellation driven by photospheric flows. Unlike the observed double-decker filaments with two clearly separated branches \citep[e.g.,][]{Liu2012, Zhu&Alexander2014}, the two ropes in this simulation are initially merged to some extent until shortly before the eruption a splitting is caused by tether-cutting reconnection with the ambient field at the HFT between the two ropes. The reconnection adds flux and twist to the upper rope while strengthening the overlying flux above the lower rope, therefore leading to a partial eruption. \citet{Xia&Keppens2016}, on the other hand, focused on the internal dynamics of a twin-layer filament set up in a gravitational stratified atmosphere with dominantly horizontal magnetic field. Their three-dimensional MHD simulations demonstrate falling Rayleigh-Taylor fingers and uprising bubbles that are in line with vertical threads and rising plumes often observed in quiescent prominences, but seldom observed in double-decker filaments. 

\citet{Awasthi2018} identified an even more complex system consisting of multiple flux ropes braiding about each other, morphologically similar to the braided thread-like structures observed in AIA 131~{\AA}. Compared with a single or double flux rope, the braiding introduces a new degree of freedom as well as additional free energy, which is manifested in multi-episodes of internal reconnection. \citet{Awasthi2018} concluded that the complex ICME subsequently observed in situ derives its complexity from the source, namely, the multi-flux-rope configuration together with internal interactions. 

\section{Concluding Remarks} \label{sec:discussion}
\begin{figure}[htp!] 
	\centering
	\includegraphics[width=0.9\textwidth]{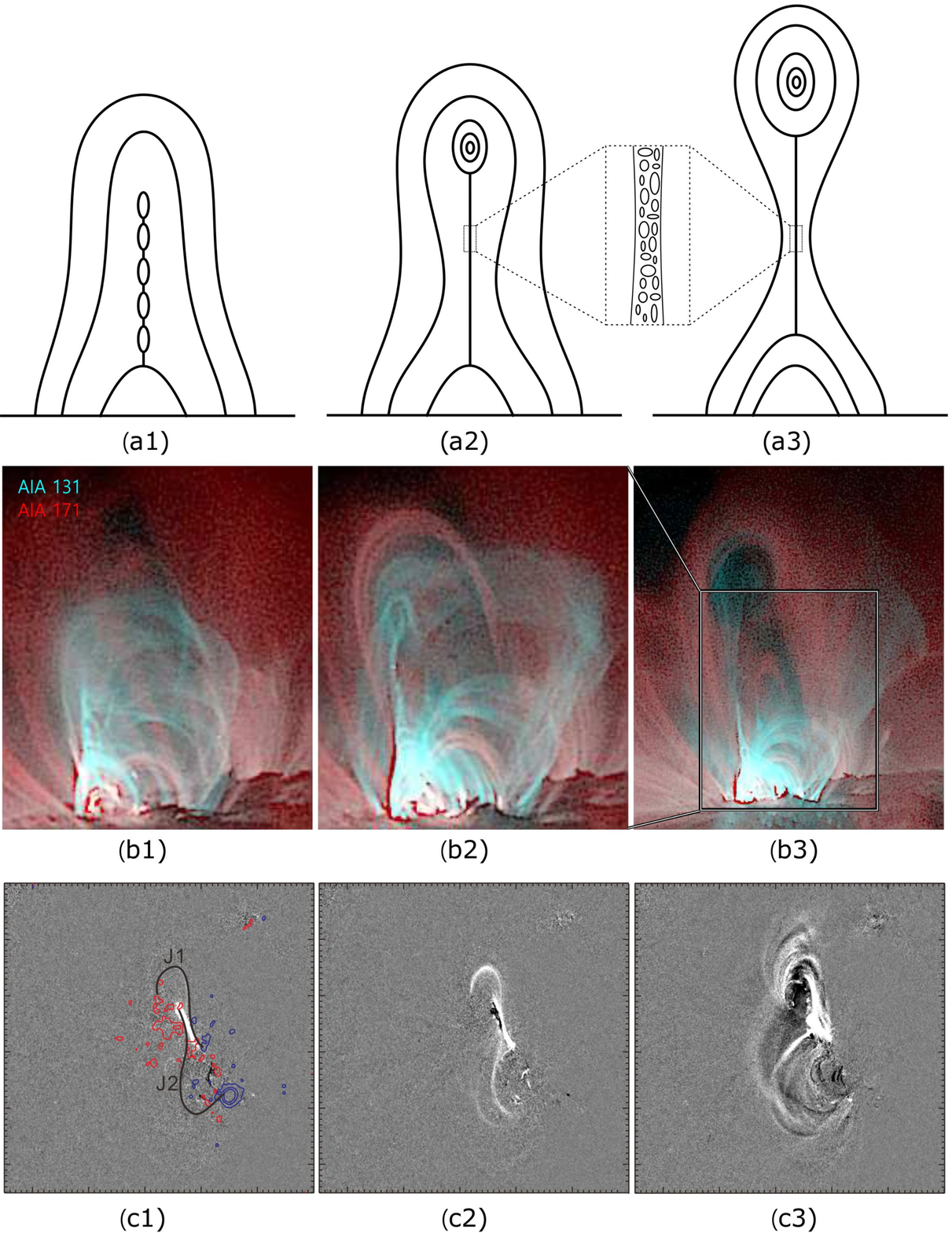}
	\caption{CME initiation through the formation a seed flux rope. \textbf{(a1--a3)} Schematic illustration of the CME initiation process \citep[from][]{Gou2019}. A vertical current sheet underneath a magnetically sheared arcade breaks up into multiple plasmoids (left). The coalescence and ejection of plasmoids initiate a seed flux rope (middle), which rises and stretches the overlying field. Consequently, fast reconnection is induced at the current sheet (right), which embeds plasmoids of various scales, as illustrated by the inset. \textbf{(b1--b3)} Composite images of AIA 131~\AA\ ($\sim\,$10~MK; cyan) and 171~\AA\ ($\sim\,$0.7~MK; red) showing the initiation of the CME on 2013 May 13 \citep[from][]{Gou2019}. The AIA images have been rotated 90 degree clockwise. The field of view in \textbf{b1} and \textbf{b2} is indicated by a rectangle in \textbf{b3}. \textbf{(c1--c3)} Running difference images of AIA 94~\AA\ ($\sim\,$6~MK) showing the initiation of the CME on 2010 August 1 \citep[adapted from][]{Liu2010tc}. `J1' and `J2' in \textbf{c1} illustrate the sheared J-shaped loops before eruption. \label{fig:cartoon}}
\end{figure}

Magnetic flux ropes are inherently scale free. This is manifested by the wide spectrum in their size and energy in the solar atmosphere and the heliosphere. Their formation and evolution is intimately associated with flares and CMEs. Thus, the flux rope is recognized as an important agent through which it is promising to understand the physical mechanisms of, and the power-law distribution of energies in, flares and CMEs \citep{Vourlidas2002,Yashiro2006,Yashiro2008}. It has also been recognized that the formation of mini flux ropes during the spontaneous current-sheet fragmentation is key to both upward and downward cascading processes that are capable of creating a hierarchical distribution over a broad range of scales \citep{Shibata&Tanuma2001,Uzdensky2010,Loureiro2012, Nishizuka&Shibata2013,Gou2019}.

Recent advances combining theory, simulation, and observation have illuminated a new evolutionary path for CMEs. This picture is predicated on the formation of a vertical current sheet in a magnetically sheared arcade before the eruption, as verified in various numerical experiments \citep[e.g.,][]{Mikic+Linker1994,Amari1996,DeVore&Antiochos2000,Aulanier2010}. The current sheet develops as magnetic energy builds up slowly in the corona, driven by photospheric flows, and breaks up into plasmoids when its length exceeds the critical wavelength for the tearing-mode instability \citep[][Figure~\ref{fig:cartoon}(a1)]{Furth&al1963,Uzdensky2010,Barta2011,Loureiro2012}. Propelled by magnetic tension force, the plasmoids move along the current sheet, while neighboring plasmoids merge into larger ones due to the coalescence instability \citep{Finn&Kaw1977,Pritchett&Wu1979}. Upward moving plasmoids eventually merge with the leading plasmoid at the upper tip of the current sheet. A coherent flux rope, i.e., the seed CME, hence starts to form (Figure~\ref{fig:cartoon}(a2,b2)), consistent with the appearance of a ``monster'' plasmoid as expected from stochastic, plasmoid-dominated reconnections \citep{Uzdensky2010,Loureiro2012}.  Projected onto the disk, this seed CME may be observed in the form of an S-shaped bright loop suddenly appearing prior to the eruption \citep[e.g.,][Figure~\ref{fig:cartoon}(c2)]{Liu2010tc}. Because of the hoop force and the upward reconnection outflows, the flux rope keeps rising to stretch the overlying field, which therefore reconnects at the current sheet as in the standard model \citep[][Figure~\ref{fig:cartoon}(a3)]{Lin2004}. While the plasmoids are building up into the CME, they are simultaneously cascading into smaller scales (illustrated by the inset of Figure~\ref{fig:cartoon}a) in a fractal fashion \citep{Shibata&Tanuma2001, Loureiro&Udensky2016,Cheng2018cs}, which results in the close coupling between the flux-rope eruption and particle acceleration \citep[e.g.,][see also \S~\ref{subsubsec:seed}]{Temmer2010,Gou2017,Gou2019}. The observed seed CME at a meso-scale of $\sim\,$10$^8$~cm is hence able to bridge the gap between the macro-scale CME ($\sim\,$10$^{11}$~cm near the Sun) and micro-scale (down to $\sim\,$10$^4$ cm) plasmoids in the current sheet across a hierarchical spectrum.

The above picture is consistent with both statistical and case studies demonstrating that the bulk of CME flux ropes is formed by flare reconnections during eruptions \citep[e.g.,][]{Qiu2007,Hu2014,WangW2017}; further, it has the capacity to accommodate a wide variety of plasma phenomena in the solar atmosphere. If the overlying field is strong enough, the flux-rope eruption can be confined \citep{Torok&Kliem2005}. As the flux rope temporarily settles down, a prominence may form at the bottom of the helical field lines via plasma condensation \citep{LiuW2012,Xia2014}. However, as the current sheet continues to spawn plasmoids and the plasmoids continue to merge into the flux rope, the accumulated flux in the rope may eventually reach the tipping point of eruption \citep{Zhang2014,Zhang2020}. Additionally, whenever open field is accessible to the plasmoids, a jet may ensue instead of a CME \citep{Shibata1999,Sterling2015}. 

At this point, however, we still need a clear understanding of how small flux ropes organize themselves into a coherent, large-scale flux rope. Intimately related to this question, it is still obscure as to how a flux rope interacts with its ambient field, how neighboring flux ropes interact with each other, and how a flux rope evolves by adjusting its topology quasi-statically before, and dynamically during, the eruption. Combining high-resolution observations with data-constrained and data driven simulations will greatly help clarify these issues.

In the past few years, significant advancements have also been made in detecting signatures of flux ropes in the lower atmosphere \citep[see the review by][]{Wang&Liu2019}, owing primarily to meter-class ground-based telescopes represented by the 1.6-meter Goode Solar Telescope at Big Bear Solar Observatory and the 1-meter New Vacuum Solar Telescope at Fuxian Solar Observatory. These telescopes are capable of achieving the diffraction-limited spatial resolution of $\sim\,$0''.1. 

In the future, coordinated multi-instrument observations will continue to be crucial in studying flux ropes on the Sun and beyond. Complemented by MHD simulations, next-generation observations obtained by the state-of-the-art instruments such as the 4-meter Daniel K. Inouye Solar Telescope, Parker Solar Probe (PSP), Solar Orbiter, as well as the Advanced Space-based Solar Observatory that is scheduled to launch in 2022 by China \citep[ASO-S;][]{Gan2019}, will help make breakthroughs in understanding the origin, structure, and evolution of magnetic flux ropes in the solar atmosphere. In particular, PSP has a great potential to traverse CMEs with the solar activity level on the rise in the next few years. This will provide unprecedented opportunities to `anatomize' flux ropes at close distances ($\gtrsim 10\ R_\odot$) from the Sun. This hopefully will reveal the plasma and magnetic structure of burgeoning flux ropes, in which the positive feedback between the flux-rope ejection and the magnetic reconnection is still ongoing. Meanwhile, increasingly high resolution observations, obtained by either ground-based or space-born telescopes, will continue to update our knowledge on the twisted magnetic fields in the solar atmosphere. Among these, spectral diagnostics of chromospheric and prominence magnetic fields and plasma properties are promising to provide valuable information about how twisted magnetic fields are structured and interact with plasmas in the chromospheric and coronal environment.

\normalem
\begin{acknowledgements}
This work was supported by NSFC (Grant No. 41761134088, 41774150, and 11925302), CAS Key Research Program (Grant No. KZZD-EW-01-4), the fundamental research funds for the central universities, and the Strategic Priority Program of the Chinese Academy of Sciences (Grant No. XDB41000000)
\end{acknowledgements}
  

\begin{thebibliography}{506}
	\providecommand\natexlab[1]{#1}
	\providecommand\JournalTitle[1]{#1}
	
	\bibitem[{Alexander} {et~al.}(2013)]{Alexander2013}
	{Alexander}, C.~E., {Walsh}, R.~W., {R{\'e}gnier}, S., {et~al.} 2013, \apjl,
	775, L32
	
	\bibitem[{Alexander} {et~al.}(2006)]{Alexander2006}
	{Alexander}, D., {Liu}, R., \& {Gilbert}, H.~R. 2006, \apj, 653, 719
	
	\bibitem[{Amari} {et~al.}(2014)]{Amari2014}
	{Amari}, T., {Canou}, A., \& {Aly}, J.-J. 2014, \nat, 514, 465
	
	\bibitem[{Amari} {et~al.}(2018)]{Amari2018}
	{Amari}, T., {Canou}, A., {Aly}, J.-J., {Delyon}, F., \& {Alauzet}, F. 2018,
	\nat, 554, 211
	
	\bibitem[{Amari} {et~al.}(2003{\natexlab{a}})]{Amari2003a}
	{Amari}, T., {Luciani}, J.~F., {Aly}, J.~J., {Mikic}, Z., \& {Linker}, J.
	2003{\natexlab{a}}, \apj, 585, 1073
	
	\bibitem[{Amari} {et~al.}(2003{\natexlab{b}})]{Amari2003b}
	{Amari}, T., {Luciani}, J.~F., {Aly}, J.~J., {Mikic}, Z., \& {Linker}, J.
	2003{\natexlab{b}}, \apj, 595, 1231
	
	\bibitem[{Amari} {et~al.}(1996)]{Amari1996}
	{Amari}, T., {Luciani}, J.~F., {Aly}, J.~J., \& {Tagger}, M. 1996, \aap, 306,
	913
	
	\bibitem[{Antiochos} {et~al.}(1999)]{Antiochos1999}
	{Antiochos}, S.~K., {DeVore}, C.~R., \& {Klimchuk}, J.~A. 1999, \apj, 510, 485
	
	\bibitem[{Anzer} \& {Heinzel}(2005)]{Anzer+Heinzel2005}
	{Anzer}, U., \& {Heinzel}, P. 2005, \apj, 622, 714
	
	\bibitem[{Archontis} \& {Hood}(2010)]{Archontis&Hood2010}
	{Archontis}, V., \& {Hood}, A.~W. 2010, \aap, 514, A56
	
	\bibitem[{Archontis} \& {Hood}(2012)]{Archontis&Hood2012}
	{Archontis}, V., \& {Hood}, A.~W. 2012, \aap, 537, A62
	
	\bibitem[{Archontis} \& {T{\"o}r{\"o}k}(2008)]{Archontis&Torok2008}
	{Archontis}, V., \& {T{\"o}r{\"o}k}, T. 2008, \aap, 492, L35
	
	\bibitem[{Asai} {et~al.}(2004)]{Asai2004}
	{Asai}, A., {Yokoyama}, T., {Shimojo}, M., \& {Shibata}, K. 2004, \apjl, 605,
	L77
	
	\bibitem[{Aschwanden}(2002)]{Aschwanden2002}
	{Aschwanden}, M.~J. 2002, \ssr, 101, 1
	
	\bibitem[{Attrill} {et~al.}(2006)]{Attrill2006}
	{Attrill}, G., {Nakwacki}, M.~S., {Harra}, L.~K., {et~al.} 2006, \solphys, 238,
	117
	
	\bibitem[{Aulanier} {et~al.}(2013)]{Aulanier2013}
	{Aulanier}, G., {D{\'e}moulin}, P., {Schrijver}, C.~J., {et~al.} 2013, \aap,
	549, A66
	
	\bibitem[{Aulanier} {et~al.}(2002)]{Aulanier2002}
	{Aulanier}, G., {DeVore}, C.~R., \& {Antiochos}, S.~K. 2002, \apjl, 567, L97
	
	\bibitem[{Aulanier} \& {Dud{\'\i}k}(2019)]{Aulanier&Dudik2019}
	{Aulanier}, G., \& {Dud{\'\i}k}, J. 2019, \aap, 621, A72
	
	\bibitem[{Aulanier} {et~al.}(2012)]{Aulanier2012}
	{Aulanier}, G., {Janvier}, M., \& {Schmieder}, B. 2012, \aap, 543, A110
	
	\bibitem[{Aulanier} {et~al.}(2010)]{Aulanier2010}
	{Aulanier}, G., {T{\"o}r{\"o}k}, T., {D{\'e}moulin}, P., \& {DeLuca}, E.~E.
	2010, \apj, 708, 314
	
	\bibitem[{Awasthi} \& {Liu}(2019)]{Awasthi&Liu2019}
	{Awasthi}, A.~K., \& {Liu}, R. 2019, Frontiers in Physics, 7, 218
	
	\bibitem[{Awasthi} {et~al.}(2018)]{Awasthi2018}
	{Awasthi}, A.~K., {Liu}, R., {Wang}, H., {Wang}, Y., \& {Shen}, C. 2018, \apj,
	857, 124
	
	\bibitem[{Awasthi} {et~al.}(2019)]{Awasthi2019}
	{Awasthi}, A.~K., {Liu}, R., \& {Wang}, Y. 2019, \apj, 872, 109
	
	\bibitem[{Bak-Steslicka} {et~al.}(2016)]{Bak-Steslicka2016}
	{Bak-Steslicka}, U., {Gibson}, S., \& {Chmielewska}, E. 2016, Frontiers in
	Astronomy and Space Sciences, 3, 7
	
	\bibitem[{B{\'a}rta} {et~al.}(2011)]{Barta2011}
	{B{\'a}rta}, M., {B{\"u}chner}, J., {Karlick{\'y}}, M., \& {Sk{\'a}la}, J.
	2011, \apj, 737, 24
	
	\bibitem[{Bateman}(1978)]{Bateman1978}
	{Bateman}, G. 1978, {MHD instabilities} (Cambridge, Massachusetts, USA: The MIT
	Press)
	
	\bibitem[{Baty}(2001)]{Baty2001}
	{Baty}, H. 2001, \aap, 367, 321
	
	\bibitem[{Baty} \& {Heyvaerts}(1996)]{Baty&Heyvaerts1996}
	{Baty}, H., \& {Heyvaerts}, J. 1996, \aap, 308, 935
	
	\bibitem[{Baumgartner} {et~al.}(2018)]{Baumgartner2018}
	{Baumgartner}, C., {Thalmann}, J.~K., \& {Veronig}, A.~M. 2018, \apj, 853, 105
	
	\bibitem[{Beck}(2012)]{Beck2012}
	{Beck}, R. 2012, \ssr, 166, 215
	
	\bibitem[{Berger}(1988)]{Berger1988}
	{Berger}, M.~A. 1988, \aap, 201, 355
	
	\bibitem[{Berger} \& {Prior}(2006)]{Berger&Prior2006}
	{Berger}, M.~A., \& {Prior}, C. 2006, Journal of Physics A Mathematical
	General, 39, 8321
	
	\bibitem[{Berger}(2012)]{Berger2012}
	{Berger}, T. 2012, in Astronomical Society of the Pacific Conference Series,
	Vol. 463, Second ATST-EAST Meeting: Magnetic Fields from the Photosphere to
	the Corona., ed. T.~R. {Rimmele}, A.~{Tritschler}, F.~{W{\"o}ger},
	M.~{Collados Vera}, H.~{Socas-Navarro}, R.~{Schlichenmaier}, M.~{Carlsson},
	T.~{Berger}, A.~{Cadavid}, P.~R. {Gilbert}, P.~R. {Goode}, \&
	M.~{Kn{\"o}lker}, 147
	
	\bibitem[{Berger} {et~al.}(2010)]{Berger2010}
	{Berger}, T.~E., {Slater}, G., {Hurlburt}, N., {et~al.} 2010, \apj, 716, 1288
	
	\bibitem[{Berger} {et~al.}(2017)]{Berger2017}
	{Berger}, T., {Hillier}, A., \& {Liu}, W. 2017, \apj, 850, 60
	
	\bibitem[{Berger} {et~al.}(2011)]{Berger2011}
	{Berger}, T., {Testa}, P., {Hillier}, A., {et~al.} 2011, \nat, 472, 197
	
	\bibitem[{B{\k{a}}k-St{\c{e}}{\'s}licka} {et~al.}(2013)]{Bak-Steslicka2013}
	{B{\k{a}}k-St{\c{e}}{\'s}licka}, U., {Gibson}, S.~E., {Fan}, Y., {et~al.} 2013,
	\apjl, 770, L28
	
	\bibitem[{Blackman}(2015)]{Blackman2015}
	{Blackman}, E.~G. 2015, \ssr, 188, 59
	
	\bibitem[{Bommier} \& {Leroy}(1998)]{Bommier&Leroy1998}
	{Bommier}, V., \& {Leroy}, J.~L. 1998, in Astronomical Society of the Pacific
	Conference Series, Vol. 150, IAU Colloq. 167: New Perspectives on Solar
	Prominences, ed. D.~F. {Webb}, B.~{Schmieder}, \& D.~M. {Rust}, 434
	
	\bibitem[{Browning} {et~al.}(2008)]{Browning2008}
	{Browning}, P.~K., {Gerrard}, C., {Hood}, A.~W., {Kevis}, R., \& {van der
		Linden}, R.~A.~M. 2008, \aap, 485, 837
	
	\bibitem[{Burlaga}(1988)]{Burlaga1988}
	{Burlaga}, L.~F. 1988, \jgr, 93, 7217
	
	\bibitem[{Burlaga} {et~al.}(1981)]{Burlaga1981}
	{Burlaga}, L., {Sittler}, E., {Mariani}, F., \& {Schwenn}, R. 1981, \jgr, 86,
	6673
	
	\bibitem[{Cane} \& {Richardson}(2003)]{Cane&Richardson2003}
	{Cane}, H.~V., \& {Richardson}, I.~G. 2003, Journal of Geophysical Research
	(Space Physics), 108, 1156
	
	\bibitem[{Canfield} {et~al.}(1999)]{Canfield1999}
	{Canfield}, R.~C., {Hudson}, H.~S., \& {McKenzie}, D.~E. 1999, \grl, 26, 627
	
	\bibitem[{Canou} \& {Amari}(2010)]{Canou&Amari2010}
	{Canou}, A., \& {Amari}, T. 2010, \apj, 715, 1566
	
	\bibitem[{Carmichael}(1964)]{Carmichael1964}
	{Carmichael}, H. 1964, {A Process for Flares}, Vol.~50, The Physics of Solar
	Flares, Proceedings of the AAS-NASA Symposium held 28-30 October, 1963 at the
	Goddard Space Flight Center, Greenbelt, MD. Edited by Wilmot N. Hess.
	Washington, DC: National Aeronautics and Space Administration, Science and
	Technical Information Division, 1964., p.451, Vol.~50, 451
	
	\bibitem[{Cartwright} \& {Moldwin}(2008)]{Cartwright&Moldwin2008}
	{Cartwright}, M.~L., \& {Moldwin}, M.~B. 2008, Journal of Geophysical Research
	(Space Physics), 113, A09105
	
	\bibitem[{Chae} {et~al.}(2017)]{Chae2017}
	{Chae}, J., {Cho}, K., {Kwon}, R.-Y., \& {Lim}, E.-K. 2017, \apj, 841, 49
	
	\bibitem[{Chandra} {et~al.}(2010)]{Chandra2010}
	{Chandra}, R., {Pariat}, E., {Schmieder}, B., {Mand rini}, C.~H., \& {Uddin},
	W. 2010, \solphys, 261, 127
	
	\bibitem[{Chandra} {et~al.}(2011)]{Chandra2011}
	{Chandra}, R., {Schmieder}, B., {Mandrini}, C.~H., {et~al.} 2011, \solphys,
	269, 83
	
	\bibitem[{Chen} {et~al.}(2020)]{ChenB2020}
	{Chen}, B., {Yu}, S., {Reeves}, K.~K., \& {Gary}, D.~E. 2020, \apjl, 895, L50
	
	\bibitem[{Chen}(1989)]{Chen1989}
	{Chen}, J. 1989, \apj, 338, 453
	
	\bibitem[{Chen}(2017)]{ChenJ2017}
	{Chen}, J. 2017, Physics of Plasmas, 24, 090501
	
	\bibitem[{Chen}(2011)]{Chen2011}
	{Chen}, P.~F. 2011, Living Reviews in Solar Physics, 8, 1
	
	\bibitem[{Chen} {et~al.}(2014)]{Chen2014}
	{Chen}, P.~F., {Harra}, L.~K., \& {Fang}, C. 2014, \apj, 784, 50
	
	\bibitem[{Chen} {et~al.}(2017)]{Chen2017}
	{Chen}, X., {Liu}, R., {Deng}, N., \& {Wang}, H. 2017, \aap, 606, A84
	
	\bibitem[{Chen} {et~al.}(2019)]{ChenY2019}
	{Chen}, Y., {Hu}, Q., \& {le Roux}, J.~A. 2019, \apj, 881, 58
	
	\bibitem[{Cheng} \& {Qiu}(2016)]{Cheng&Qiu2016}
	{Cheng}, J.~X., \& {Qiu}, J. 2016, \apj, 825, 37
	
	\bibitem[{Cheng} \& {Ding}(2016)]{Cheng&Ding2016}
	{Cheng}, X., \& {Ding}, M.~D. 2016, \apjs, 225, 16
	
	\bibitem[{Cheng} {et~al.}(2014{\natexlab{a}})]{Cheng2014prominence}
	{Cheng}, X., {Ding}, M.~D., {Zhang}, J., {et~al.} 2014{\natexlab{a}}, \apjl,
	789, L35
	
	\bibitem[{Cheng} {et~al.}(2014{\natexlab{b}})]{Cheng2014double}
	{Cheng}, X., {Ding}, M.~D., {Zhang}, J., {et~al.} 2014{\natexlab{b}}, \apj,
	789, 93
	
	\bibitem[{Cheng} {et~al.}(2017)]{Cheng2017}
	{Cheng}, X., {Guo}, Y., \& {Ding}, M. 2017, Science China Earth Sciences, 60,
	1383
	
	\bibitem[{Cheng} {et~al.}(2018{\natexlab{a}})]{Cheng2018}
	{Cheng}, X., {Kliem}, B., \& {Ding}, M.~D. 2018{\natexlab{a}}, \apj, 856, 48
	
	\bibitem[{Cheng} {et~al.}(2018{\natexlab{b}})]{Cheng2018cs}
	{Cheng}, X., {Li}, Y., {Wan}, L.~F., {et~al.} 2018{\natexlab{b}}, \apj, 866, 64
	
	\bibitem[{Cheng} {et~al.}(2011{\natexlab{a}})]{Cheng2011torus}
	{Cheng}, X., {Zhang}, J., {Ding}, M.~D., {Guo}, Y., \& {Su}, J.~T.
	2011{\natexlab{a}}, \apj, 732, 87
	
	\bibitem[{Cheng} {et~al.}(2013{\natexlab{a}})]{Cheng2013driver}
	{Cheng}, X., {Zhang}, J., {Ding}, M.~D., {Liu}, Y., \& {Poomvises}, W.
	2013{\natexlab{a}}, \apj, 763, 43
	
	\bibitem[{Cheng} {et~al.}(2013{\natexlab{b}})]{Cheng2013successive}
	{Cheng}, X., {Zhang}, J., {Ding}, M.~D., {et~al.} 2013{\natexlab{b}}, \apjl,
	769, L25
	
	\bibitem[{Cheng} {et~al.}(2020)]{Cheng2020}
	{Cheng}, X., {Zhang}, J., {Kliem}, B., {et~al.} 2020, \apj, 894, 85
	
	\bibitem[{Cheng} {et~al.}(2011{\natexlab{b}})]{Cheng2011blob}
	{Cheng}, X., {Zhang}, J., {Liu}, Y., \& {Ding}, M.~D. 2011{\natexlab{b}},
	\apjl, 732, L25
	
	\bibitem[{Cheng} {et~al.}(2012)]{Cheng2012}
	{Cheng}, X., {Zhang}, J., {Saar}, S.~H., \& {Ding}, M.~D. 2012, \apj, 761, 62
	
	\bibitem[{Cheng} {et~al.}(2014{\natexlab{c}})]{Cheng2014track}
	{Cheng}, X., {Ding}, M.~D., {Guo}, Y., {et~al.} 2014{\natexlab{c}}, \apj, 780,
	28
	
	\bibitem[{Cheung} \& {Isobe}(2014)]{Cheung&Isobe2014}
	{Cheung}, M. C.~M., \& {Isobe}, H. 2014, Living Reviews in Solar Physics, 11, 3
	
	\bibitem[{Chi} {et~al.}(2016)]{Chi2016}
	{Chi}, Y., {Shen}, C., {Wang}, Y., {et~al.} 2016, \solphys, 291, 2419
	
	\bibitem[{Chintzoglou} {et~al.}(2015)]{Chintzoglou2015}
	{Chintzoglou}, G., {Patsourakos}, S., \& {Vourlidas}, A. 2015, arXiv:1507.01165
	
	\bibitem[{Cho} {et~al.}(2009)]{Cho2009}
	{Cho}, K.-S., {Lee}, J., {Bong}, S.-C., {et~al.} 2009, \apj, 703, 1
	
	\bibitem[{Dahlin} {et~al.}(2019)]{Dahlin2019}
	{Dahlin}, J.~T., {Antiochos}, S.~K., \& {DeVore}, C.~R. 2019, \apj, 879, 96
	
	\bibitem[{Dalmasse} {et~al.}(2015)]{Dalmasse2015}
	{Dalmasse}, K., {Aulanier}, G., {D{\'e}moulin}, P., {et~al.} 2015, \apj, 810,
	17
	
	\bibitem[{Dasso} {et~al.}(2006)]{Dasso2006}
	{Dasso}, S., {Mandrini}, C.~H., {D{\'e}moulin}, P., \& {Luoni}, M.~L. 2006,
	\aap, 455, 349
	
	\bibitem[{Daughton} {et~al.}(2011)]{Daughton2011}
	{Daughton}, W., {Roytershteyn}, V., {Karimabadi}, H., {et~al.} 2011, Nature
	Physics, 7, 539
	
	\bibitem[{D{\'e}moulin}(2006)]{Demoulin2006}
	{D{\'e}moulin}, P. 2006, Advances in Space Research, 37, 1269
	
	\bibitem[{D{\'e}moulin}(2008)]{Demoulin2008}
	{D{\'e}moulin}, P. 2008, Annales Geophysicae, 26, 3113
	
	\bibitem[{D{\'e}moulin} \& {Aulanier}(2010)]{Demoulin&Aulanier2010}
	{D{\'e}moulin}, P., \& {Aulanier}, G. 2010, \apj, 718, 1388
	
	\bibitem[{Demoulin} {et~al.}(1997)]{Demoulin1997}
	{Demoulin}, P., {Bagala}, L.~G., {Mandrini}, C.~H., {Henoux}, J.~C., \&
	{Rovira}, M.~G. 1997, \aap, 325, 305
	
	\bibitem[{Demoulin} {et~al.}(1996)]{Demoulin1996qsl}
	{Demoulin}, P., {Henoux}, J.~C., {Priest}, E.~R., \& {Mand rini}, C.~H. 1996,
	\aap, 308, 643
	
	\bibitem[{D{\'e}moulin} {et~al.}(1996)]{Demoulin1996rope}
	{D{\'e}moulin}, P., {Priest}, E.~R., \& {Lonie}, D.~P. 1996, \jgr, 101, 7631
	
	\bibitem[{Dere} {et~al.}(1999)]{Dere1999}
	{Dere}, K.~P., {Brueckner}, G.~E., {Howard}, R.~A., {Michels}, D.~J., \&
	{Delaboudiniere}, J.~P. 1999, \apj, 516, 465
	
	\bibitem[{DeVore} \& {Antiochos}(2000)]{DeVore&Antiochos2000}
	{DeVore}, C.~R., \& {Antiochos}, S.~K. 2000, \apj, 539, 954
	
	\bibitem[{Dhakal} {et~al.}(2018)]{Dhakal2018}
	{Dhakal}, S.~K., {Chintzoglou}, G., \& {Zhang}, J. 2018, \apj, 860, 35
	
	\bibitem[{Ding} {et~al.}(2006)]{Ding2006}
	{Ding}, J.~Y., {Hu}, Y.~Q., \& {Wang}, J.~X. 2006, \solphys, 235, 223
	
	\bibitem[{Dissauer} {et~al.}(2018)]{Dissauer2018}
	{Dissauer}, K., {Veronig}, A.~M., {Temmer}, M., {Podladchikova}, T., \&
	{Vanninathan}, K. 2018, \apj, 855, 137
	
	\bibitem[{Drake} {et~al.}(2006)]{Drake2006}
	{Drake}, J.~F., {Swisdak}, M., {Che}, H., \& {Shay}, M.~A. 2006, \nat, 443, 553
	
	\bibitem[{Drake} {et~al.}(2013)]{Drake2013}
	{Drake}, J.~F., {Swisdak}, M., \& {Fermo}, R. 2013, \apjl, 763, L5
	
	\bibitem[{Duan} {et~al.}(2019)]{Duan2019}
	{Duan}, A., {Jiang}, C., {He}, W., {et~al.} 2019, \apj, 884, 73
	
	\bibitem[{Dud{\'{\i}}k} {et~al.}(2008)]{Dudik2008}
	{Dud{\'{\i}}k}, J., {Aulanier}, G., {Schmieder}, B., {Bommier}, V., \&
	{Roudier}, T. 2008, \solphys, 248, 29
	
	\bibitem[{Dud{\'\i}k} {et~al.}(2012)]{Dudik2012}
	{Dud{\'\i}k}, J., {Aulanier}, G., {Schmieder}, B., {Zapi{\'o}r}, M., \&
	{Heinzel}, P. 2012, \apj, 761, 9
	
	\bibitem[{Einaudi} \& {van Hoven}(1983)]{Einaudi&vanHoven1983}
	{Einaudi}, G., \& {van Hoven}, G. 1983, \solphys, 88, 163
	
	\bibitem[{Fan}(2001)]{Fan2001}
	{Fan}, Y. 2001, \apjl, 554, L111
	
	\bibitem[{Fan}(2009)]{Fan2009}
	{Fan}, Y. 2009, \apj, 697, 1529
	
	\bibitem[{Fan}(2010)]{Fan2010}
	{Fan}, Y. 2010, \apj, 719, 728
	
	\bibitem[{Fan}(2012)]{Fan2012}
	{Fan}, Y. 2012, \apj, 758, 60
	
	\bibitem[{Fan}(2018)]{Fan2018}
	{Fan}, Y. 2018, \apj, 862, 54
	
	\bibitem[{Fan} \& {Gibson}(2003)]{Fan&Gibson2003}
	{Fan}, Y., \& {Gibson}, S.~E. 2003, \apjl, 589, L105
	
	\bibitem[{Fan} \& {Gibson}(2004)]{Fan&Gibson2004}
	{Fan}, Y., \& {Gibson}, S.~E. 2004, \apj, 609, 1123
	
	\bibitem[{Fan} \& {Gibson}(2007)]{Fan&Gibson2007}
	{Fan}, Y., \& {Gibson}, S.~E. 2007, \apj, 668, 1232
	
	\bibitem[Fan \& Liu(2019)]{Fan&Liu2019}
	Fan, Y., \& Liu, T. 2019, Frontiers in Astronomy and Space Sciences, 6, 27
	
	\bibitem[{Farrugia} {et~al.}(1999)]{Farrugia1999}
	{Farrugia}, C.~J., {Janoo}, L.~A., {Torbert}, R.~B., {et~al.} 1999, in American
	Institute of Physics Conference Series, Vol. 471, American Institute of
	Physics Conference Series, ed. S.~R. {Habbal}, R.~{Esser}, J.~V. {Hollweg},
	\& P.~A. {Isenberg}, 745
	
	\bibitem[{Feng} {et~al.}(2019)]{Feng2019}
	{Feng}, H., {Zhao}, Y., {Zhao}, G., {Liu}, Q., \& {Wu}, D. 2019, \grl, 46, 5
	
	\bibitem[{Filippov} {et~al.}(2015)]{Filippov2015}
	{Filippov}, B., {Martsenyuk}, O., {Srivastava}, A.~K., \& {Uddin}, W. 2015,
	Journal of Astrophysics and Astronomy, 36, 157
	
	\bibitem[{Finn} \& {Kaw}(1977)]{Finn&Kaw1977}
	{Finn}, J.~M., \& {Kaw}, P.~K. 1977, Physics of Fluids, 20, 72
	
	\bibitem[{Finn} {et~al.}(1981)]{Finn1981}
	{Finn}, J.~M., {Manheimer}, W.~M., \& {Ott}, E. 1981, Physics of Fluids, 24,
	1336
	
	\bibitem[{Forbes}(2010)]{Forbes2010}
	{Forbes}, T. 2010, {Models of coronal mass ejections and flares}, ed. C.~J.
	{Schrijver} \& G.~L. {Siscoe}, Heliophysics: Space Storms and Radiation:
	Causes and Effects, ed. C.~J. {Schrijver} \& G.~L. {Siscoe}, 159
	
	\bibitem[{Forbes}(2000)]{Forbes2000}
	{Forbes}, T.~G. 2000, \jgr, 105, 23153
	
	\bibitem[{Forbes} \& {Lin}(2000)]{Forbes&Lin2000}
	{Forbes}, T.~G., \& {Lin}, J. 2000, Journal of Atmospheric and
	Solar-Terrestrial Physics, 62, 1499
	
	\bibitem[{Forbes} \& {Priest}(1984)]{Forbes&Priest1984}
	{Forbes}, T.~G., \& {Priest}, E.~R. 1984, \solphys, 94, 315
	
	\bibitem[{Forbes} \& {Priest}(1995)]{Forbes&Priest1995}
	{Forbes}, T.~G., \& {Priest}, E.~R. 1995, \apj, 446, 377
	
	\bibitem[{Forbes} {et~al.}(2006)]{Forbes2006}
	{Forbes}, T.~G., {Linker}, J.~A., {Chen}, J., {et~al.} 2006, \ssr, 123, 251
	
	\bibitem[{Forland} {et~al.}(2013)]{Forland2013}
	{Forland}, B.~C., {Gibson}, S.~E., {Dove}, J.~B., {Rachmeler}, L.~A., \& {Fan},
	Y. 2013, \solphys, 288, 603
	
	\bibitem[{Furth} {et~al.}(1963)]{Furth&al1963}
	{Furth}, H.~P., {Killeen}, J., \& {Rosenbluth}, M.~N. 1963, Physics of Fluids,
	6, 459
	
	\bibitem[{Gaizauskas}(1998)]{Gaizauskas1998}
	{Gaizauskas}, V. 1998, Astronomical Society of the Pacific Conference Series,
	Vol. 150, {Filament Channels: Essential Ingredients for Filament Formation
		(Review)}, ed. D.~F. {Webb}, B.~{Schmieder}, \& D.~M. {Rust}, Astronomical
	Society of the Pacific Conference Series, Vol. 150, IAU Colloq. 167: New
	Perspectives on Solar Prominences, ed. D.~F. {Webb}, B.~{Schmieder}, \& D.~M.
	{Rust}, 257
	
	\bibitem[{Galsgaard} \& {Nordlund}(1997)]{Galsgaard&Nordlund1997}
	{Galsgaard}, K., \& {Nordlund}, {\r{A}}. 1997, \jgr, 102, 219
	
	\bibitem[{Gan} {et~al.}(2019)]{Gan2019}
	{Gan}, W.-Q., {Zhu}, C., {Deng}, Y.-Y., {et~al.} 2019, Research in Astronomy
	and Astrophysics, 19, 156
	
	\bibitem[{Gary} \& {Moore}(2004)]{Gary&Moore2004}
	{Gary}, G.~A., \& {Moore}, R.~L. 2004, \apj, 611, 545
	
	\bibitem[{Georgoulis} {et~al.}(2012)]{Georgoulis2012}
	{Georgoulis}, M.~K., {Titov}, V.~S., \& {Miki{\'c}}, Z. 2012, \apj, 761, 61
	
	\bibitem[{Gibson}(2015)]{Gibson2015}
	{Gibson}, S. 2015, in Astrophysics and Space Science Library, Vol. 415, Solar
	Prominences, ed. J.-C. {Vial} \& O.~{Engvold} (Springer), 323
	
	\bibitem[Gibson(2018)]{Gibson2018}
	Gibson, S.~E. 2018, Living Reviews in Solar Physics, 15, 7
	
	\bibitem[{Gibson} \& {Fan}(2006{\natexlab{a}})]{Gibson&Fan2006mfr}
	{Gibson}, S.~E., \& {Fan}, Y. 2006{\natexlab{a}}, Journal of Geophysical
	Research (Space Physics), 111, A12103
	
	\bibitem[{Gibson} \& {Fan}(2006{\natexlab{b}})]{Gibson&Fan2006partial}
	{Gibson}, S.~E., \& {Fan}, Y. 2006{\natexlab{b}}, \apj, 637, L65
	
	\bibitem[{Gibson} {et~al.}(2004)]{Gibson2004}
	{Gibson}, S.~E., {Fan}, Y., {Mandrini}, C., {Fisher}, G., \& {Demoulin}, P.
	2004, \apj, 617, 600
	
	\bibitem[{Gibson} {et~al.}(2006)]{Gibson2006}
	{Gibson}, S.~E., {Foster}, D., {Burkepile}, J., {de Toma}, G., \& {Stanger}, A.
	2006, \apj, 641, 590
	
	\bibitem[{Gilbert} {et~al.}(2007)]{Gilbert2007}
	{Gilbert}, H.~R., {Alexander}, D., \& {Liu}, R. 2007, \solphys, 245, 287
	
	\bibitem[{Gilbert} {et~al.}(2000)]{Gilbert2000}
	{Gilbert}, H.~R., {Holzer}, T.~E., {Burkepile}, J.~T., \& {Hundhausen}, A.~J.
	2000, \apj, 537, 503
	
	\bibitem[Goedbloed {et~al.}(2019)]{Goedbloed2019}
	Goedbloed, H., Keppens, R., \& Poedts, S. 2019, Magnetohydrodynamics: Of
	Laboratory and Astrophysical Plasmas (Cambridge University Press)
	
	\bibitem[{Gold} \& {Hoyle}(1960)]{Gold&Hoyle1960}
	{Gold}, T., \& {Hoyle}, F. 1960, \mnras, 120, 89
	
	\bibitem[{Gopalswamy} {et~al.}(2013)]{Gopalswamy2013}
	{Gopalswamy}, N., {M{\"a}kel{\"a}}, P., {Akiyama}, S., {et~al.} 2013, \solphys,
	284, 17
	
	\bibitem[{Gosling} {et~al.}(2005)]{Gosling2005}
	{Gosling}, J.~T., {Skoug}, R.~M., {McComas}, D.~J., \& {Smith}, C.~W. 2005,
	Journal of Geophysical Research (Space Physics), 110, A01107
	
	\bibitem[{Gou} {et~al.}(2019)]{Gou2019}
	{Gou}, T., {Liu}, R., {Kliem}, B., {Wang}, Y., \& {Veronig}, A.~M. 2019,
	Science Advances, 5, 7004
	
	\bibitem[{Gou} {et~al.}(2017)]{Gou2017}
	{Gou}, T., {Veronig}, A.~M., {Dickson}, E.~C., {Hernand ez-Perez}, A., \&
	{Liu}, R. 2017, \apjl, 845, L1
	
	\bibitem[{Green} \& {Kliem}(2009)]{Green&Kliem2009}
	{Green}, L.~M., \& {Kliem}, B. 2009, \apjl, 700, L83
	
	\bibitem[{Green} \& {Kliem}(2014)]{Green&Kliem2014}
	{Green}, L.~M., \& {Kliem}, B. 2014, in IAU Symposium, Vol. 300, Nature of
	Prominences and their Role in Space Weather, ed. B.~{Schmieder}, J.-M.
	{Malherbe}, \& S.~T. {Wu}, 209
	
	\bibitem[{Green} {et~al.}(2007)]{Green2007}
	{Green}, L.~M., {Kliem}, B., {T{\"o}r{\"o}k}, T., {van Driel-Gesztelyi}, L., \&
	{Attrill}, G.~D.~R. 2007, \solphys, 246, 365
	
	\bibitem[{Green} {et~al.}(2011)]{Green2011}
	{Green}, L.~M., {Kliem}, B., \& {Wallace}, A.~J. 2011, \aap, 526, A2
	
	\bibitem[{Gun{\'a}r} {et~al.}(2014)]{Gunar2014}
	{Gun{\'a}r}, S., {Schwartz}, P., {Dud{\'\i}k}, J., {et~al.} 2014, \aap, 567,
	A123
	
	\bibitem[{Guo} {et~al.}(2013)]{Guo2013}
	{Guo}, Y., {Ding}, M.~D., {Cheng}, X., {Zhao}, J., \& {Pariat}, E. 2013, \apj,
	779, 157
	
	\bibitem[{Guo} {et~al.}(2012)]{Guo2012}
	{Guo}, Y., {Ding}, M.~D., {Schmieder}, B., {D{\'e}moulin}, P., \& {Li}, H.
	2012, \apj, 746, 17
	
	\bibitem[{Guo} {et~al.}(2010{\natexlab{a}})]{Guo2010confined}
	{Guo}, Y., {Ding}, M.~D., {Schmieder}, B., {et~al.} 2010{\natexlab{a}}, \apjl,
	725, L38
	
	\bibitem[{Guo} {et~al.}(2010{\natexlab{b}})]{Guo2010rope}
	{Guo}, Y., {Schmieder}, B., {D{\'e}moulin}, P., {et~al.} 2010{\natexlab{b}},
	\apj, 714, 343
	
	\bibitem[{Guo} {et~al.}(2019{\natexlab{a}})]{Guo2019simulation}
	{Guo}, Y., {Xia}, C., {Keppens}, R., {Ding}, M.~D., \& {Chen}, P.~F.
	2019{\natexlab{a}}, \apjl, 870, L21
	
	\bibitem[{Guo} {et~al.}(2019{\natexlab{b}})]{Guo2019reconstruction}
	{Guo}, Y., {Xu}, Y., {Ding}, M.~D., {et~al.} 2019{\natexlab{b}}, \apjl, 884, L1
	
	\bibitem[{Guo} {et~al.}(2017)]{Guo2017}
	{Guo}, Y., {Pariat}, E., {Valori}, G., {et~al.} 2017, \apj, 840, 40
	
	\bibitem[{Hale}(1927)]{Hale1927}
	{Hale}, G.~E. 1927, \nat, 119, 708
	
	\bibitem[{Hannah} \& {Kontar}(2013)]{Hannah&Kontar2013}
	{Hannah}, I.~G., \& {Kontar}, E.~P. 2013, \aap, 553, A10
	
	\bibitem[{Harra} \& {Sterling}(2001)]{Harra&Sterling2001}
	{Harra}, L.~K., \& {Sterling}, A.~C. 2001, \apjl, 561, L215
	
	\bibitem[{Hassanin} \& {Kliem}(2016)]{Hassanin&Kliem2016}
	{Hassanin}, A., \& {Kliem}, B. 2016, \apj, 832, 106
	
	\bibitem[{Haynes} \& {Arber}(2007)]{Haynes&Arber2007}
	{Haynes}, M., \& {Arber}, T.~D. 2007, \aap, 467, 327
	
	\bibitem[{Hirayama}(1974)]{Hirayama1974}
	{Hirayama}, T. 1974, \solphys, 34, 323
	
	\bibitem[{Hood} {et~al.}(2009)]{Hood2009}
	{Hood}, A.~W., {Browning}, P.~K., \& {van der Linden}, R.~A.~M. 2009, \aap,
	506, 913
	
	\bibitem[{Hood} \& {Priest}(1979)]{Hood&Priest1979}
	{Hood}, A.~W., \& {Priest}, E.~R. 1979, \solphys, 64, 303
	
	\bibitem[{Hood} \& {Priest}(1981)]{Hood&Priest1981}
	{Hood}, A.~W., \& {Priest}, E.~R. 1981, Geophysical and Astrophysical Fluid
	Dynamics, 17, 297
	
	\bibitem[{Hou} {et~al.}(2018)]{Hou2018}
	{Hou}, Y.~J., {Zhang}, J., {Li}, T., {Yang}, S.~H., \& {Li}, X.~H. 2018, \aap,
	619, A100
	
	\bibitem[{Howard} {et~al.}(2017)]{Howard2017}
	{Howard}, T.~A., {DeForest}, C.~E., {Schneck}, U.~G., \& {Alden}, C.~R. 2017,
	\apj, 834, 86
	
	\bibitem[Hu(2017)]{Hu2017rev}
	Hu, Q. 2017, Science China Earth Sciences, 60, 1466
	
	\bibitem[{Hu} {et~al.}(2014)]{Hu2014}
	{Hu}, Q., {Qiu}, J., {Dasgupta}, B., {Khare}, A., \& {Webb}, G.~M. 2014, \apj,
	793, 53
	
	\bibitem[{Hu} {et~al.}(2015)]{Hu2015}
	{Hu}, Q., {Qiu}, J., \& {Krucker}, S. 2015, Journal of Geophysical Research
	(Space Physics), 120, 5266
	
	\bibitem[{Hu} \& {Sonnerup}(2002)]{Hu&Sonnerup2002}
	{Hu}, Q., \& {Sonnerup}, B.~U.~{\"O}. 2002, Journal of Geophysical Research
	(Space Physics), 107, 1142
	
	\bibitem[{Hudson} {et~al.}(1999)]{Hudson1999}
	{Hudson}, H.~S., {Acton}, L.~W., {Harvey}, K.~L., \& {McKenzie}, D.~E. 1999,
	\apjl, 513, L83
	
	\bibitem[{Hudson} {et~al.}(2001)]{Hudson2001}
	{Hudson}, H.~S., {Kosugi}, T., {Nitta}, N.~V., \& {Shimojo}, M. 2001, \apjl,
	561, L211
	
	\bibitem[{Hundhausen}(1987)]{Hundhausen1987}
	{Hundhausen}, A.~J. 1987, in Sixth International Solar Wind Conference, ed.
	V.~J. {Pizzo}, T.~{Holzer}, \& D.~G. {Sime}, Vol.~2, 181
	
	\bibitem[{Hyder}(1966)]{Hyder1966}
	{Hyder}, C.~L. 1966, \zap, 63, 78
	
	\bibitem[{Illing} \& {Hundhausen}(1986)]{Illing&Hundhausen1986}
	{Illing}, R.~M.~E., \& {Hundhausen}, A.~J. 1986, \jgr, 91, 10951
	
	\bibitem[{Inoue}(2016)]{Inoue2016}
	{Inoue}, S. 2016, Progress in Earth and Planetary Science, 3, 19
	
	\bibitem[{Inoue} {et~al.}(2011)]{Inoue2011}
	{Inoue}, S., {Kusano}, K., {Magara}, T., {Shiota}, D., \& {Yamamoto}, T.~T.
	2011, \apj, 738, 161
	
	\bibitem[{Isenberg} \& {Forbes}(2007)]{Isenberg&Forbes2007}
	{Isenberg}, P.~A., \& {Forbes}, T.~G. 2007, \apj, 670, 1453
	
	\bibitem[{James} {et~al.}(2018)]{James2018}
	{James}, A.~W., {Valori}, G., {Green}, L.~M., {et~al.} 2018, \apjl, 855, L16
	
	\bibitem[{Janvier} {et~al.}(2014)]{Janvier2014}
	{Janvier}, M., {Aulanier}, G., {Bommier}, V., {et~al.} 2014, \apj, 788, 60
	
	\bibitem[{Janvier} {et~al.}(2015)]{Janvier2015}
	{Janvier}, M., {Aulanier}, G., \& {D{\'e}moulin}, P. 2015, \solphys, 290, 3425
	
	\bibitem[{Janvier} {et~al.}(2013)]{Janvier2013}
	{Janvier}, M., {Aulanier}, G., {Pariat}, E., \& {D{\'e}moulin}, P. 2013, \aap,
	555, A77
	
	\bibitem[{Janvier} {et~al.}(2016)]{Janvier2016}
	{Janvier}, M., {Savcheva}, A., {Pariat}, E., {et~al.} 2016, \aap, 591, A141
	
	\bibitem[{Jel{\'\i}nek} {et~al.}(2020)]{Jelinek2020}
	{Jel{\'\i}nek}, P., {Karlick{\'y}}, M., {Smirnova}, V.~V., \& {Solov'ev}, A.~A.
	2020, \aap, 637, A42
	
	\bibitem[{Ji} {et~al.}(2003)]{Ji2003}
	{Ji}, H., {Wang}, H., {Schmahl}, E.~J., {Moon}, Y.-J., \& {Jiang}, Y. 2003,
	\apjl, 595, L135
	
	\bibitem[{Jian} {et~al.}(2006)]{Jian2006}
	{Jian}, L., {Russell}, C.~T., {Luhmann}, J.~G., \& {Skoug}, R.~M. 2006,
	\solphys, 239, 393
	
	\bibitem[{Jiang} {et~al.}(2019)]{Jiang2019}
	{Jiang}, C., {Duan}, A., {Feng}, X., {et~al.} 2019, Frontiers in Astronomy and
	Space Sciences, 6, 63
	
	\bibitem[{Jiang} {et~al.}(2018)]{Jiang2018}
	{Jiang}, C., {Zou}, P., {Feng}, X., {et~al.} 2018, \apj, 869, 13
	
	\bibitem[{Jiang} {et~al.}(2013)]{Jiang2013}
	{Jiang}, Y., {Hong}, J., {Yang}, J., {et~al.} 2013, \apj, 764, 68
	
	\bibitem[{Jiang} {et~al.}(2008)]{Jiang2008}
	{Jiang}, Y., {Shen}, Y., {Yi}, B., {Yang}, J., \& {Wang}, J. 2008, \apj, 677,
	699
	
	\bibitem[{Jiang} {et~al.}(2011)]{Jiang2011}
	{Jiang}, Y., {Yang}, J., {Hong}, J., {Bi}, Y., \& {Zheng}, R. 2011, \apj, 738,
	179
	
	\bibitem[{Jing} {et~al.}(2006)]{Jing2006}
	{Jing}, J., {Lee}, J., {Spirock}, T.~J., \& {Wang}, H. 2006, \solphys, 236, 97
	
	\bibitem[{Jing} {et~al.}(2003)]{Jing2003}
	{Jing}, J., {Lee}, J., {Spirock}, T.~J., {et~al.} 2003, \apjl, 584, L103
	
	\bibitem[{Jing} {et~al.}(2018)]{Jing2018}
	{Jing}, J., {Liu}, C., {Lee}, J., {et~al.} 2018, \apj, 864, 138
	
	\bibitem[{Jing} {et~al.}(2010)]{Jing2010}
	{Jing}, J., {Yuan}, Y., {Wiegelmann}, T., {et~al.} 2010, \apjl, 719, L56
	
	\bibitem[{Joshi} {et~al.}(2017)]{Joshi2017}
	{Joshi}, B., {Kushwaha}, U., {Veronig}, A.~M., {et~al.} 2017, \apj, 834, 42
	
	\bibitem[{Joshi} {et~al.}(2016)]{Joshi2016}
	{Joshi}, N.~C., {Schmieder}, B., {Magara}, T., {Guo}, Y., \& {Aulanier}, G.
	2016, \apj, 820, 126
	
	\bibitem[{Kahler} {et~al.}(2011)]{Kahler2011JGR}
	{Kahler}, S.~W., {Krucker}, S., \& {Szabo}, A. 2011, Journal of Geophysical
	Research (Space Physics), 116, A01104
	
	\bibitem[{Karlick{\'y}} \& {B{\'a}rta}(2007)]{Karlicky&Barta2007}
	{Karlick{\'y}}, M., \& {B{\'a}rta}, M. 2007, \aap, 464, 735
	
	\bibitem[{Karlick{\'y}} \& {B{\'a}rta}(2011)]{Karlicky&Barta2011}
	{Karlick{\'y}}, M., \& {B{\'a}rta}, M. 2011, \apj, 733, 107
	
	\bibitem[{Karlick{\'y}} \& {Kliem}(2010)]{Karlicky&Kliem2010}
	{Karlick{\'y}}, M., \& {Kliem}, B. 2010, \solphys, 266, 71
	
	\bibitem[{Karpen} {et~al.}(2012)]{Karpen2012}
	{Karpen}, J.~T., {Antiochos}, S.~K., \& {DeVore}, C.~R. 2012, \apj, 760, 81
	
	\bibitem[{Karpen} {et~al.}(2006)]{Karpen2006}
	{Karpen}, J.~T., {Antiochos}, S.~K., \& {Klimchuk}, J.~A. 2006, \apj, 637, 531
	
	\bibitem[{Keppens} {et~al.}(2019)]{Keppens2019}
	{Keppens}, R., {Guo}, Y., {Makwana}, K., {et~al.} 2019, Reviews of Modern
	Plasma Physics, 3, 14
	
	\bibitem[{Keppens} {et~al.}(2014)]{Keppens2014}
	{Keppens}, R., {Porth}, O., \& {Xia}, C. 2014, \apj, 795, 77
	
	\bibitem[{Kilpua} {et~al.}(2019)]{Kilpua2019cme}
	{Kilpua}, E. K.~J., {Good}, S.~W., {Palmerio}, E., {et~al.} 2019, Frontiers in
	Astronomy and Space Sciences, 6, 50
	
	\bibitem[{Kippenhahn} \& {Schl{\"u}ter}(1957)]{Kippenhahn&Schluter1957}
	{Kippenhahn}, R., \& {Schl{\"u}ter}, A. 1957, \zap, 43, 36
	
	\bibitem[{Kliem} {et~al.}(2000)]{Kliem2000}
	{Kliem}, B., {Karlick{\'y}}, M., \& {Benz}, A.~O. 2000, \aap, 360, 715
	
	\bibitem[{Kliem} {et~al.}(2014{\natexlab{a}})]{Kliem2014instability}
	{Kliem}, B., {Lin}, J., {Forbes}, T.~G., {Priest}, E.~R., \& {T{\"o}r{\"o}k},
	T. 2014{\natexlab{a}}, \apj, 789, 46
	
	\bibitem[{Kliem} {et~al.}(2010)]{Kliem2010}
	{Kliem}, B., {Linton}, M.~G., {T{\"o}r{\"o}k}, T., \& {Karlick{\'y}}, M. 2010,
	\solphys, 266, 91
	
	\bibitem[{Kliem} {et~al.}(2013)]{Kliem2013}
	{Kliem}, B., {Su}, Y.~N., {van Ballegooijen}, A.~A., \& {DeLuca}, E.~E. 2013,
	\apj, 779, 129
	
	\bibitem[{Kliem} {et~al.}(2004)]{Kliem2004}
	{Kliem}, B., {Titov}, V.~S., \& {T{\"o}r{\"o}k}, T. 2004, \aap, 413, L23
	
	\bibitem[{Kliem} \& {T{\"o}r{\"o}k}(2006)]{Kliem&Torok2006}
	{Kliem}, B., \& {T{\"o}r{\"o}k}, T. 2006, \prl, 96, 255002
	
	\bibitem[{Kliem} {et~al.}(2012)]{Kliem2012}
	{Kliem}, B., {T{\"o}r{\"o}k}, T., \& {Thompson}, W.~T. 2012, \solphys, 281, 137
	
	\bibitem[{Kliem} {et~al.}(2014{\natexlab{b}})]{Kliem2014}
	{Kliem}, B., {T{\"o}r{\"o}k}, T., {Titov}, V.~S., {et~al.} 2014{\natexlab{b}},
	\apj, 792, 107
	
	\bibitem[{Klimchuk} \& {Sturrock}(1992)]{Klimchuk&Sturrock1992}
	{Klimchuk}, J.~A., \& {Sturrock}, P.~A. 1992, \apj, 385, 344
	
	\bibitem[{Kontogiannis} {et~al.}(2017)]{Kontogiannis2017}
	{Kontogiannis}, I., {Georgoulis}, M.~K., {Park}, S.-H., \& {Guerra}, J.~A.
	2017, \solphys, 292, 159
	
	\bibitem[{Kopp} \& {Pneuman}(1976)]{Kopp&Pneuman1976}
	{Kopp}, R.~A., \& {Pneuman}, G.~W. 1976, \solphys, 50, 85
	
	\bibitem[{Krall}(2007)]{Krall2007}
	{Krall}, J. 2007, \apj, 657, 559
	
	\bibitem[{Kruskal} \& {Tuck}(1958)]{Kruskal&Tuck1958}
	{Kruskal}, M., \& {Tuck}, J.~L. 1958, Proceedings of the Royal Society of
	London Series A, 245, 222
	
	\bibitem[{Kumar} \& {Cho}(2014)]{Kumar&Cho2014}
	{Kumar}, P., \& {Cho}, K.-S. 2014, \aap, 572, A83
	
	\bibitem[{Kumar} {et~al.}(2012)]{Kumar2012}
	{Kumar}, P., {Cho}, K.-S., {Bong}, S.-C., {Park}, S.-H., \& {Kim}, Y.~H. 2012,
	\apj, 746, 67
	
	\bibitem[{Kumar} {et~al.}(2010)]{Kumar2010}
	{Kumar}, P., {Manoharan}, P.~K., \& {Uddin}, W. 2010, \apj, 710, 1195
	
	\bibitem[{Kundu} {et~al.}(2001)]{Kundu2001}
	{Kundu}, M.~R., {Nindos}, A., {Vilmer}, N., {et~al.} 2001, \apj, 559, 443
	
	\bibitem[{Kuperus} \& {Raadu}(1974)]{Kuperus&Raadu1974}
	{Kuperus}, M., \& {Raadu}, M.~A. 1974, \aap, 31, 189
	
	\bibitem[{Labrosse} {et~al.}(2010)]{Labrosse2010}
	{Labrosse}, N., {Heinzel}, P., {Vial}, J.-C., {et~al.} 2010, \ssr, 151, 243
	
	\bibitem[{Larson} {et~al.}(1997)]{Larson1997}
	{Larson}, D.~E., {Lin}, R.~P., {McTiernan}, J.~M., {et~al.} 1997, \grl, 24,
	1911
	
	\bibitem[{Lazarian} {et~al.}(2012)]{Lazarian2012}
	{Lazarian}, A., {Vlahos}, L., {Kowal}, G., {et~al.} 2012, \ssr, 173, 557
	
	\bibitem[{Leake} {et~al.}(2014)]{Leake2014}
	{Leake}, J.~E., {Linton}, M.~G., \& {Antiochos}, S.~K. 2014, \apj, 787, 46
	
	\bibitem[{Leake} {et~al.}(2013)]{Leake2013}
	{Leake}, J.~E., {Linton}, M.~G., \& {T{\"o}r{\"o}k}, T. 2013, \apj, 778, 99
	
	\bibitem[{Leamon} {et~al.}(2003)]{Leamon2003}
	{Leamon}, R.~J., {Canfield}, R.~C., {Blehm}, Z., \& {Pevtsov}, A.~A. 2003,
	\apjl, 596, L255
	
	\bibitem[{Leka} {et~al.}(1996)]{Leka1996}
	{Leka}, K.~D., {Canfield}, R.~C., {McClymont}, A.~N., \& {van Driel-Gesztelyi},
	L. 1996, \apj, 462, 547
	
	\bibitem[{Leka} {et~al.}(2005)]{Leka2005}
	{Leka}, K.~D., {Fan}, Y., \& {Barnes}, G. 2005, \apj, 626, 1091
	
	\bibitem[{Leroy}(1989)]{Leroy1989}
	{Leroy}, J.~L. 1989, in Astrophysics and Space Science Library, Vol. 150,
	Dynamics and Structure of Quiescent Solar Prominences, ed. E.~R. {Priest}, 77
	
	\bibitem[{Leroy} {et~al.}(1984)]{Leroy1984}
	{Leroy}, J.~L., {Bommier}, V., \& {Sahal-Brechot}, S. 1984, \aap, 131, 33
	
	\bibitem[{Levens} {et~al.}(2016)]{Levens2016}
	{Levens}, P.~J., {Schmieder}, B., {L{\'o}pez Ariste}, A., {et~al.} 2016, \apj,
	826, 164
	
	\bibitem[{Li} {et~al.}(2016)]{Li2016}
	{Li}, L., {Zhang}, J., {Peter}, H., {et~al.} 2016, Nature Physics, 12, 847
	
	\bibitem[{Li} {et~al.}(2012)]{Li2012}
	{Li}, X., {Morgan}, H., {Leonard}, D., \& {Jeska}, L. 2012, \apjl, 752, L22
	
	\bibitem[{Lin} {et~al.}(2008)]{Lin2008}
	{Lin}, J., {Cranmer}, S.~R., \& {Farrugia}, C.~J. 2008, Journal of Geophysical
	Research (Space Physics), 113, A11107
	
	\bibitem[{Lin} \& {Forbes}(2000)]{Lin&Forbes2000}
	{Lin}, J., \& {Forbes}, T.~G. 2000, \jgr, 105, 2375
	
	\bibitem[{Lin} {et~al.}(2005)]{Lin2005}
	{Lin}, J., {Ko}, Y.-K., {Sui}, L., {et~al.} 2005, \apj, 622, 1251
	
	\bibitem[{Lin} {et~al.}(2004)]{Lin2004}
	{Lin}, J., {Raymond}, J.~C., \& {van Ballegooijen}, A.~A. 2004, \apj, 602, 422
	
	\bibitem[{Lin} {et~al.}(2003)]{LinY2003}
	{Lin}, Y., {Engvold}, O.~R., \& {Wiik}, J.~E. 2003, \solphys, 216, 109
	
	\bibitem[{Linton} {et~al.}(2001)]{Linton2001}
	{Linton}, M.~G., {Dahlburg}, R.~B., \& {Antiochos}, S.~K. 2001, \apj, 553, 905
	
	\bibitem[{Linton} \& {Longcope}(2006)]{Linton&Longcope2006}
	{Linton}, M.~G., \& {Longcope}, D.~W. 2006, \apj, 642, 1177
	
	\bibitem[{Linton} \& {Moldwin}(2009)]{Linton&Moldwin2009}
	{Linton}, M.~G., \& {Moldwin}, M.~B. 2009, Journal of Geophysical Research
	(Space Physics), 114, A00B09
	
	\bibitem[{Lites} {et~al.}(2010)]{Lites2010}
	{Lites}, B.~W., {Kubo}, M., {Berger}, T., {et~al.} 2010, \apj, 718, 474
	
	\bibitem[{Liu} {et~al.}(2007{\natexlab{a}})]{LiuC2007}
	{Liu}, C., {Lee}, J., {Gary}, D.~E., \& {Wang}, H. 2007{\natexlab{a}}, \apjl,
	658, L127
	
	\bibitem[{Liu} {et~al.}(2009{\natexlab{a}})]{LiuC2009}
	{Liu}, C., {Lee}, J., {Karlick{\'y}}, M., {et~al.} 2009{\natexlab{a}}, \apj,
	703, 757
	
	\bibitem[{Liu} {et~al.}(2017{\natexlab{a}})]{LiuL2017}
	{Liu}, L., {Wang}, Y., {Liu}, R., {et~al.} 2017{\natexlab{a}}, \apj, 844, 141
	
	\bibitem[{Liu}(2013)]{Liu2013}
	{Liu}, R. 2013, \mnras, 434, 1309
	
	\bibitem[{Liu} \& {Alexander}(2009)]{Liu&Alexander2009}
	{Liu}, R., \& {Alexander}, D. 2009, \apj, 697, 999
	
	\bibitem[{Liu} {et~al.}(2007{\natexlab{b}})]{Liu2007}
	{Liu}, R., {Alexander}, D., \& {Gilbert}, H.~R. 2007{\natexlab{b}}, \apj, 661,
	1260
	
	\bibitem[{Liu} {et~al.}(2009{\natexlab{b}})]{Liu2009}
	{Liu}, R., {Alexander}, D., \& {Gilbert}, H.~R. 2009{\natexlab{b}}, \apj, 691,
	1079
	
	\bibitem[{Liu} {et~al.}(2018)]{Liu2018}
	{Liu}, R., {Chen}, J., \& {Wang}, Y. 2018, Science China Physics, Mechanics,
	and Astronomy, 61, 69611
	
	\bibitem[{Liu} {et~al.}(2016{\natexlab{a}})]{Liu2016SR}
	{Liu}, R., {Chen}, J., {Wang}, Y., \& {Liu}, K. 2016{\natexlab{a}}, Scientific
	Reports, 6, 34021
	
	\bibitem[{Liu} {et~al.}(2008)]{Liu2008}
	{Liu}, R., {Gilbert}, H.~R., {Alexander}, D., \& {Su}, Y. 2008, \apj, 680, 1508
	
	\bibitem[{Liu} {et~al.}(2012{\natexlab{a}})]{Liu2012}
	{Liu}, R., {Kliem}, B., {T{\"o}r{\"o}k}, T., {et~al.} 2012{\natexlab{a}}, \apj,
	756, 59
	
	\bibitem[{Liu} {et~al.}(2010{\natexlab{a}})]{Liu2010arcade}
	{Liu}, R., {Liu}, C., {Park}, S.-H., \& {Wang}, H. 2010{\natexlab{a}}, \apj,
	723, 229
	
	\bibitem[{Liu} {et~al.}(2010{\natexlab{b}})]{Liu2010tc}
	{Liu}, R., {Liu}, C., {Wang}, S., {Deng}, N., \& {Wang}, H. 2010{\natexlab{b}},
	\apjl, 725, L84
	
	\bibitem[{Liu} {et~al.}(2014)]{Liu2014}
	{Liu}, R., {Titov}, V.~S., {Gou}, T., {et~al.} 2014, \apj, 790, 8
	
	\bibitem[{Liu} {et~al.}(2016{\natexlab{b}})]{Liu2016}
	{Liu}, R., {Kliem}, B., {Titov}, V.~S., {et~al.} 2016{\natexlab{b}}, \apj, 818,
	148
	
	\bibitem[{Liu} {et~al.}(2012{\natexlab{b}})]{LiuW2012}
	{Liu}, W., {Berger}, T.~E., \& {Low}, B.~C. 2012{\natexlab{b}}, \apjl, 745, L21
	
	\bibitem[{Liu} {et~al.}(2013)]{LiuW2013}
	{Liu}, W., {Chen}, Q., \& {Petrosian}, V. 2013, \apj, 767, 168
	
	\bibitem[Liu(2008)]{LiuY2008}
	Liu, Y. 2008, The Astrophysical Journal Letters, 679, L151
	
	\bibitem[{Liu} {et~al.}(2017{\natexlab{b}})]{LiuY2017}
	{Liu}, Y., {Sun}, X., {T{\"o}r{\"o}k}, T., {Titov}, V.~S., \& {Leake}, J.~E.
	2017{\natexlab{b}}, \apjl, 846, L6
	
	\bibitem[{Longcope} \& {Beveridge}(2007)]{Longcope&Beveridge2007}
	{Longcope}, D.~W., \& {Beveridge}, C. 2007, \apj, 669, 621
	
	\bibitem[{Longcope} \& {Welsch}(2000)]{Longcope&Welsch2000}
	{Longcope}, D.~W., \& {Welsch}, B.~T. 2000, \apj, 545, 1089
	
	\bibitem[{Loureiro} {et~al.}(2012)]{Loureiro2012}
	{Loureiro}, N.~F., {Samtaney}, R., {Schekochihin}, A.~A., \& {Uzdensky}, D.~A.
	2012, Physics of Plasmas, 19, 042303
	
	\bibitem[{Loureiro} \& {Uzdensky}(2016)]{Loureiro&Udensky2016}
	{Loureiro}, N.~F., \& {Uzdensky}, D.~A. 2016, Plasma Physics and Controlled
	Fusion, 58, 014021
	
	\bibitem[{Low}(1987)]{Low1987}
	{Low}, B.~C. 1987, \apj, 323, 358
	
	\bibitem[{Low}(1996)]{Low1996}
	{Low}, B.~C. 1996, \solphys, 167, 217
	
	\bibitem[{Low}(2001)]{Low2001}
	{Low}, B.~C. 2001, \jgr, 106, 25141
	
	\bibitem[{Lowder} \& {Yeates}(2017)]{Lowder+Yeates2017}
	{Lowder}, C., \& {Yeates}, A. 2017, \apj, 846, 106
	
	\bibitem[{Lugaz} {et~al.}(2017)]{Lugaz2017}
	{Lugaz}, N., {Temmer}, M., {Wang}, Y., \& {Farrugia}, C.~J. 2017, \solphys,
	292, 64
	
	\bibitem[{Luna} \& {Karpen}(2012)]{Luna+Karpen2012}
	{Luna}, M., \& {Karpen}, J. 2012, \apjl, 750, L1
	
	\bibitem[{Luna} {et~al.}(2012)]{Luna2012apjl}
	{Luna}, M., {Karpen}, J.~T., \& {DeVore}, C.~R. 2012, \apj, 746, 30
	
	\bibitem[{Luna} {et~al.}(2014)]{Luna2014}
	{Luna}, M., {Knizhnik}, K., {Muglach}, K., {et~al.} 2014, \apj, 785, 79
	
	\bibitem[Lundquist(1950)]{Lundquist1950}
	Lundquist, S. 1950, Ark. Fys, 2, 361
	
	\bibitem[{Lynch} {et~al.}(2004)]{Lynch2004}
	{Lynch}, B.~J., {Antiochos}, S.~K., {MacNeice}, P.~J., {Zurbuchen}, T.~H., \&
	{Fisk}, L.~A. 2004, \apj, 617, 589
	
	\bibitem[{Lynch} \& {Edmondson}(2013)]{Lynch+Edmondson2013}
	{Lynch}, B.~J., \& {Edmondson}, J.~K. 2013, \apj, 764, 87
	
	\bibitem[{Mackay} {et~al.}(2010)]{Mackay2010}
	{Mackay}, D.~H., {Karpen}, J.~T., {Ballester}, J.~L., {Schmieder}, B., \&
	{Aulanier}, G. 2010, \ssr, 151, 333
	
	\bibitem[{MacTaggart} \& {Haynes}(2014)]{MacTaggart&Haynes2014}
	{MacTaggart}, D., \& {Haynes}, A.~L. 2014, \mnras, 438, 1500
	
	\bibitem[{MacTaggart} \& {Hood}(2010)]{MacTaggart&Hood2010}
	{MacTaggart}, D., \& {Hood}, A.~W. 2010, \apjl, 716, L219
	
	\bibitem[{Magara}(2006)]{Magara2006}
	{Magara}, T. 2006, \apj, 653, 1499
	
	\bibitem[{Manchester} {et~al.}(2004)]{Manchester2004}
	{Manchester}, W., I., {Gombosi}, T., {DeZeeuw}, D., \& {Fan}, Y. 2004, \apj,
	610, 588
	
	\bibitem[{Mandrini} {et~al.}(2005)]{Mandrini2005}
	{Mandrini}, C.~H., {Pohjolainen}, S., {Dasso}, S., {et~al.} 2005, \aap, 434,
	725
	
	\bibitem[{Marscher} {et~al.}(2008)]{Marscher2008}
	{Marscher}, A.~P., {Jorstad}, S.~G., {D'Arcangelo}, F.~D., {et~al.} 2008, \nat,
	452, 966
	
	\bibitem[{Martin}(1998)]{Martin1998}
	{Martin}, S.~F. 1998, \solphys, 182, 107
	
	\bibitem[Martin \& Echols(1994)]{Martin&Echols1994}
	Martin, S.~F., \& Echols, C.~R. 1994, An Observational and Conceptual Model of
	the Magnetic Field of a Filament, ed. R.~J. Rutten \& C.~J. Schrijver, Solar
	Surface Magnetism, ed. R.~J. Rutten \& C.~J. Schrijver (Dordrecht: Springer
	Netherlands), 339
	
	\bibitem[{Marubashi}(2000)]{Marubashi2000}
	{Marubashi}, K. 2000, Advances in Space Research, 26, 55
	
	\bibitem[{Marubashi} \& {Lepping}(2007)]{Marubashi&Lepping2007}
	{Marubashi}, K., \& {Lepping}, R.~P. 2007, Annales Geophysicae, 25, 2453
	
	\bibitem[{McAteer} {et~al.}(2007)]{McAteer2007}
	{McAteer}, R.~T.~J., {Young}, C.~A., {Ireland}, J., \& {Gallagher}, P.~T. 2007,
	\apj, 662, 691
	
	\bibitem[{McCauley} {et~al.}(2015)]{McCauley2015}
	{McCauley}, P.~I., {Su}, Y.~N., {Schanche}, N., {et~al.} 2015, \solphys, 290,
	1703
	
	\bibitem[{McComas} {et~al.}(1994)]{McComas1994}
	{McComas}, D.~J., {Gosling}, J.~T., {Hammond}, C.~M., {et~al.} 1994, \grl, 21,
	1751
	
	\bibitem[{McKenzie} \& {Canfield}(2008)]{McKenzie&Canfield2008}
	{McKenzie}, D.~E., \& {Canfield}, R.~C. 2008, \aap, 481, L65
	
	\bibitem[{McKenzie} \& {Hudson}(1999)]{McKenzie&Hudson1999}
	{McKenzie}, D.~E., \& {Hudson}, H.~S. 1999, \apjl, 519, L93
	
	\bibitem[{McKenzie} \& {Savage}(2009)]{McKenzie&Savage2009}
	{McKenzie}, D.~E., \& {Savage}, S.~L. 2009, \apj, 697, 1569
	
	\bibitem[{Mei} {et~al.}(2017)]{Mei2017}
	{Mei}, Z.~X., {Keppens}, R., {Roussev}, I.~I., \& {Lin}, J. 2017, \aap, 604, L7
	
	\bibitem[{Mei} {et~al.}(2018)]{Mei2018}
	{Mei}, Z.~X., {Keppens}, R., {Roussev}, I.~I., \& {Lin}, J. 2018, \aap, 609, A2
	
	\bibitem[{Melrose}(1995)]{Melrose1995}
	{Melrose}, D.~B. 1995, \apj, 451, 391
	
	\bibitem[{Melrose}(1996)]{Melrose1996}
	{Melrose}, D.~B. 1996, \apj, 471, 497
	
	\bibitem[{Melrose}(2017)]{Melrose2017}
	{Melrose}, D.~B. 2017, Journal of Geophysical Research (Space Physics), 122,
	7963
	
	\bibitem[{Mikic} \& {Linker}(1994)]{Mikic+Linker1994}
	{Mikic}, Z., \& {Linker}, J.~A. 1994, \apj, 430, 898
	
	\bibitem[{Miklenic} {et~al.}(2011)]{Miklenic2011}
	{Miklenic}, C., {Veronig}, A.~M., {Temmer}, M., {M{\"o}stl}, C., \& {Biernat},
	H.~K. 2011, \solphys, 273, 125
	
	\bibitem[{Milligan} {et~al.}(2010)]{Milligan2010}
	{Milligan}, R.~O., {McAteer}, R.~T.~J., {Dennis}, B.~R., \& {Young}, C.~A.
	2010, \apj, 713, 1292
	
	\bibitem[{Mishra} {et~al.}(2017)]{Mishra2017}
	{Mishra}, W., {Wang}, Y., {Srivastava}, N., \& {Shen}, C. 2017, \apjs, 232, 5
	
	\bibitem[{Moffatt}(1969)]{Moffatt1969}
	{Moffatt}, H.~K. 1969, Journal of Fluid Mechanics, 35, 117
	
	\bibitem[{Moore} {et~al.}(2001)]{Moore2001}
	{Moore}, R.~L., {Sterling}, A.~C., {Hudson}, H.~S., \& {Lemen}, J.~R. 2001,
	\apj, 552, 833
	
	\bibitem[{Mulligan} \& {Russell}(2001)]{Mulligan&Russell2001}
	{Mulligan}, T., \& {Russell}, C.~T. 2001, \jgr, 106, 10581
	
	\bibitem[{Myers} {et~al.}(2015)]{Myers2015}
	{Myers}, C.~E., {Yamada}, M., {Ji}, H., {et~al.} 2015, \nat, 528, 526
	
	\bibitem[{Myshyakov} \& {Tsvetkov}(2020)]{Myshyakov&Tsvetkov2020}
	{Myshyakov}, I., \& {Tsvetkov}, T. 2020, \apj, 889, 28
	
	\bibitem[{Ni} {et~al.}(2017)]{Ni2017}
	{Ni}, L., {Zhang}, Q.-M., {Murphy}, N.~A., \& {Lin}, J. 2017, \apj, 841, 27
	
	\bibitem[{Nindos} {et~al.}(2015)]{Nindos2015}
	{Nindos}, A., {Patsourakos}, S., {Vourlidas}, A., \& {Tagikas}, C. 2015, \apj,
	808, 117
	
	\bibitem[{Nishida} {et~al.}(2013)]{Nishida2013}
	{Nishida}, K., {Nishizuka}, N., \& {Shibata}, K. 2013, \apjl, 775, L39
	
	\bibitem[{Nishizuka} {et~al.}(2009)]{Nishizuka2009}
	{Nishizuka}, N., {Asai}, A., {Takasaki}, H., {Kurokawa}, H., \& {Shibata}, K.
	2009, \apjl, 694, L74
	
	\bibitem[{Nishizuka} {et~al.}(2015)]{Nishizuka2015}
	{Nishizuka}, N., {Karlick{\'y}}, M., {Janvier}, M., \& {B{\'a}rta}, M. 2015,
	\apj, 799, 126
	
	\bibitem[{Nishizuka} \& {Shibata}(2013)]{Nishizuka&Shibata2013}
	{Nishizuka}, N., \& {Shibata}, K. 2013, \prl, 110, 051101
	
	\bibitem[{Nishizuka} {et~al.}(2010)]{Nishizuka2010}
	{Nishizuka}, N., {Takasaki}, H., {Asai}, A., \& {Shibata}, K. 2010, \apj, 711,
	1062
	
	\bibitem[{Ohyama} \& {Shibata}(1997)]{Ohyama&Shibata1997}
	{Ohyama}, M., \& {Shibata}, K. 1997, \pasj, 49, 249
	
	\bibitem[{Ohyama} \& {Shibata}(1998)]{Ohyama&Shibata1998}
	{Ohyama}, M., \& {Shibata}, K. 1998, \apj, 499, 934
	
	\bibitem[{Oka} {et~al.}(2010)]{Oka2010}
	{Oka}, M., {Phan}, T.~D., {Krucker}, S., {Fujimoto}, M., \& {Shinohara}, I.
	2010, \apj, 714, 915
	
	\bibitem[{Okamoto} {et~al.}(2016)]{Okamoto2016}
	{Okamoto}, T.~J., {Liu}, W., \& {Tsuneta}, S. 2016, \apj, 831, 126
	
	\bibitem[{Okamoto} {et~al.}(2008)]{Okamoto2008}
	{Okamoto}, T.~J., {Tsuneta}, S., {Lites}, B.~W., {et~al.} 2008, \apjl, 673,
	L215
	
	\bibitem[{Okamoto} {et~al.}(2009)]{Okamoto2009}
	{Okamoto}, T.~J., {Tsuneta}, S., {Lites}, B.~W., {et~al.} 2009, \apj, 697, 913
	
	\bibitem[{Ouyang} {et~al.}(2017)]{Ouyang2017}
	{Ouyang}, Y., {Zhou}, Y.~H., {Chen}, P.~F., \& {Fang}, C. 2017, \apj, 835, 94
	
	\bibitem[{Panesar} {et~al.}(2013)]{Panesar2013}
	{Panesar}, N.~K., {Innes}, D.~E., {Tiwari}, S.~K., \& {Low}, B.~C. 2013, \aap,
	549, A105
	
	\bibitem[{Parenti}(2014)]{Parenti2014}
	{Parenti}, S. 2014, Living Reviews in Solar Physics, 11, 1
	
	\bibitem[{Pariat} {et~al.}(2009)]{Pariat2009}
	{Pariat}, E., {Antiochos}, S.~K., \& {DeVore}, C.~R. 2009, \apj, 691, 61
	
	\bibitem[{Pariat} \& {D{\'e}moulin}(2012)]{Pariat&Demoulin2012}
	{Pariat}, E., \& {D{\'e}moulin}, P. 2012, \aap, 541, A78
	
	\bibitem[{Parker}(1996)]{Parker1996}
	{Parker}, E.~N. 1996, \apj, 471, 489
	
	\bibitem[{Patsourakos} {et~al.}(2008)]{Patsourakos2008}
	{Patsourakos}, S., {Pariat}, E., {Vourlidas}, A., {Antiochos}, S.~K., \&
	{Wuelser}, J.~P. 2008, \apjl, 680, L73
	
	\bibitem[{Patsourakos} {et~al.}(2013)]{Patsourakos2013}
	{Patsourakos}, S., {Vourlidas}, A., \& {Stenborg}, G. 2013, \apj, 764, 125
	
	\bibitem[{Pevtsov}(2002)]{Pevtsov2002}
	{Pevtsov}, A.~A. 2002, \solphys, 207, 111
	
	\bibitem[{Pevtsov} \& {Balasubramaniam}(2003)]{Pevtsov&Balasubramaniam2003}
	{Pevtsov}, A.~A., \& {Balasubramaniam}, K.~S. 2003, Advances in Space Research,
	32, 1867
	
	\bibitem[{Pevtsov} {et~al.}(2003)]{Pevtsov2003}
	{Pevtsov}, A.~A., {Balasubramaniam}, K.~S., \& {Rogers}, J.~W. 2003, \apj, 595,
	500
	
	\bibitem[Pevtsov {et~al.}(2014)]{Pevtsov2014}
	Pevtsov, A.~A., Berger, M.~A., Nindos, A., Norton, A.~A., \& van
	Driel-Gesztelyi, L. 2014, Space Science Reviews, 186, 285
	
	\bibitem[{Pevtsov} {et~al.}(1997)]{Pevtsov1997}
	{Pevtsov}, A.~A., {Canfield}, R.~C., \& {McClymont}, A. e.~N. 1997, \apj, 481,
	973
	
	\bibitem[{Priest} \& {Forbes}(2002)]{Priest&Forbes2002}
	{Priest}, E.~R., \& {Forbes}, T.~G. 2002, \aapr, 10, 313
	
	\bibitem[{Priest} \& {Longcope}(2017)]{Priest&Longcope2017}
	{Priest}, E.~R., \& {Longcope}, D.~W. 2017, \solphys, 292, 25
	
	\bibitem[{Pritchett} \& {Wu}(1979)]{Pritchett&Wu1979}
	{Pritchett}, P.~L., \& {Wu}, C.~C. 1979, Physics of Fluids, 22, 2140
	
	\bibitem[{Qiu} \& {Cheng}(2017)]{Qiu&Cheng2017}
	{Qiu}, J., \& {Cheng}, J. 2017, \apjl, 838, L6
	
	\bibitem[{Qiu} {et~al.}(2007)]{Qiu2007}
	{Qiu}, J., {Hu}, Q., {Howard}, T.~A., \& {Yurchyshyn}, V.~B. 2007, \apj, 659,
	758
	
	\bibitem[{Qiu} {et~al.}(2004)]{Qiu2004}
	{Qiu}, J., {Wang}, H., {Cheng}, C.~Z., \& {Gary}, D.~E. 2004, \apj, 604, 900
	
	\bibitem[{Rachmeler} {et~al.}(2013)]{Rachmeler2013}
	{Rachmeler}, L.~A., {Gibson}, S.~E., {Dove}, J.~B., {DeVore}, C.~R., \& {Fan},
	Y. 2013, \solphys, 288, 617
	
	\bibitem[{Reeves} {et~al.}(2012)]{Reeves2012}
	{Reeves}, K.~K., {Gibson}, S.~E., {Kucera}, T.~A., {Hudson}, H.~S., \& {Kano},
	R. 2012, \apj, 746, 146
	
	\bibitem[{Reeves} \& {Golub}(2011)]{Reeves&Golub2011}
	{Reeves}, K.~K., \& {Golub}, L. 2011, \apjl, 727, L52
	
	\bibitem[{R{\'e}gnier} \& {Amari}(2004)]{Regnier&Amari2004}
	{R{\'e}gnier}, S., \& {Amari}, T. 2004, \aap, 425, 345
	
	\bibitem[{R{\'e}gnier} {et~al.}(2002)]{Regnier2002}
	{R{\'e}gnier}, S., {Amari}, T., \& {Kersal{\'e}}, E. 2002, \aap, 392, 1119
	
	\bibitem[{R{\'e}gnier} {et~al.}(2011)]{Regnier2011}
	{R{\'e}gnier}, S., {Walsh}, R.~W., \& {Alexander}, C.~E. 2011, \aap, 533, L1
	
	\bibitem[{Richard} {et~al.}(1990)]{Richard1990}
	{Richard}, R.~L., {Sydora}, R.~D., \& {Ashour-Abdalla}, M. 1990, Physics of
	Fluids B, 2, 488
	
	\bibitem[{Romano} {et~al.}(2003)]{Romano2003}
	{Romano}, P., {Contarino}, L., \& {Zuccarello}, F. 2003, \solphys, 214, 313
	
	\bibitem[{Romano} {et~al.}(2005)]{Romano2005}
	{Romano}, P., {Contarino}, L., \& {Zuccarello}, F. 2005, \aap, 433, 683
	
	\bibitem[{Rouillard} {et~al.}(2010)]{Rouillard2010}
	{Rouillard}, A.~P., {Davies}, J.~A., {Lavraud}, B., {et~al.} 2010, Journal of
	Geophysical Research (Space Physics), 115, A04103
	
	\bibitem[{Rouillard} {et~al.}(2011)]{Rouillard2011}
	{Rouillard}, A.~P., {Sheeley}, N.~R., J., {Cooper}, T.~J., {et~al.} 2011, \apj,
	734, 7
	
	\bibitem[{Rouppe van der Voort} {et~al.}(2017)]{Rouppe2017}
	{Rouppe van der Voort}, L., {De Pontieu}, B., {Scharmer}, G.~B., {et~al.} 2017,
	\apjl, 851, L6
	
	\bibitem[{Roussev} {et~al.}(2012)]{Roussev2012}
	{Roussev}, I.~I., {Galsgaard}, K., {Downs}, C., {et~al.} 2012, Nature Physics,
	8, 845
	
	\bibitem[{Ruffenach} {et~al.}(2015)]{Ruffenach2015}
	{Ruffenach}, A., {Lavraud}, B., {Farrugia}, C.~J., {et~al.} 2015, Journal of
	Geophysical Research (Space Physics), 120, 43
	
	\bibitem[{Russell} \& {Elphic}(1979)]{Russell&Elphic1979}
	{Russell}, C.~T., \& {Elphic}, R.~C. 1979, \nat, 279, 616
	
	\bibitem[Russell {et~al.}(1990)]{Russell1990}
	Russell, C.~T., Priest, E.~R., \& Lee, L.-C., eds. 1990, Geophysical Monograph
	Series, Vol.~58, Physics of magnetic flux ropes, American Geophysical Union,
	Washington DC
	
	\bibitem[{Rust}(1994)]{Rust1994}
	{Rust}, D.~M. 1994, \grl, 21, 241
	
	\bibitem[{Rust} \& {Kumar}(1994)]{Rust&Kumar1994}
	{Rust}, D.~M., \& {Kumar}, A. 1994, \solphys, 155, 69
	
	\bibitem[{Rust} \& {Kumar}(1996)]{Rust&Kumar1996}
	{Rust}, D.~M., \& {Kumar}, A. 1996, \apjl, 464, L199
	
	\bibitem[{Rust} \& {LaBonte}(2005)]{Rust&Labonte2005}
	{Rust}, D.~M., \& {LaBonte}, B.~J. 2005, \apj, 622, L69
	
	\bibitem[{Ryutova} {et~al.}(2010)]{Ryutova2010}
	{Ryutova}, M., {Berger}, T., {Frank}, Z., {Tarbell}, T., \& {Title}, A. 2010,
	\solphys, 267, 75
	
	\bibitem[{Sakurai}(1981)]{Sakurai1981}
	{Sakurai}, T. 1981, \solphys, 69, 343
	
	\bibitem[{Sakurai} {et~al.}(1992)]{Sakurai1992}
	{Sakurai}, T., {Shibata}, K., {Ichimoto}, K., {Tsuneta}, S., \& {Acton}, L.~W.
	1992, \pasj, 44, L123
	
	\bibitem[{Sasso} {et~al.}(2014)]{Sasso2014}
	{Sasso}, C., {Lagg}, A., \& {Solanki}, S.~K. 2014, \aap, 561, A98
	
	\bibitem[{Savcheva} \& {van Ballegooijen}(2009)]{Savcheva&vanBallegooijen2009}
	{Savcheva}, A., \& {van Ballegooijen}, A. 2009, \apj, 703, 1766
	
	\bibitem[{Schmieder} {et~al.}(2004)]{Schmieder2004}
	{Schmieder}, B., {Mein}, N., {Deng}, Y., {et~al.} 2004, \solphys, 223, 119
	
	\bibitem[{Schmieder} {et~al.}(1991)]{Schmiede1991}
	{Schmieder}, B., {Raadu}, M.~A., \& {Wiik}, J.~E. 1991, \aap, 252, 353
	
	\bibitem[{Schmit} \& {Gibson}(2013)]{Schmit&Gibson2013}
	{Schmit}, D.~J., \& {Gibson}, S. 2013, \apj, 770, 35
	
	\bibitem[{Schrijver} \& {Title}(2011)]{Schrijver&Title2011}
	{Schrijver}, C.~J., \& {Title}, A.~M. 2011, Journal of Geophysical Research
	(Space Physics), 116, A04108
	
	\bibitem[{Schrijver} {et~al.}(2013)]{Schrijver2013}
	{Schrijver}, C.~J., {Title}, A.~M., {Yeates}, A.~R., \& {DeRosa}, M.~L. 2013,
	\apj, 773, 93
	
	\bibitem[{Schrijver} \& {Zwaan}(2000)]{Schrijver&Zwaan2000}
	{Schrijver}, C.~J., \& {Zwaan}, C. 2000, {Solar and Stellar Magnetic Activity}
	
	\bibitem[{Scott} {et~al.}(2017)]{Scott2017}
	{Scott}, R.~B., {Pontin}, D.~I., \& {Hornig}, G. 2017, \apj, 848, 117
	
	\bibitem[{Shafranov}(1958)]{Shafranov1958}
	{Shafranov}, V.~D. 1958, Soviet Journal of Experimental and Theoretical
	Physics, 6, 545
	
	\bibitem[{Sheeley} {et~al.}(2009)]{Sheeley2009}
	{Sheeley}, N.~R., J., {Lee}, D.~D.~H., {Casto}, K.~P., {Wang}, Y.~M., \&
	{Rich}, N.~B. 2009, \apj, 694, 1471
	
	\bibitem[{Shen} {et~al.}(2012{\natexlab{a}})]{ShenC2012}
	{Shen}, C., {Wang}, Y., {Wang}, S., {et~al.} 2012{\natexlab{a}}, Nature
	Physics, 8, 923
	
	\bibitem[{Shen} {et~al.}(2017)]{ShenF2017}
	{Shen}, F., {Wang}, Y., {Shen}, C., \& {Feng}, X. 2017, \solphys, 292, 104
	
	\bibitem[{Shen} {et~al.}(2014)]{Shen2014}
	{Shen}, Y., {Liu}, Y.~D., {Chen}, P.~F., \& {Ichimoto}, K. 2014, \apj, 795, 130
	
	\bibitem[{Shen} {et~al.}(2015)]{Shen2015}
	{Shen}, Y., {Liu}, Y., {Liu}, Y.~D., {et~al.} 2015, \apjl, 814, L17
	
	\bibitem[{Shen} {et~al.}(2012{\natexlab{b}})]{Shen2012}
	{Shen}, Y., {Liu}, Y., \& {Su}, J. 2012{\natexlab{b}}, \apj, 750, 12
	
	\bibitem[{Shibata}(1999)]{Shibata1999}
	{Shibata}, K. 1999, \apss, 264, 129
	
	\bibitem[{Shibata} {et~al.}(1995)]{Shibata1995}
	{Shibata}, K., {Masuda}, S., {Shimojo}, M., {et~al.} 1995, \apjl, 451, L83
	
	\bibitem[{Shibata} \& {Tanuma}(2001)]{Shibata&Tanuma2001}
	{Shibata}, K., \& {Tanuma}, S. 2001, Earth, Planets, and Space, 53, 473
	
	\bibitem[{Shibata} {et~al.}(2007)]{Shibata2007}
	{Shibata}, K., {Nakamura}, T., {Matsumoto}, T., {et~al.} 2007, Science, 318,
	1591
	
	\bibitem[{Shimizu} {et~al.}(2008)]{Shimizu2008}
	{Shimizu}, M., {Nishida}, K., {Takasaki}, H., {et~al.} 2008, \apjl, 683, L203
	
	\bibitem[{Shiota} {et~al.}(2010)]{Shiota2010}
	{Shiota}, D., {Kusano}, K., {Miyoshi}, T., \& {Shibata}, K. 2010, \apj, 718,
	1305
	
	\bibitem[{Slavin} {et~al.}(2003)]{Slavin2003}
	{Slavin}, J.~A., {Lepping}, R.~P., {Gjerloev}, J., {et~al.} 2003, Journal of
	Geophysical Research (Space Physics), 108, 1015
	
	\bibitem[{Song} {et~al.}(2014)]{Song2014}
	{Song}, H.~Q., {Zhang}, J., {Chen}, Y., \& {Cheng}, X. 2014, \apjl, 792, L40
	
	\bibitem[{Song} {et~al.}(2019)]{Song2019}
	{Song}, H.~Q., {Zhang}, J., {Li}, L.~P., {et~al.} 2019, \apj, 887, 124
	
	\bibitem[{Srivastava} {et~al.}(2010)]{Srivastava2010}
	{Srivastava}, A.~K., {Zaqarashvili}, T.~V., {Kumar}, P., \& {Khodachenko},
	M.~L. 2010, \apj, 715, 292
	
	\bibitem[{Stein}(2012)]{Stein2012}
	{Stein}, R.~F. 2012, Living Reviews in Solar Physics, 9, 4
	
	\bibitem[{Sterling} {et~al.}(2000)]{Sterling2000}
	{Sterling}, A.~C., {Hudson}, H.~S., {Thompson}, B.~J., \& {Zarro}, D.~M. 2000,
	\apj, 532, 628
	
	\bibitem[{Sterling} {et~al.}(2015)]{Sterling2015}
	{Sterling}, A.~C., {Moore}, R.~L., {Falconer}, D.~A., \& {Adams}, M. 2015,
	\nat, 523, 437
	
	\bibitem[{Sturrock}(1966)]{Sturrock1966}
	{Sturrock}, P.~A. 1966, \nat, 211, 695
	
	\bibitem[{Sturrock} {et~al.}(2015)]{Sturrock2015}
	{Sturrock}, Z., {Hood}, A.~W., {Archontis}, V., \& {McNeill}, C.~M. 2015, \aap,
	582, A76
	
	\bibitem[{Su} {et~al.}(2018)]{Su2018}
	{Su}, Y., {Liu}, R., {Li}, S., {et~al.} 2018, \apj, 855, 77
	
	\bibitem[{Su} {et~al.}(2011)]{Su2011}
	{Su}, Y., {Surges}, V., {van Ballegooijen}, A., {DeLuca}, E., \& {Golub}, L.
	2011, \apj, 734, 53
	
	\bibitem[{Su} \& {van Ballegooijen}(2012)]{Su&vanBallegooijen2012}
	{Su}, Y., \& {van Ballegooijen}, A. 2012, \apj, 757, 168
	
	\bibitem[{Su} \& {van Ballegooijen}(2013)]{Su&vanBallegooijen2013}
	{Su}, Y., \& {van Ballegooijen}, A. 2013, \apj, 764, 91
	
	\bibitem[{Su} {et~al.}(2015)]{Su2015}
	{Su}, Y., {van Ballegooijen}, A., {McCauley}, P., {et~al.} 2015, \apj, 807, 144
	
	\bibitem[{Sui} \& {Holman}(2003)]{Sui&Holman2003}
	{Sui}, L., \& {Holman}, G.~D. 2003, \apjl, 596, L251
	
	\bibitem[{Takasao} {et~al.}(2012)]{Takasao2012}
	{Takasao}, S., {Asai}, A., {Isobe}, H., \& {Shibata}, K. 2012, \apjl, 745, L6
	
	\bibitem[{Takasao} {et~al.}(2016)]{Takasao2016}
	{Takasao}, S., {Asai}, A., {Isobe}, H., \& {Shibata}, K. 2016, \apj, 828, 103
	
	\bibitem[{Tandberg-Hanssen}(1995)]{Tandberg-Hanssen1995}
	{Tandberg-Hanssen}, E. 1995, Astrophysics and Space Science Library book series
	(ASSL, volume 199), Vol. 199, {The nature of solar prominences} (Springer
	Science+Business Media Dordrecht: Kluwer Academic Publishers)
	
	\bibitem[{Tassev} \& {Savcheva}(2017)]{Tassev&Savcheva2017}
	{Tassev}, S., \& {Savcheva}, A. 2017, \apj, 840, 89
	
	\bibitem[{Taylor}(1974)]{Taylor1974}
	{Taylor}, J.~B. 1974, Physical Review Letters, 33, 1139
	
	\bibitem[{Taylor}(1986)]{Taylor1986}
	{Taylor}, J.~B. 1986, Reviews of Modern Physics, 58, 741
	
	\bibitem[{Temmer} {et~al.}(2017)]{Temmer2017}
	{Temmer}, M., {Thalmann}, J.~K., {Dissauer}, K., {et~al.} 2017, \solphys, 292,
	93
	
	\bibitem[{Temmer} {et~al.}(2010)]{Temmer2010}
	{Temmer}, M., {Veronig}, A.~M., {Kontar}, E.~P., {Krucker}, S., \& {Vr{\v
			s}nak}, B. 2010, \apj, 712, 1410
	
	\bibitem[{Threlfall} {et~al.}(2018)]{Threlfall2018}
	{Threlfall}, J., {Hood}, A.~W., \& {Priest}, E.~R. 2018, \solphys, 293, 98
	
	\bibitem[{Tian} {et~al.}(2012)]{Tian2012}
	{Tian}, H., {McIntosh}, S.~W., {Xia}, L., {He}, J., \& {Wang}, X. 2012, \apj,
	748, 106
	
	\bibitem[{Tian} {et~al.}(2010)]{Tian2010}
	{Tian}, H., {Yao}, S., {Zong}, Q., {He}, J., \& {Qi}, Y. 2010, \apj, 720, 454
	
	\bibitem[{Tian} {et~al.}(2018)]{Tian2018}
	{Tian}, Z., {Shen}, Y., \& {Liu}, Y. 2018, \na, 65, 7
	
	\bibitem[{Titov}(2007)]{Titov2007}
	{Titov}, V.~S. 2007, \apj, 660, 863
	
	\bibitem[{Titov} \& {D{\'e}moulin}(1999)]{Titov&Demoulin1999}
	{Titov}, V.~S., \& {D{\'e}moulin}, P. 1999, \aap, 351, 707
	
	\bibitem[{Titov} {et~al.}(2003)]{Titov2003}
	{Titov}, V.~S., {Galsgaard}, K., \& {Neukirch}, T. 2003, \apj, 582, 1172
	
	\bibitem[{Titov} {et~al.}(2002)]{Titov2002}
	{Titov}, V.~S., {Hornig}, G., \& {D{\'e}moulin}, P. 2002, Journal of
	Geophysical Research (Space Physics), 107, 1164
	
	\bibitem[{Titov} {et~al.}(2012)]{Titov2012}
	{Titov}, V.~S., {Mikic}, Z., {T{\"o}r{\"o}k}, T., {Linker}, J.~A., \&
	{Panasenco}, O. 2012, \apj, 759, 70
	
	\bibitem[{Titov} {et~al.}(1993)]{Titov1993}
	{Titov}, V.~S., {Priest}, E.~R., \& {Demoulin}, P. 1993, \aap, 276, 564
	
	\bibitem[Toriumi \& Wang(2019)]{Toriumi&Wang2019}
	Toriumi, S., \& Wang, H. 2019, Living Reviews in Solar Physics, 16, 3
	
	\bibitem[{T{\"o}r{\"o}k} {et~al.}(2010)]{Torok2010}
	{T{\"o}r{\"o}k}, T., {Berger}, M.~A., \& {Kliem}, B. 2010, \aap, 516, A49
	
	\bibitem[{T{\"o}r{\"o}k} \& {Kliem}(2003)]{Torok&Kliem2003}
	{T{\"o}r{\"o}k}, T., \& {Kliem}, B. 2003, \aap, 406, 1043
	
	\bibitem[{T{\"o}r{\"o}k} \& {Kliem}(2005)]{Torok&Kliem2005}
	{T{\"o}r{\"o}k}, T., \& {Kliem}, B. 2005, \apjl, 630, L97
	
	\bibitem[{T{\"o}r{\"o}k} {et~al.}(2004)]{Torok2004}
	{T{\"o}r{\"o}k}, T., {Kliem}, B., \& {Titov}, V.~S. 2004, \aap, 413, L27
	
	\bibitem[{T{\"o}r{\"o}k} {et~al.}(2011)]{Torok2011}
	{T{\"o}r{\"o}k}, T., {Panasenco}, O., {Titov}, V.~S., {et~al.} 2011, \apjl,
	739, L63
	
	\bibitem[{T{\"o}r{\"o}k} {et~al.}(2014)]{Torok2014}
	{T{\"o}r{\"o}k}, T., {Leake}, J.~E., {Titov}, V.~S., {et~al.} 2014, \apjl, 782,
	L10
	
	\bibitem[{Tripathi} {et~al.}(2009{\natexlab{a}})]{Tripathi2009partial}
	{Tripathi}, D., {Gibson}, S.~E., {Qiu}, J., {et~al.} 2009{\natexlab{a}}, \aap,
	498, 295
	
	\bibitem[{Tripathi} {et~al.}(2009{\natexlab{b}})]{Tripathi2009sigmoid}
	{Tripathi}, D., {Kliem}, B., {Mason}, H.~E., {Young}, P.~R., \& {Green}, L.~M.
	2009{\natexlab{b}}, \apjl, 698, L27
	
	\bibitem[{Tsuneta}(1997)]{Tsuneta1997}
	{Tsuneta}, S. 1997, \apj, 483, 507
	
	\bibitem[{Uzdensky} {et~al.}(2010)]{Uzdensky2010}
	{Uzdensky}, D.~A., {Loureiro}, N.~F., \& {Schekochihin}, A.~A. 2010, \prl, 105,
	235002
	
	\bibitem[{van Ballegooijen} \& {Cranmer}(2010)]{vanBallegooijen&Cranmer2010}
	{van Ballegooijen}, A.~A., \& {Cranmer}, S.~R. 2010, \apj, 711, 164
	
	\bibitem[{van Ballegooijen} \& {Mackay}(2007)]{vanBallegooijen&Mackay2007}
	{van Ballegooijen}, A.~A., \& {Mackay}, D.~H. 2007, \apj, 659, 1713
	
	\bibitem[{van Ballegooijen} \& {Martens}(1989)]{vanBallegooijen&Martens1989}
	{van Ballegooijen}, A.~A., \& {Martens}, P.~C.~H. 1989, \apj, 343, 971
	
	\bibitem[{Vargas Dom{\'\i}nguez} {et~al.}(2012)]{VargasDominguez2012}
	{Vargas Dom{\'\i}nguez}, S., {MacTaggart}, D., {Green}, L., {van
		Driel-Gesztelyi}, L., \& {Hood}, A.~W. 2012, \solphys, 278, 33
	
	\bibitem[{Vasantharaju} {et~al.}(2019)]{Vasantharaju2019}
	{Vasantharaju}, N., {Vemareddy}, P., {Ravindra}, B., \& {Doddamani}, V.~H.
	2019, \apj, 885, 89
	
	\bibitem[{Vemareddy} \& {D{\'e}moulin}(2017)]{Vemareddy&Demoulin2017}
	{Vemareddy}, P., \& {D{\'e}moulin}, P. 2017, \aap, 597, A104
	
	\bibitem[{Veronig} {et~al.}(2019)]{Veronig2019}
	{Veronig}, A.~M., {G{\"o}m{\"o}ry}, P., {Dissauer}, K., {Temmer}, M., \&
	{Vanninathan}, K. 2019, \apj, 879, 85
	
	\bibitem[{Veronig} {et~al.}(2018)]{Veronig2018}
	{Veronig}, A.~M., {Podladchikova}, T., {Dissauer}, K., {et~al.} 2018, \apj,
	868, 107
	
	\bibitem[{Vourlidas} {et~al.}(2002)]{Vourlidas2002}
	{Vourlidas}, A., {Buzasi}, D., {Howard}, R.~A., \& {Esfand iari}, E. 2002, in
	ESA Special Publication, Vol.~1, Solar Variability: From Core to Outer
	Frontiers, ed. A.~{Wilson}, 91
	
	\bibitem[{Vourlidas} {et~al.}(2011)]{Vourlidas2011}
	{Vourlidas}, A., {Colaninno}, R., {Nieves-Chinchilla}, T., \& {Stenborg}, G.
	2011, \apjl, 733, L23
	
	\bibitem[{Vourlidas} {et~al.}(2013)]{Vourlidas2013}
	{Vourlidas}, A., {Lynch}, B.~J., {Howard}, R.~A., \& {Li}, Y. 2013, \solphys,
	284, 179
	
	\bibitem[{Vourlidas} \& {Webb}(2018)]{Vourlidas&Webb2018}
	{Vourlidas}, A., \& {Webb}, D.~F. 2018, \apj, 861, 103
	
	\bibitem[{Vrsnak} {et~al.}(1991)]{Vrsnak1991}
	{Vrsnak}, B., {Ruzdjak}, V., \& {Rompolt}, B. 1991, \solphys, 136, 151
	
	\bibitem[{Vrsnak} {et~al.}(1993)]{Vrsnak1993}
	{Vrsnak}, B., {Ruzdjak}, V., {Rompolt}, B., {Rosa}, D., \& {Zlobec}, P. 1993,
	\solphys, 146, 147
	
	\bibitem[{Vr{\v s}nak} {et~al.}(2007)]{Vrvsnak2007}
	{Vr{\v s}nak}, B., {Veronig}, A.~M., {Thalmann}, J.~K., \& {{\v Z}ic}, T. 2007,
	\aap, 471, 295
	
	\bibitem[{Wang} {et~al.}(2018{\natexlab{a}})]{WangD2018}
	{Wang}, D., {Liu}, R., {Wang}, Y., {et~al.} 2018{\natexlab{a}}, \apj, 869, 177
	
	\bibitem[{Wang} {et~al.}(2017{\natexlab{a}})]{WangD2017}
	{Wang}, D., {Liu}, R., {Wang}, Y., {et~al.} 2017{\natexlab{a}}, \apjl, 843, L9
	
	\bibitem[{Wang} {et~al.}(2015{\natexlab{a}})]{WangH2015}
	{Wang}, H., {Cao}, W., {Liu}, C., {et~al.} 2015{\natexlab{a}}, Nature
	Communications, 6, 7008
	
	\bibitem[{Wang} \& {Liu}(2019)]{Wang&Liu2019}
	{Wang}, H., \& {Liu}, C. 2019, Frontiers in Astronomy and Space Sciences, 6, 18
	
	\bibitem[{Wang} {et~al.}(2017{\natexlab{b}})]{WangH2017}
	{Wang}, H., {Liu}, C., {Ahn}, K., {et~al.} 2017{\natexlab{b}}, Nature
	Astronomy, 1, 0085
	
	\bibitem[{Wang} {et~al.}(2018{\natexlab{b}})]{WangH2018}
	{Wang}, H., {Liu}, R., {Li}, Q., {et~al.} 2018{\natexlab{b}}, \apjl, 852, L18
	
	\bibitem[{Wang} {et~al.}(2016{\natexlab{a}})]{WangR2016}
	{Wang}, R., {Liu}, Y.~D., {Zimovets}, I., {et~al.} 2016{\natexlab{a}}, \apjl,
	827, L12
	
	\bibitem[{Wang} {et~al.}(2017{\natexlab{d}})]{WangW2017apj}
	{Wang}, W., {Liu}, R., \& {Wang}, Y. 2017{\natexlab{d}}, \apj, 834, 38
	
	\bibitem[{Wang} {et~al.}(2017{\natexlab{c}})]{WangW2017}
	{Wang}, W., {Liu}, R., {Wang}, Y., {et~al.} 2017{\natexlab{c}}, Nature
	Communications, 8, 1330
	
	\bibitem[{Wang} {et~al.}(2019)]{WangW2019}
	{Wang}, W., {Zhu}, C., {Qiu}, J., {et~al.} 2019, \apj, 871, 25
	
	\bibitem[{Wang} \& {Stenborg}(2010)]{Wang&Stenborg2010}
	{Wang}, Y.~M., \& {Stenborg}, G. 2010, \apjl, 719, L181
	
	\bibitem[{Wang} {et~al.}(2015{\natexlab{b}})]{WangY2015}
	{Wang}, Y., {Zhou}, Z., {Shen}, C., {Liu}, R., \& {Wang}, S.
	2015{\natexlab{b}}, Journal of Geophysical Research (Space Physics), 120,
	1543
	
	\bibitem[{Wang} {et~al.}(2016{\natexlab{b}})]{WangY2016}
	{Wang}, Y., {Zhuang}, B., {Hu}, Q., {et~al.} 2016{\natexlab{b}}, Journal of
	Geophysical Research: Space Physics
	
	\bibitem[{Wang} {et~al.}(2018{\natexlab{c}})]{WangY2018}
	{Wang}, Y., {Shen}, C., {Liu}, R., {et~al.} 2018{\natexlab{c}}, Journal of
	Geophysical Research (Space Physics), 123, 3238
	
	\bibitem[{Webb} \& {Howard}(2012)]{Webb&Howard2012}
	{Webb}, D.~F., \& {Howard}, T.~A. 2012, Living Reviews in Solar Physics, 9, 3
	
	\bibitem[{Webb} {et~al.}(2000)]{Webb2000}
	{Webb}, D.~F., {Lepping}, R.~P., {Burlaga}, L.~F., {et~al.} 2000, \jgr, 105,
	27251
	
	\bibitem[{Wei} {et~al.}(2003{\natexlab{a}})]{Wei2003JGR}
	{Wei}, F., {Liu}, R., {Fan}, Q., \& {Feng}, X. 2003{\natexlab{a}}, Journal of
	Geophysical Research (Space Physics), 108, 1263
	
	\bibitem[{Wei} {et~al.}(2003{\natexlab{b}})]{Wei2003GRL}
	{Wei}, F., {Liu}, R., {Feng}, X., {Zhong}, D., \& {Yang}, F.
	2003{\natexlab{b}}, \grl, 30, 2283
	
	\bibitem[{Wheatland}(2000)]{Wheatland2000}
	{Wheatland}, M.~S. 2000, \apj, 532, 616
	
	\bibitem[{Wiegelmann} {et~al.}(2017)]{Wiegelmann2017}
	{Wiegelmann}, T., {Petrie}, G. J.~D., \& {Riley}, P. 2017, \ssr, 210, 249
	
	\bibitem[{Williams} {et~al.}(2009)]{Williams2009}
	{Williams}, D.~R., {Harra}, L.~K., {Brooks}, D.~H., {Imada}, S., \& {Hansteen},
	V.~H. 2009, \pasj, 61, 493
	
	\bibitem[{Williams} {et~al.}(2005)]{Williams2005}
	{Williams}, D.~R., {T{\"o}r{\"o}k}, T., {D{\'e}moulin}, P., {van
		Driel-Gesztelyi}, L., \& {Kliem}, B. 2005, \apjl, 628, L163
	
	\bibitem[{Wyper} {et~al.}(2017)]{Wyper2017}
	{Wyper}, P.~F., {Antiochos}, S.~K., \& {DeVore}, C.~R. 2017, \nat, 544, 452
	
	\bibitem[{Xia} \& {Keppens}(2016)]{Xia&Keppens2016}
	{Xia}, C., \& {Keppens}, R. 2016, \apjl, 825, L29
	
	\bibitem[{Xia} {et~al.}(2014)]{Xia2014}
	{Xia}, C., {Keppens}, R., {Antolin}, P., \& {Porth}, O. 2014, \apjl, 792, L38
	
	\bibitem[{Xie} {et~al.}(2013)]{Xie2013}
	{Xie}, H., {Gopalswamy}, N., \& {St. Cyr}, O.~C. 2013, \solphys, 284, 47
	
	\bibitem[{Xing} {et~al.}(2020)]{Xing2020}
	{Xing}, C., {Cheng}, X., {Qiu}, J., {et~al.} 2020, \apj, 889, 125
	
	\bibitem[{Xue} {et~al.}(2017)]{Xue2017}
	{Xue}, Z., {Yan}, X., {Yang}, L., {Wang}, J., \& {Zhao}, L. 2017, \apjl, 840,
	L23
	
	\bibitem[{Xue} {et~al.}(2016)]{Xue2016}
	{Xue}, Z., {Yan}, X., {Cheng}, X., {et~al.} 2016, Nature Communications, 7,
	11837
	
	\bibitem[{Yan} {et~al.}(2014)]{Yan2014}
	{Yan}, X.~L., {Xue}, Z.~K., {Liu}, J.~H., {Kong}, D.~F., \& {Xu}, C.~L. 2014,
	\apj, 797, 52
	
	\bibitem[{Yan} {et~al.}(2015)]{Yan2015}
	{Yan}, X.~L., {Xue}, Z.~K., {Pan}, G.~M., {et~al.} 2015, \apjs, 219, 17
	
	\bibitem[{Yan} {et~al.}(2018)]{Yan2018}
	{Yan}, X.~L., {Yang}, L.~H., {Xue}, Z.~K., {et~al.} 2018, \apjl, 853, L18
	
	\bibitem[{Yang} {et~al.}(2012)]{Yang2012}
	{Yang}, J., {Jiang}, Y., {Bi}, Y., {et~al.} 2012, \apj, 749, 12
	
	\bibitem[{Yang} {et~al.}(2016)]{Yang2016}
	{Yang}, K., {Guo}, Y., \& {Ding}, M.~D. 2016, \apj, 824, 148
	
	\bibitem[{Yashiro} {et~al.}(2006)]{Yashiro2006}
	{Yashiro}, S., {Akiyama}, S., {Gopalswamy}, N., \& {Howard}, R.~A. 2006, \apjl,
	650, L143
	
	\bibitem[{Yashiro} {et~al.}(2008)]{Yashiro2008}
	{Yashiro}, S., {Michalek}, G., \& {Gopalswamy}, N. 2008, Annales Geophysicae,
	26, 3103
	
	\bibitem[{Yeates} \& {Hornig}(2016)]{Yeates+Hornig2016}
	{Yeates}, A.~R., \& {Hornig}, G. 2016, \aap, 594, A98
	
	\bibitem[{Yeates} {et~al.}(2007)]{Yeates2007}
	{Yeates}, A.~R., {Mackay}, D.~H., \& {van Ballegooijen}, A.~A. 2007, \solphys,
	245, 87
	
	\bibitem[{Zhang} {et~al.}(2012{\natexlab{a}})]{Zhang2012}
	{Zhang}, J., {Cheng}, X., \& {Ding}, M.-D. 2012{\natexlab{a}}, Nature
	Communications, 3, id. 747
	
	\bibitem[{Zhang} {et~al.}(2001)]{Zhang2001}
	{Zhang}, J., {Dere}, K.~P., {Howard}, R.~A., {Kundu}, M.~R., \& {White}, S.~M.
	2001, \apj, 559, 452
	
	\bibitem[{Zhang} {et~al.}(2015)]{Zhang2015}
	{Zhang}, J., {Yang}, S.~H., \& {Li}, T. 2015, \aap, 580, A2
	
	\bibitem[Zhang {et~al.}(2014)]{Zhang2014}
	Zhang, Q., Liu, R., Wang, Y., {et~al.} 2014, The Astrophysical Journal, 789,
	133
	
	\bibitem[{Zhang} {et~al.}(2012{\natexlab{b}})]{ZhangQM2012}
	{Zhang}, Q.~M., {Chen}, P.~F., {Xia}, C., \& {Keppens}, R. 2012{\natexlab{b}},
	\aap, 542, A52
	
	\bibitem[{Zhang} \& {Ji}(2014)]{Zhang&Ji2014}
	{Zhang}, Q.~M., \& {Ji}, H.~S. 2014, \aap, 567, A11
	
	\bibitem[{Zhang} {et~al.}(2017)]{ZhangQM2017}
	{Zhang}, Q.~M., {Li}, D., \& {Ning}, Z.~J. 2017, \apj, 851, 47
	
	\bibitem[Zhang {et~al.}(2020)]{Zhang2020}
	Zhang, Q., Wang, Y., Liu, R., {et~al.} 2020, \apjl, submitted
	
	\bibitem[Zhao {et~al.}(2019)]{Zhao2019}
	Zhao, Y., Feng, H., Liu, Q., \& Zhao, G. 2019, Frontiers in Physics, 7, 151
	
	\bibitem[{Zhou} {et~al.}(2020{\natexlab{a}})]{ZhouY2020}
	{Zhou}, Y.~H., {Chen}, P.~F., {Hong}, J., \& {Fang}, C. 2020{\natexlab{a}},
	Nature Astronomy
	
	\bibitem[{Zhou} {et~al.}(2018)]{ZhouY2018}
	{Zhou}, Y.-H., {Xia}, C., {Keppens}, R., {Fang}, C., \& {Chen}, P.~F. 2018,
	\apj, 856, 179
	
	\bibitem[{Zhou} {et~al.}(2017{\natexlab{a}})]{ZhouY2017}
	{Zhou}, Y.-H., {Zhang}, L.-Y., {Ouyang}, Y., {Chen}, P.~F., \& {Fang}, C.
	2017{\natexlab{a}}, \apj, 839, 9
	
	\bibitem[{Zhou} {et~al.}(2019)]{ZhouZ2019}
	{Zhou}, Z., {Cheng}, X., {Zhang}, J., {et~al.} 2019, \apjl, 877, L28
	
	\bibitem[{Zhou} {et~al.}(2020{\natexlab{b}})]{ZhouZ2020}
	{Zhou}, Z., {Liu}, R., {Cheng}, X., {et~al.} 2020{\natexlab{b}}, \apj, 891, 180
	
	\bibitem[{Zhou} {et~al.}(2017{\natexlab{b}})]{ZhouZ2017}
	{Zhou}, Z., {Zhang}, J., {Wang}, Y., {Liu}, R., \& {Chintzoglou}, G.
	2017{\natexlab{b}}, \apj, 851, 133
	
	\bibitem[{Zhu} \& {Alexander}(2014)]{Zhu&Alexander2014}
	{Zhu}, C., \& {Alexander}, D. 2014, \solphys, 289, 279
	
	\bibitem[{Zhu} {et~al.}(2016)]{Zhu2016}
	{Zhu}, C., {Liu}, R., {Alexander}, D., \& {McAteer}, R.~T.~J. 2016, \apjl, 821,
	L29
	
	\bibitem[{Zhu} {et~al.}(2015)]{Zhu2015}
	{Zhu}, C., {Liu}, R., {Alexander}, D., {Sun}, X., \& {McAteer}, R.~T.~J. 2015,
	\apj, 813, 60
	
	\bibitem[{Zhu} {et~al.}(2017)]{Zhu2017}
	{Zhu}, X., {Wang}, H., {Cheng}, X., \& {Huang}, C. 2017, \apjl, 844, L20
	
	\bibitem[{Zirker} {et~al.}(1998)]{Zirker1998}
	{Zirker}, J.~B., {Engvold}, O., \& {Martin}, S.~F. 1998, \nat, 396, 440
	
	\bibitem[{Zou} {et~al.}(2019{\natexlab{a}})]{Zou2019twostep}
	{Zou}, P., {Jiang}, C., {Feng}, X., {et~al.} 2019{\natexlab{a}}, \apj, 870, 97
	
	\bibitem[{Zou} {et~al.}(2019{\natexlab{b}})]{Zou2019}
	{Zou}, P., {Jiang}, C., {Wei}, F., {Zuo}, P., \& {Wang}, Y. 2019{\natexlab{b}},
	\apj, 884, 157
	
	\bibitem[Zuccarello {et~al.}(2016)]{Zuccarello2016}
	Zuccarello, F., Aulanier, G., \& Gilchrist, S. 2016, The Astrophysical journal
	letters, 821, L23
	
	\bibitem[{Zuccarello} {et~al.}(2015)]{Zuccarello2015}
	{Zuccarello}, F.~P., {Aulanier}, G., \& {Gilchrist}, S.~A. 2015, \apj, 814, 126
	
	\bibitem[{Zuccarello} {et~al.}(2014)]{Zuccarello2014}
	{Zuccarello}, F.~P., {Seaton}, D.~B., {Mierla}, M., {et~al.} 2014, \apj, 785,
	88
	
	\bibitem[{Zuccarello} {et~al.}(2009)]{Zuccarello2009}
	{Zuccarello}, F., {Romano}, P., {Farnik}, F., {et~al.} 2009, \aap, 493, 629
	
\end{thebibliography}

\end{document}